\documentclass[onecolumn]{aastex63}

\usepackage{times}
\usepackage{amsmath,amssymb,amstext}
\hypersetup{linkcolor=red,citecolor=blue,filecolor=cyan,urlcolor=blue}
\usepackage{tabularx}
\usepackage{longtable}
\usepackage{float}
\usepackage{natbib}
\usepackage{graphicx,color}
\usepackage{txfonts}

\usepackage{multirow}
\usepackage{array}
\usepackage{rotating}
\usepackage{sidecap}
\usepackage{hyperref}
\usepackage{epstopdf}
\usepackage{footnote}
\usepackage{tabularx}
\usepackage{booktabs}
\usepackage{longtable}
\usepackage{tabu}
\usepackage{longtable}

\newcommand{\kms}{km\,s$^{-1}$}

\begin{document}

\title{{\large{\bf From Pseudostreamer Jets to CMEs: Observations of the Breakout Continuum}}}
\author{Pankaj Kumar\altaffiliation{1,2}}

\affiliation{Department of Physics, American University, Washington, DC 20016, USA}
\affiliation{Heliophysics Science Division, NASA Goddard Space Flight Center, Greenbelt, MD, 20771, USA}

\author{Judith T.\ Karpen}
\affiliation{Heliophysics Science Division, NASA Goddard Space Flight Center, Greenbelt, MD, 20771, USA}

\author{Spiro K.\ Antiochos}
\affiliation{Heliophysics Science Division, NASA Goddard Space Flight Center, Greenbelt, MD, 20771, USA}

\author{Peter F.\ Wyper}
\affiliation{Department of Mathematical Sciences, Durham University, Durham DH1 3LE, UK}

\author{C.\ Richard DeVore}
\affiliation{Heliophysics Science Division, NASA Goddard Space Flight Center, Greenbelt, MD, 20771, USA}

\author{Benjamin J.\ Lynch}
\affiliation{Space Sciences Laboratory, University of California, Berkeley, CA 94720, USA}

\email{pankaj.kumar@nasa.gov}

%*****************************************************************************
\begin{abstract}
  The magnetic breakout model, in which reconnection in the corona leads to destabilization of a filament channel, explains numerous features of eruptive solar events, from small-scale jets to global-scale coronal mass ejections (CMEs). The underlying multipolar topology, pre-eruption activities, and sequence of magnetic-reconnection onsets (first breakout, then flare) of many observed fast CMEs/eruptive flares are fully consistent with the model. Recently, we have demonstrated that most observed coronal-hole jets in fan/spine topologies also are induced by breakout reconnection at the null point above a filament channel (with or without a filament). For these two types of eruptions occurring in similar topologies, the key question is, why do some events generate jets while others form CMEs? We focused on the initiation of eruptions in large bright points/small active regions that were located in coronal holes and clearly exhibited null-point (fan/spine) topologies: such configurations are referred to as pseudostreamers. We analyzed and compared \emph{SDO}/AIA, \emph{SOHO}/LASCO, and \emph{RHESSI} observations of three events. Our analysis of the events revealed two new observable signatures of breakout reconnection prior to the explosive jet/CME outflows and flare onset: coronal dimming and the opening-up of field lines above the breakout current sheet. Most key properties were similar among the selected erupting structures, thereby eliminating region size, photospheric field strength, magnetic configuration, and pre-eruptive evolution as discriminating factors between jets and CMEs. We consider the factors that contribute to the different types of dynamic behavior, and conclude that the main determining factor is the ratio of the magnetic free energy associated with the filament channel compared to the energy associated with the overlying flux inside and outside the pseudostreamer dome.
\end{abstract}
%*****************************************************************************

%*****************************************************************************
%%%%%% Section 1 %%%%%%%%%%%%%%%%%%%%%%%%%%%%%%%%%%%%%%%%%%%%%%%%%%%%%%%%%%
\section{INTRODUCTION}\label{intro}

Coronal mass ejections (CMEs) and coronal jets are two of the best studied forms of solar eruptions. CMEs and jets seem to be quite different physically. In coronagraph images, CMEs typically appear as large, bright, magnetic flux ropes expelled through and beyond the corona \citep[e.g., ][]{rust2001,gopalswamy2006,xie2013,chen2020}; in the heliosphere, they appear in in-situ observations as so-called magnetic clouds \citep[e.g., ][]{burlaga1984,marubashi1986,lepping1990,bothmer1998,demoulin2008,gulisano2010,li2018}.  CMEs are known to inject a large amount of new unsigned flux into the heliosphere, in some cases up to $10^{21}$ Mx \citep{lynch2005}. 
 Coronal jets, on the other hand, apparently consist of a stream of plasma traveling outward from the corona along open (or remotely closed) field lines
\citep[e.g.,][]{shibata1992,cirtain2007,savcheva2007,patsourakos2008}. Jets are neither observed nor expected to add new magnetic flux to the heliosphere, only new plasma. Both CMEs and jets contribute helicity to the heliosphere, however.

CMEs and jets were long believed to have very different physical origins, with CMEs due to the eruption of the strongly sheared field of a filament channel  \citep[see reviews by][]{klimchuk2001,forbes2006} and jets due to ``interchange'' reconnection between open and closed flux \citep[e.g.,][]{shibata1996,pariat2009}. Filaments form within filament channels, which are narrow zones of high magnetic shear located along polarity inversion lines (PILs) \citep{martin1990,martin1998}. Filament-channel field lines are nearly aligned to PILs, as indicated by the orientation of dark fibrils seen in H$\alpha$ or 304 \AA\ observations \citep{gaizauskas1997,wang2007a}. We use the term ``filament channel" to represent the entire magnetic structure rooted within the channel, not just the surface manifestations, to facilitate physically meaningful comparisons between observed and simulated eruptions. In recent years, however, observations and simulations have established a much closer physical connection between CMEs and jets.  High-resolution images, especially from \emph{SDO}, have shown that most jets are due to the eruption of the stressed magnetic flux in filament channels \citep{sterling2015,kumar2019a}. Although filament channels in jet sources are orders of magnitude smaller than for a large CME, the underlying physics of the eruption seems identical \citep{wyper2017}. Given this newly discovered correspondence between CMEs and jets, the question arises as to what determines whether an eruption will be a CME or a jet.

\begin{figure*}
\centering{
\includegraphics[width=18cm]{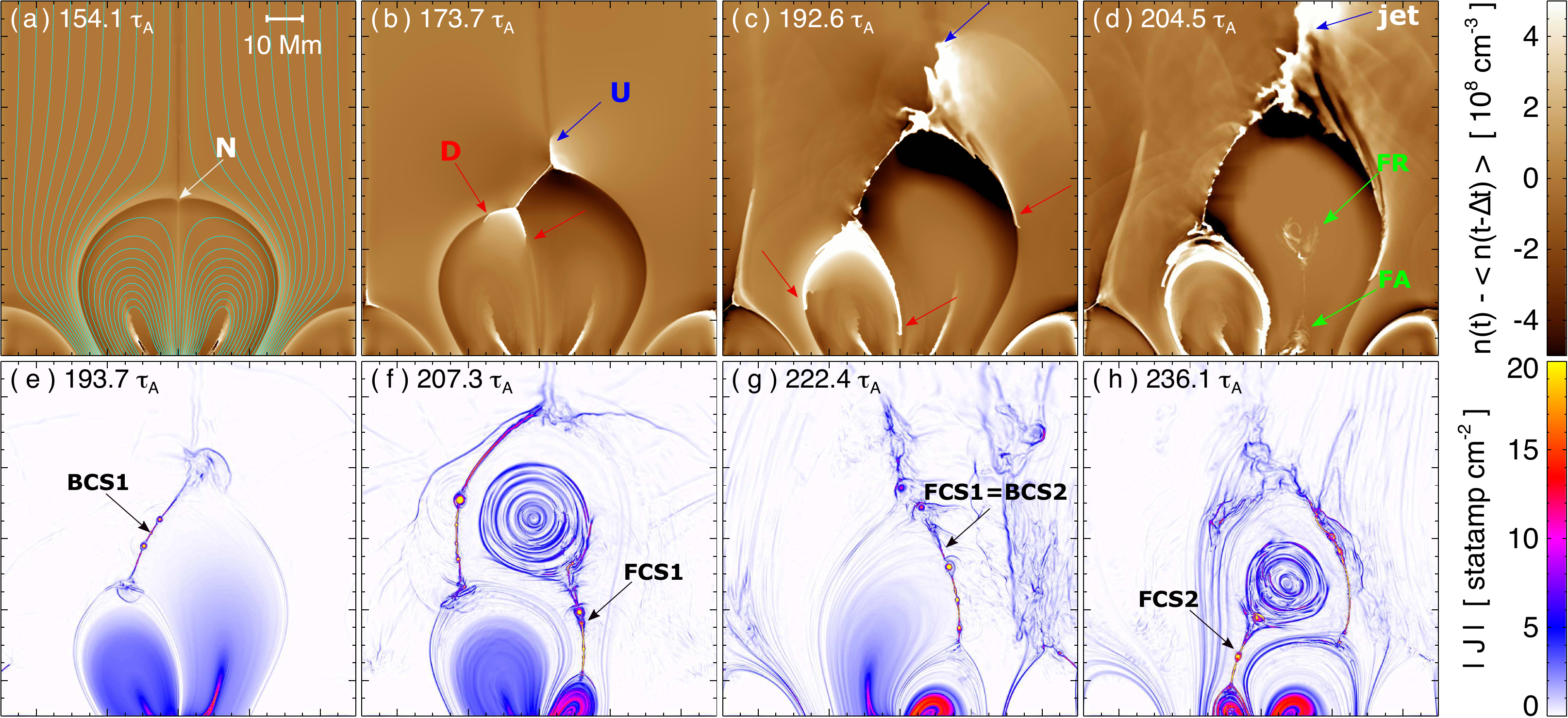}
}
 \caption{Simulated evolution of sympathetic breakout eruptions from a pseudostreamer. (a-d) Differenced plasma number density during the pre-eruption phase.  Panel (a) is a difference image created by subtracting a smoothed version of itself; the subsequent panels (b-d) are difference images made by subtracting the prior image (a-c, respectively) smoothed in the same way as panel (a). Selected magnetic field lines are superposed on panel (a). N = null, FR = flux rope, and FA = flare arcade. Blue (red) arrows indicate the upflow U (downflow D) signatures of breakout reconnection. (e-h) Current density magnitude, $\left\vert J \right\vert$. The simulation times correspond to: (e) reconnection at the first breakout current sheet (BCS1) after  the shearing flows are turned off; (f) the impulsive phase of reconnection at the first eruptive flare current sheet (FCS1) and associated flux-rope formation; (g) during the interim phase when the first flare current sheet acts as the breakout current sheet for the second eruption (FCS1 = BCS2); and (h) the impulsive phase of reconnection at the second flare current sheet (FCS2) and flux-rope formation. \citep[Adapted from][]{lynch2013,lynch2016}.} 
\label{fig1}
\end{figure*}

We address this question observationally by studying three intermediate-scale events at the jet-to-CME transition. The source regions for these events are so-called pseudostreamers \citep{wang2007}, which are common, distinctive structures appearing in coronagraph images as bright stalks emanating from large closed regions inside unipolar coronal holes (CHs) \citep{hundhausen1972,zhao2003}. The pseudostreamer topology consists of one or more 3D nulls and a separatrix dome above a minority-polarity intrusion (Fig.\ \ref{fig1}(a)), which separates coronal holes of the majority polarity \citep{wang2007}.
 This is the well-known embedded-bipole topology \citep{antiochos1990,antiochos2011,titov2011}. 
Pseudostreamers can inject mass and energy into the solar wind both quiescently and through explosive eruptions \citep{wang2007,wang2012,panasenco2013,wang2019}. Smaller pseudostreamers are rooted in bright points and generate coronal jets \citep{sterling2015,kumar2018,kumar2019a}; larger pseudostreamers are rooted in active regions (ARs) and can generate jets, slow CMEs, fast CMEs/eruptive flares, and combinations thereof \citep{panasenco2013,wang2015,yang2015,kumar2017,kumar2018,kumar2019a,kumar2019b}. This continuum of activity strongly suggests that there is a universal eruption mechanism, but it also poses a puzzle: why do some pseudostreamers yield jets while others yield CMEs? As we will demonstrate in this paper, size alone is not the determining factor.

The embedded-bipole topology that is common to all three pseudostreamers in our study is perfectly suited to the magnetic breakout mechanism for solar eruptions \citep{antiochos1999,wyper2017,wyper2018,kumar2018,masson2019}. To illustrate the model and set the framework for the observations described below, results from a high-resolution, translationally symmetric magnetohydrodynamic (MHD) simulation of sympathetic breakout CMEs from a pseudostreamer \citep{lynch2013,lynch2016} are shown in Figure \ref{fig1}. The system was energized by successively imposing shear flows at the footpoints of the pseudostreamer arcades: simultaneously at the beginning, then only in the right arcade, and finally only in the left.  The pseudostreamer expands gradually in response to the slow accumulation of magnetic energy, producing a breakout current sheet (BCS1) above the right arcade (Fig.\ \ref{fig1}(e,f)). The first eruption is initiated as the onset of interchange reconnection across BCS1 forms multiple plasmoids and transfers the restraining flux overlying the right arcade to both the left arcade and the background open field \citep{antiochos1999}. As the expansion speeds up, a flare current sheet (FCS1) forms within the right arcade and builds a circular flux rope through initially slow reconnection (Fig.\ \ref{fig1}(f,g)). The onset of explosive flare reconnection generates multiple plasmoids in FCS1, accelerates the rising flux rope (FR), and forms a flare arcade (FA) below it (Fig.\ \ref{fig1}(d)). Up to this point, the eruption has proceeded as does any single breakout event. However, the addition of magnetic shear to the left arcade enables further energy buildup there, and the pseudostreamer topology supports its eventual explosive release in a second, sympathetic eruption. After the flux rope is ejected, the flare current sheet of the first eruption in the right arcade (FCS1) assumes a new role as the breakout current sheet for the left arcade (BCS2) (Fig.\ \ref{fig1}(g)). The ongoing (now breakout) reconnection across BCS2 quickly produces a second eruption, when explosive reconnection sets in across the newly formed flare current sheet FCS2 (Fig.\ \ref{fig1}(h)),  as in the first eruption.

The breakout model predicts distinctive pre-eruption dynamics and associated observational signatures that should precede the onset of explosive flare reconnection in pseudostreamers. For example, bidirectional upflow and downflow (U/D) breakout-reconnection jets are expected along both the fan surface and the inner and outer spines (Fig.\ \ref{fig1}(b,c)). These flows and associated beams of energetic electrons could form bright ribbons at the chromospheric footprint of the magnetic separatrix, as observed previously \citep[e.g.,][]{masson2009,wang2012,lee2020}. As the null point evolves into a breakout current sheet (BCS) during the expansion of the pseudostreamer, field lines near the null should begin to open through slow breakout reconnection there (Fig.\ \ref{fig1}(b,c)). The combination of pseudostreamer expansion and the reconnection-jet outflows depletes the mass density at and above the BCS, which should show up as coronal dimming at the outer edge of the expanding flux system (Fig.\ \ref{fig1}(c,d)). In this paper, we report the first direct observations of these field-opening and coronal-dimming signatures of breakout reconnection prior to pseudostreamer eruptions. 

Coronal rain has been detected recently in small pseudostreamers rooted in decaying active regions near CH boundaries \citep{mason2019}. These persistent episodes of rain are not associated with previous flares or eruptions, in contrast to the events studied by \citet{liu2012} and \citet{panasenco2019}. The rain plasma initially appears near the null and drains back along the separatrix surface and enclosed fan loops, serving as a good proxy for the magnetic topology.  The connection between pseudostreamers and coronal rain is intriguing because the dense, cool rain might be produced by repeated interchange reconnection at the null \citep{mason2019}.  We have also observed coronal rain 3 hours before a coronal-hole jet; the rain subsided shortly before the onset of explosive flare/breakout reconnection \citep{kumar2019b}. The coronal rain observed prior to the events discussed in the present work, coupled with our detailed analysis of the eruption-associated reconnection signatures, sheds light on the underlying physical processes common to both coronal rain and pseudostreamer eruption.

In our first event (E1) at the limb, faint narrow jets preceded a slow narrow CME. The second event (E2), also at the limb, consisted of a pair of sympathetic fast eruptions from an AR near the source of event E1. The fast CMEs originated in adjacent portions of the AR filament channel and traveled with speeds that differed by a factor of two. Our final event (E3), a single fast CME, came from an AR within a low-latitude CH near disk center. All three source regions had initial magnetic structures of comparable width and peak magnetic-field strengths of about 1 kG. The eruptions were accompanied by modest-strength C-class flares, and preceded by faint narrow jets 1 to 3 hr before the onset of explosive ejection. By selecting eruptions from pseudostreamers whose key properties were comparable, we eliminated source size, peak photospheric field strength, magnetic environment, and pre-eruptive evolution as factors discriminating between the sources of jets and CMEs. As discussed below, we conclude that the ratio of the magnetic energy in the filament channel to the magnetic energy above/outside the channel (both inside and outside the separatrix) is the most likely determinant of whether the flux rope is destroyed or is partially or fully ejected when it encounters the breakout sheet. In the first case, a jet would be generated; in the second, a jet-CME hybrid or a ``pure" CME would be produced.

%%%%%%%%%%%%%%%%%%%%%%%%%%%%%%%%%%%%%%%%%%%%%%%%%%%%%%%%%%%%%%%%%%%%%

%%%%%%%%%%%%%%%%%%%%%%%%%%%%%%%%%%%%%%%%%%%%%%%%%%%%%%%%%%%%%%%%%%%%%%

%%%%%%%%%%%%%%%%%%%%%%%%%%%%%%%%%%%%%%%%%%%%%%%%%%%%%%%%%%%%%%%%%%%%%
\begin{figure*}
\centering{
\includegraphics[width=15cm]{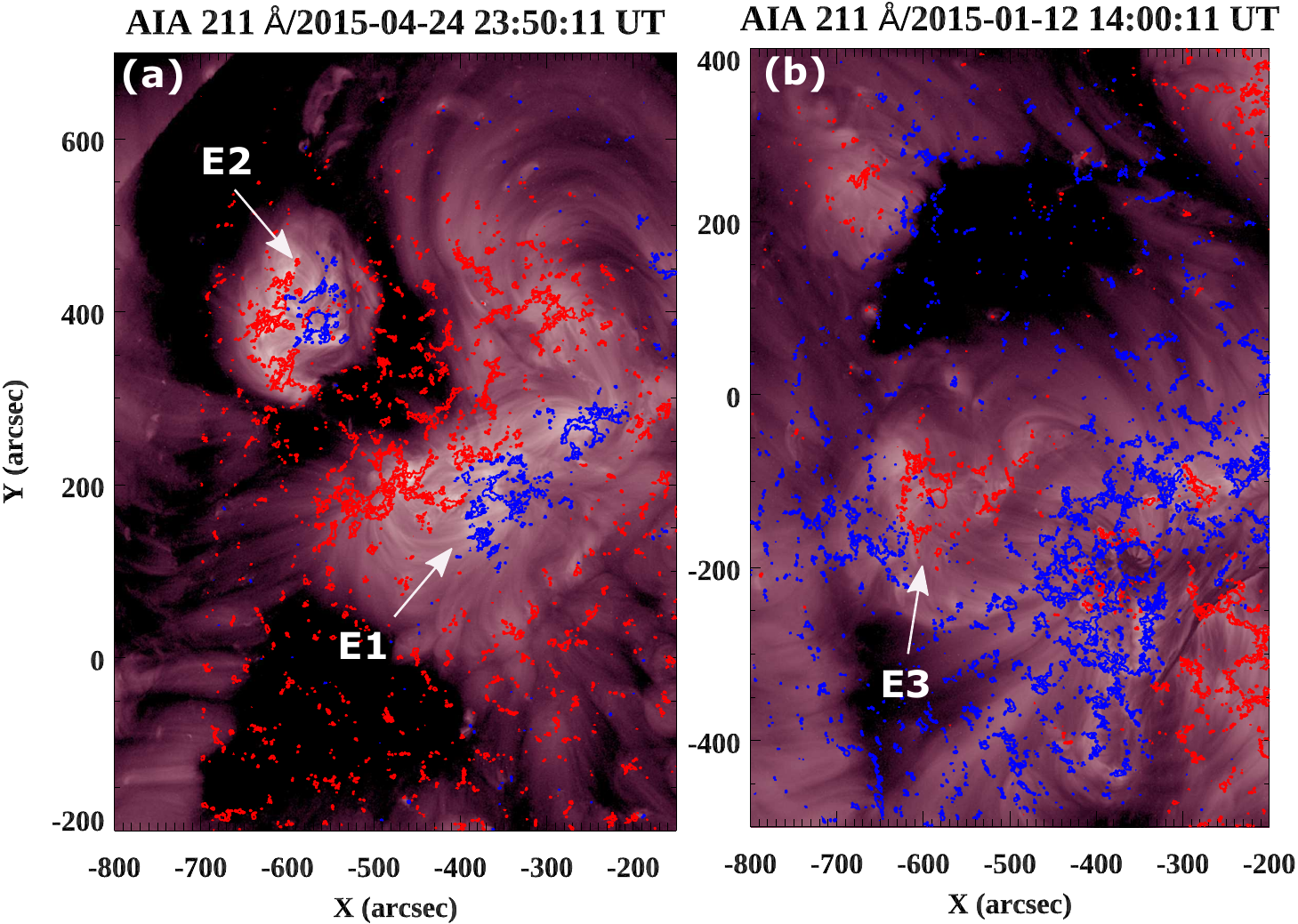}
}
\caption{AIA 211 \AA\ images overlaid by the cotemporal HMI magnetogram contours of positive (red) and negative (blue) polarities (levels = $\pm100$ G)  showing the locations and photospheric flux distributions of the source regions for events E1, E2, and E3. CME source regions for E1 and E2 are shown four days after eruption; the E3 source is shown shortly before eruption.} 
\label{fig2}
\end{figure*}

%%%%%%%%%%%%%%%%%%%%%%%%%%%%%%%%%%%%%%%%%%%%%%%%%%%%%%%%%%%%%%%%%%%%%%%%%%%%%%%%%%%%%%%%%%%%%%%%%%%%%%%%%%%%%%%%%%%%%%%%%%%%%%%%%%%%%
\begin{figure*}
\centering{
\includegraphics[width=5.9cm]{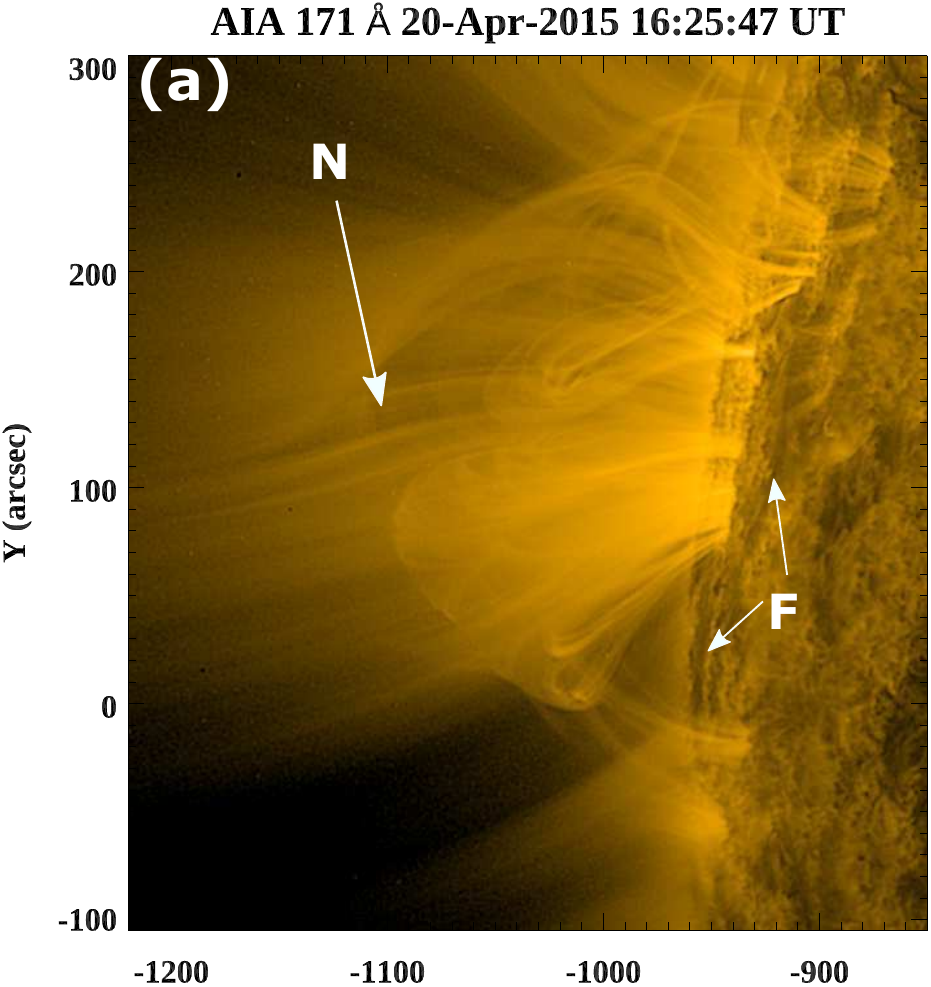}
\includegraphics[width=11.5cm]{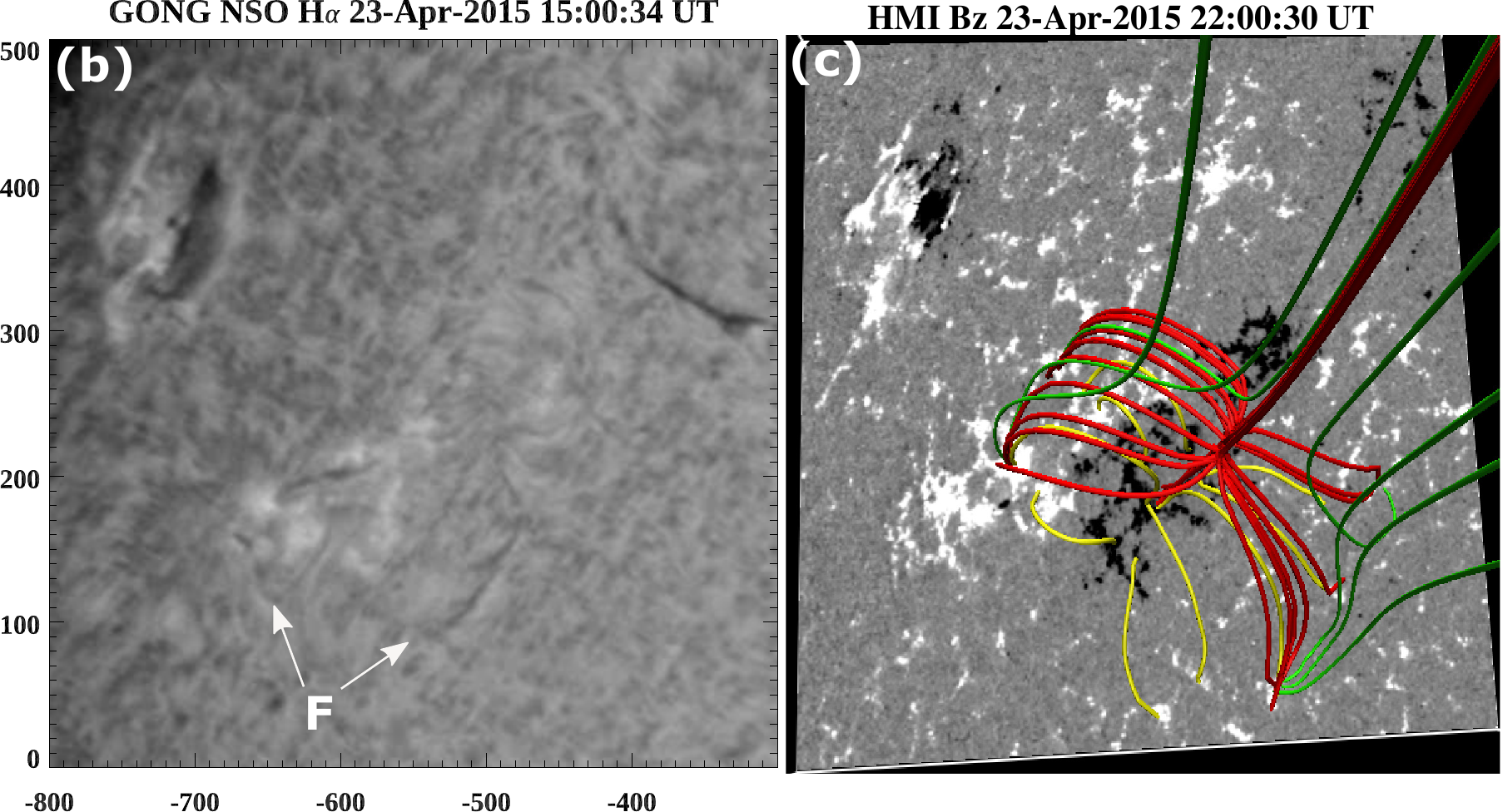}

\includegraphics[width=6.4cm]{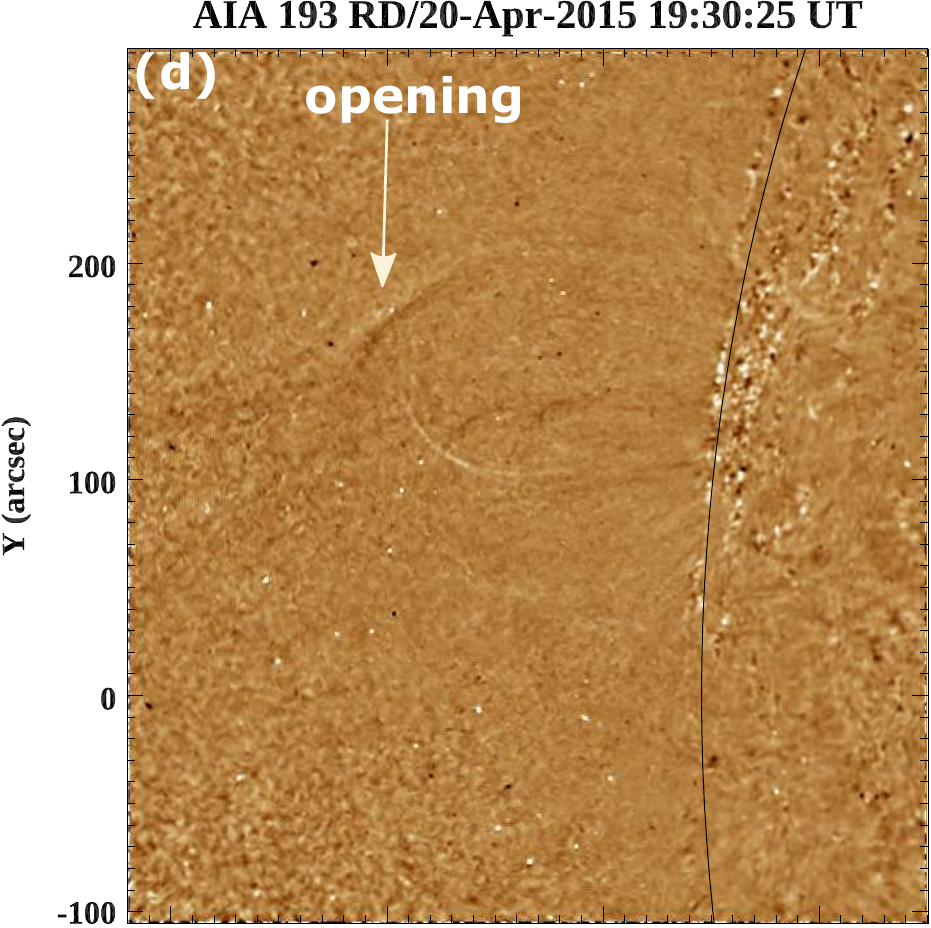}
\includegraphics[width=5.5cm]{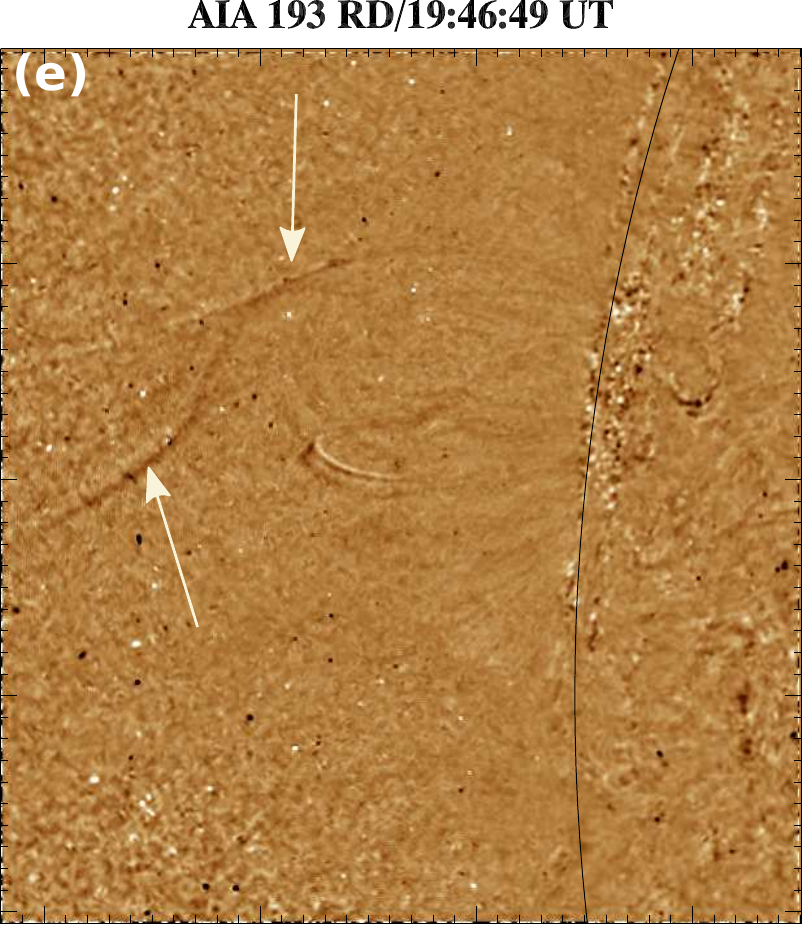}
\includegraphics[width=5.5cm]{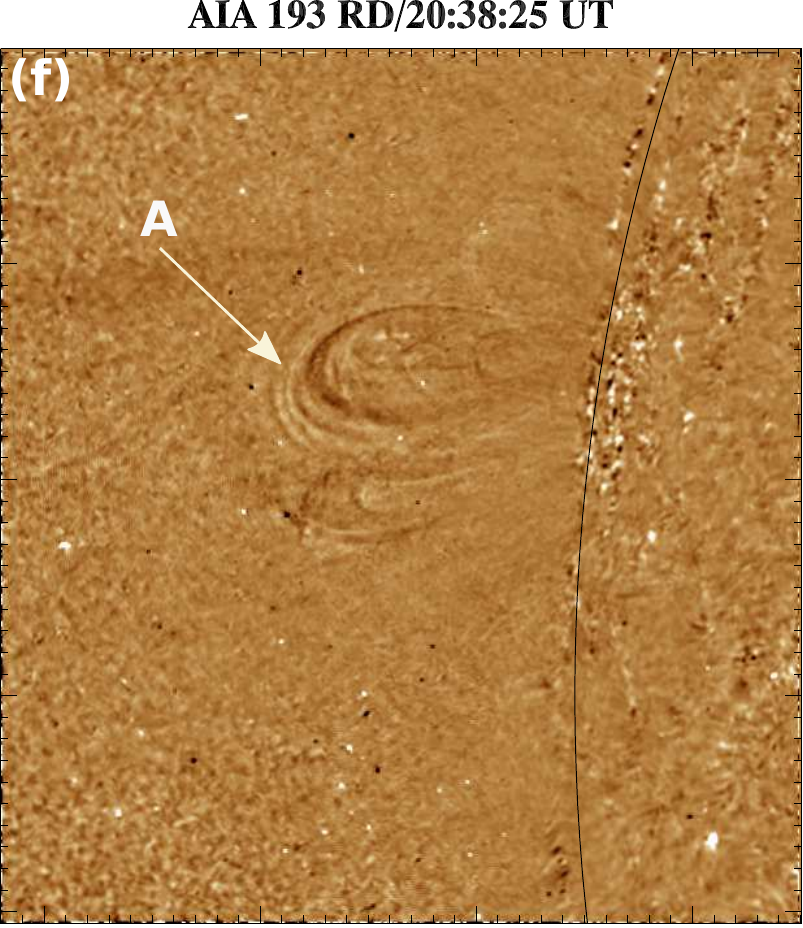}

\includegraphics[width=6.4cm]{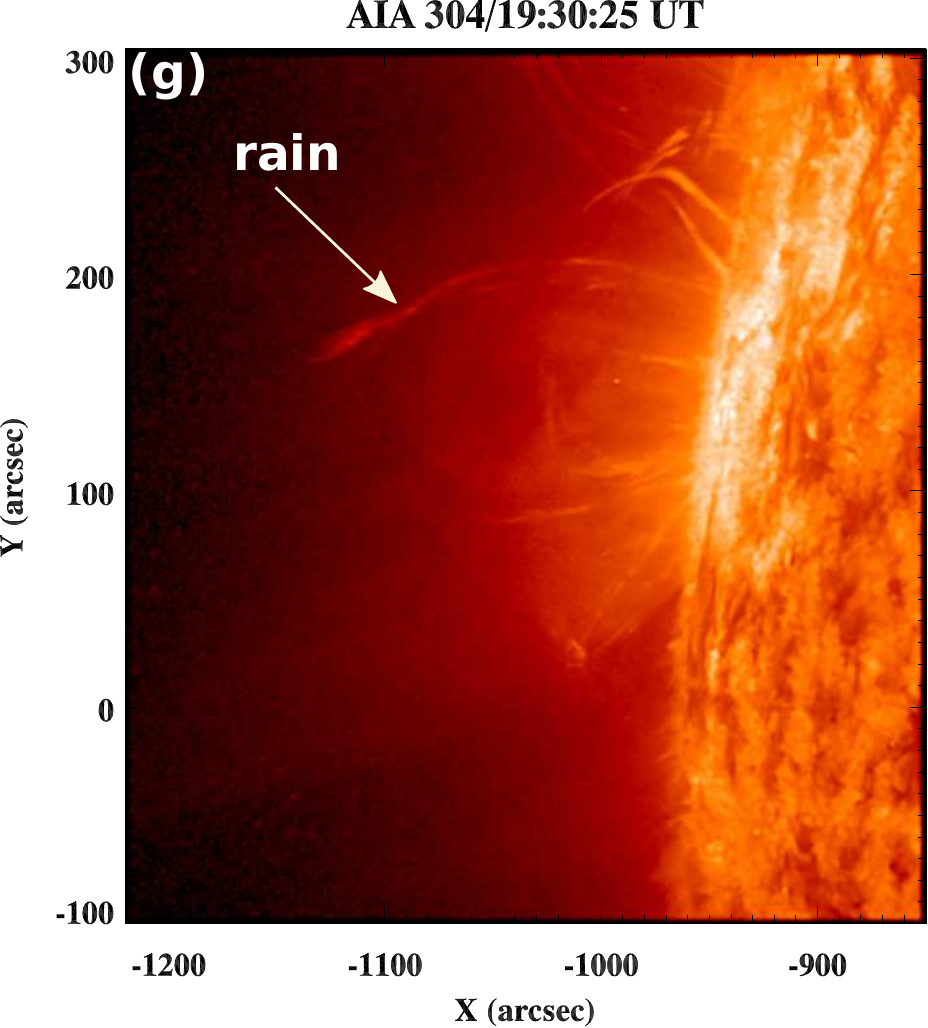}
\includegraphics[width=5.5cm]{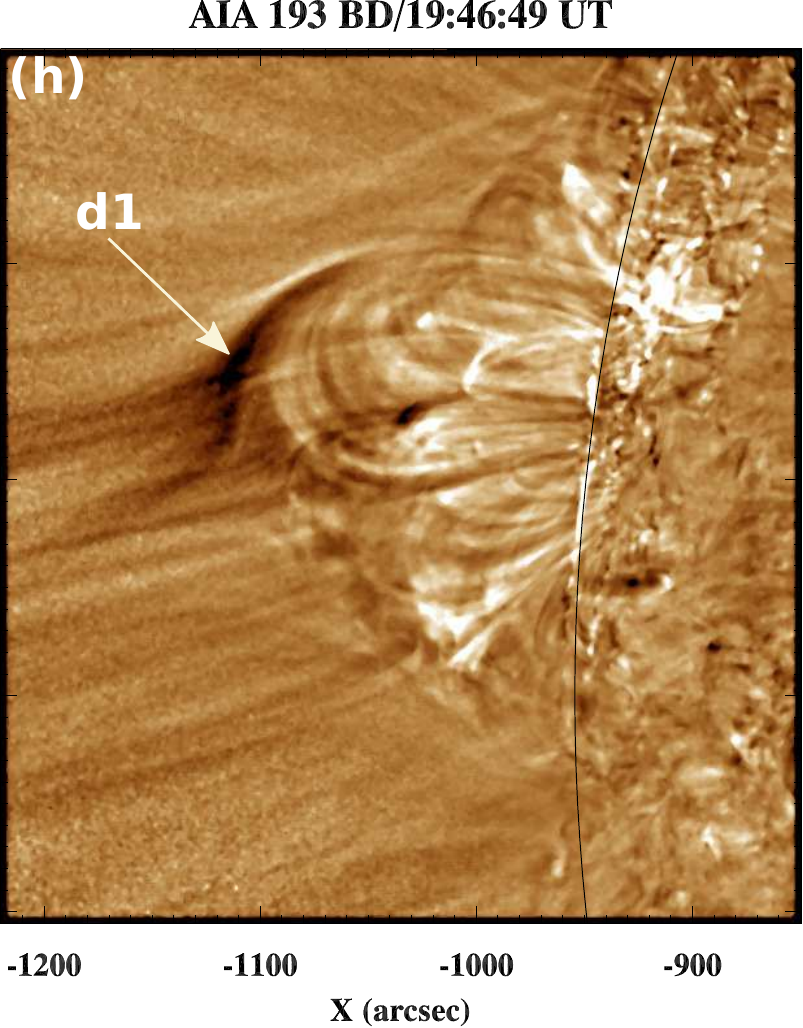}
\includegraphics[width=5.5cm]{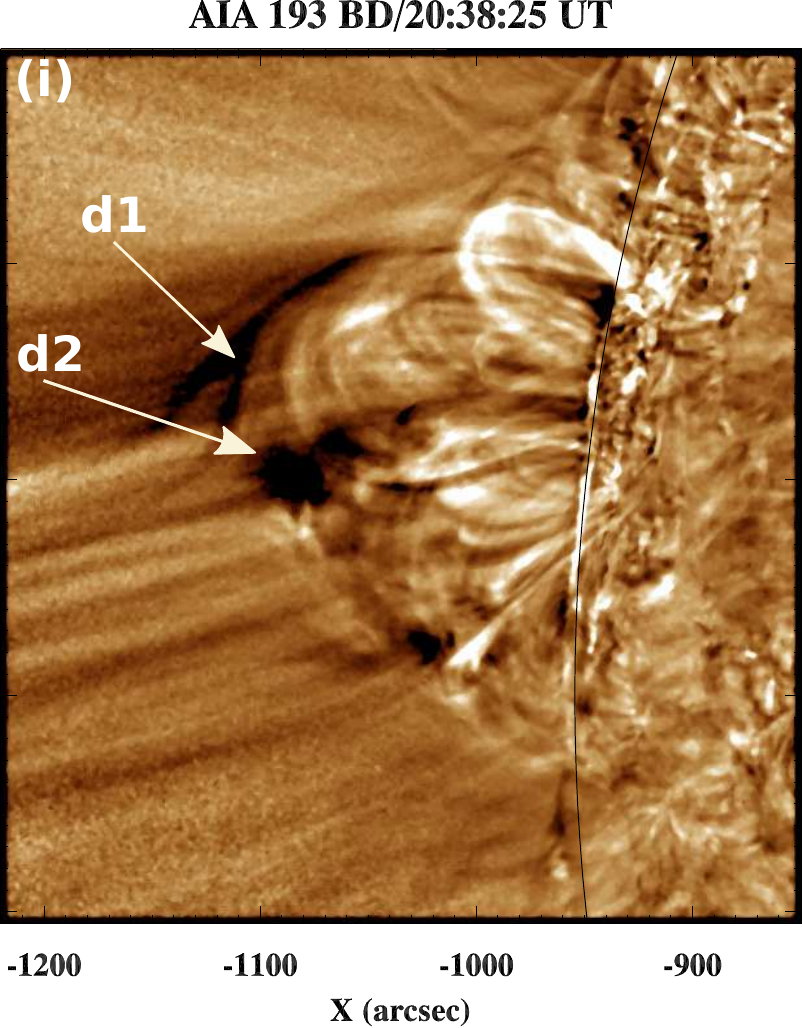}

}
\caption{(a) AIA 171 \AA\ image of the source region for event E1 on 2015 April 20. N = approximate location of the null, F=filament. (b)  H$\alpha$ image showing filament F on 2015 April 23. (c) Potential-field extrapolation from the source region at 22:00:30 UT on 2015 April 23. (d-f) AIA 193 \AA\ running-difference images ($\Delta t = 1$ min) on 2015 April 20 showing pre-eruption opening; A is a rising arc-shaped structure. (g) AIA 304 \AA\ image showing coronal rain. (h-i) AIA 193 \AA\ base-difference images; d1 and d2 are dimming regions. Note that the panels in the middle and bottom rows are co-temporal. (An animation of this Figure is available online.)} 
\label{fig3}
\end{figure*}

%%%%%%%%%%%%%%%%%%%%%%%%%%%%%%%%%%%%%%%%%%%%%%%%%%%%%%%%%%%%%%%%%%%%%%
\begin{figure*}
\centering{
\includegraphics[width=8.5cm]{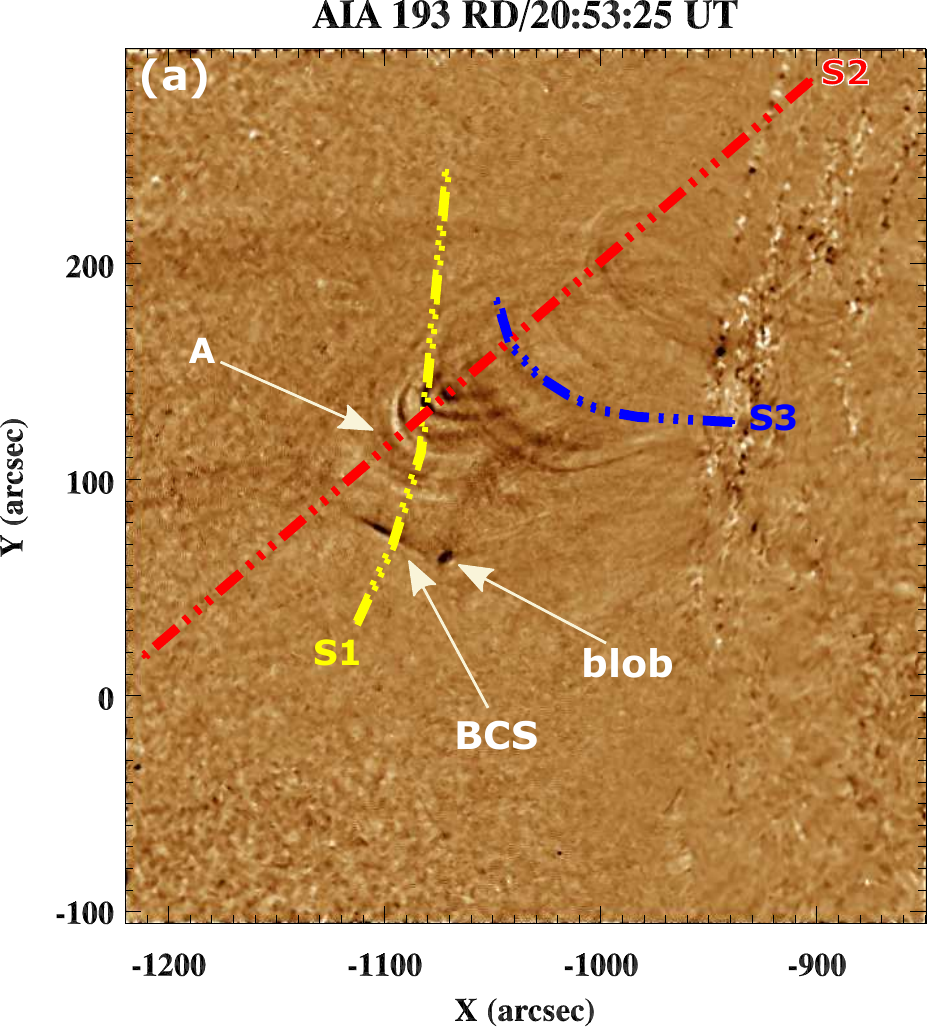}
\includegraphics[width=7.5cm]{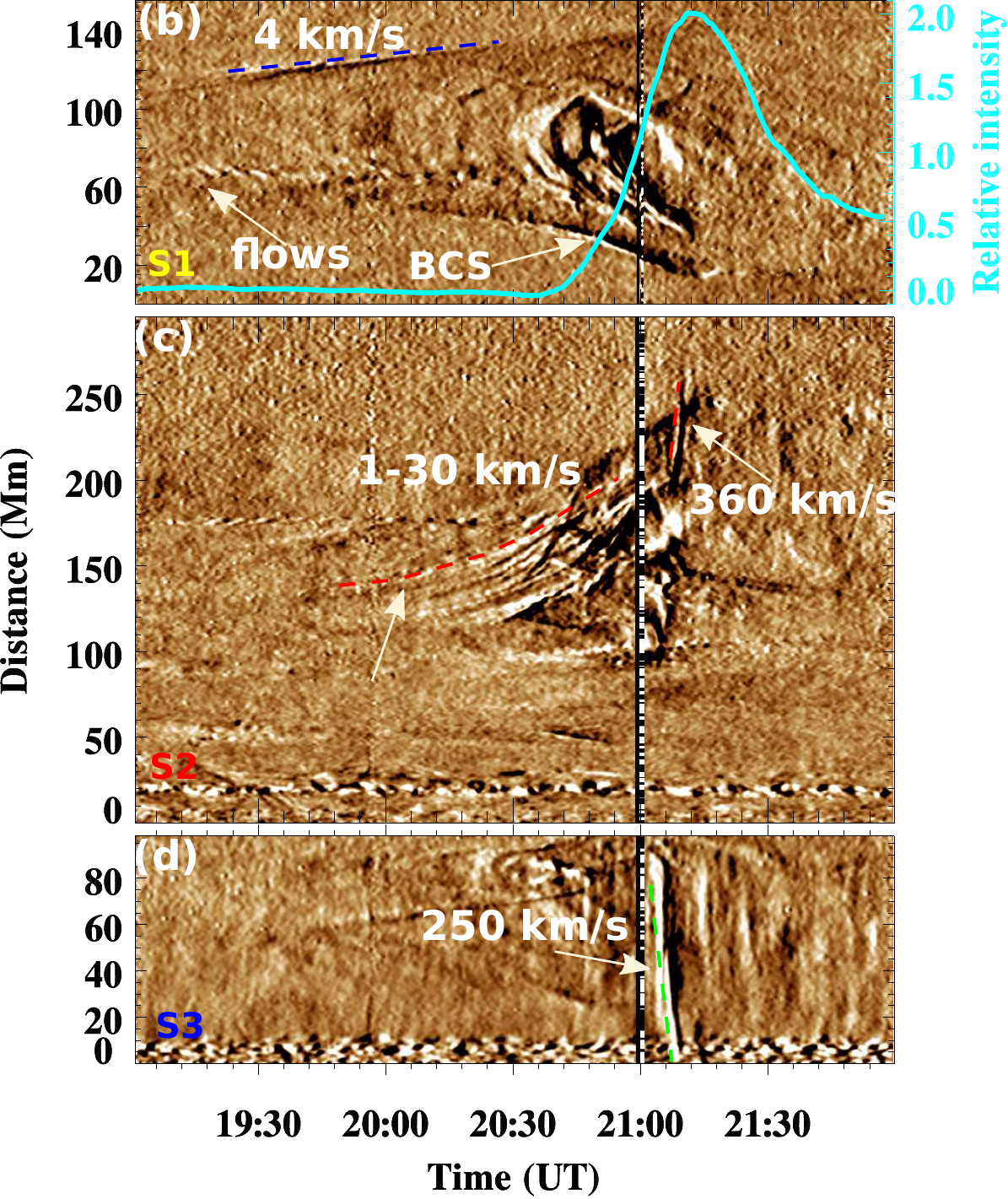}

\includegraphics[width=15cm]{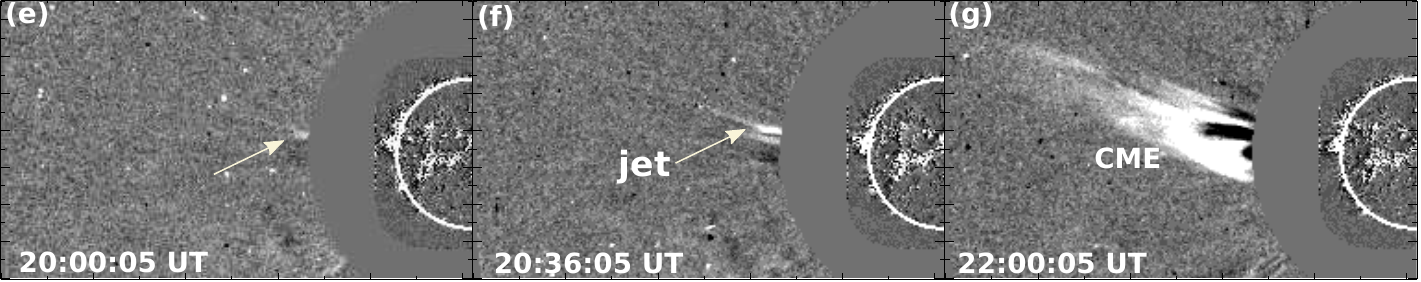}
}
\caption{(a) AIA 193 \AA\ running-difference images ($\Delta t = 1$ min) during the E1 eruption. BCS = breakout current sheet. (b-d) Time-distance running-difference intensity plots along slices S1, S2, and S3 shown in panel a. The vertical dashed line indicates the onset of explosive breakout reconnection. The cyan curve in panel b is the AIA 131 \AA\ relative intensity profile extracted from the box marked by a red dotted line in Figure \ref{fig5}c. (e-g) LASCO C2 coronagraph images showing the pre-eruption jet and the CME. (An animation of this Figure is available online.)} 
\label{fig4}
\end{figure*}

%%%%%%%%%%%%%%%%%%%%%%%%%%%%%%%%%%%%%%%%%%%%%%%%%%%%%%%%%%%%%%%%%%%%%%
%%%%%%%%%%%%%%%%%%%%%%%%%%%%%%%%%%%%%%%%%%%%%%%%%%%%%%%%%%%%%%%%%%%%%%
\begin{figure*}
\centering{
\includegraphics[width=5.05cm]{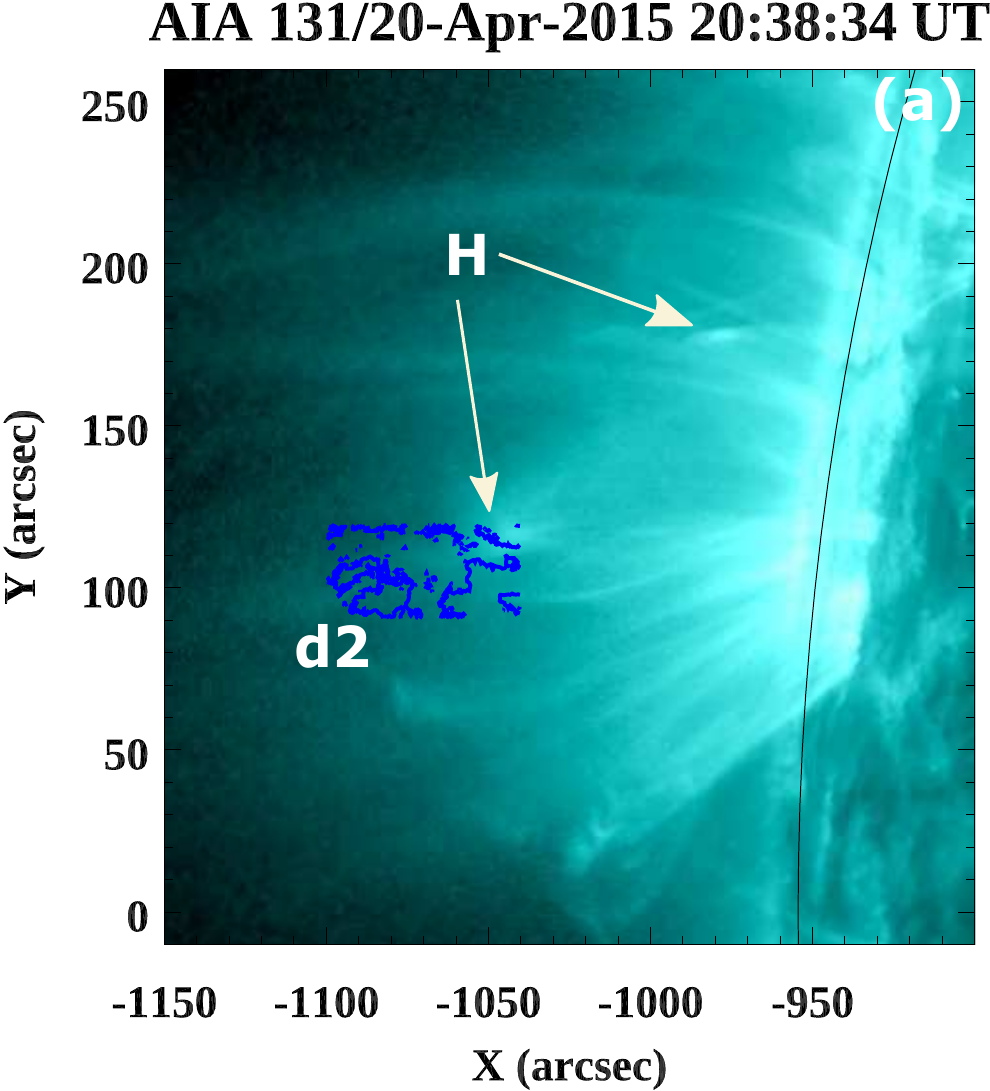}
\includegraphics[width=4.1cm]{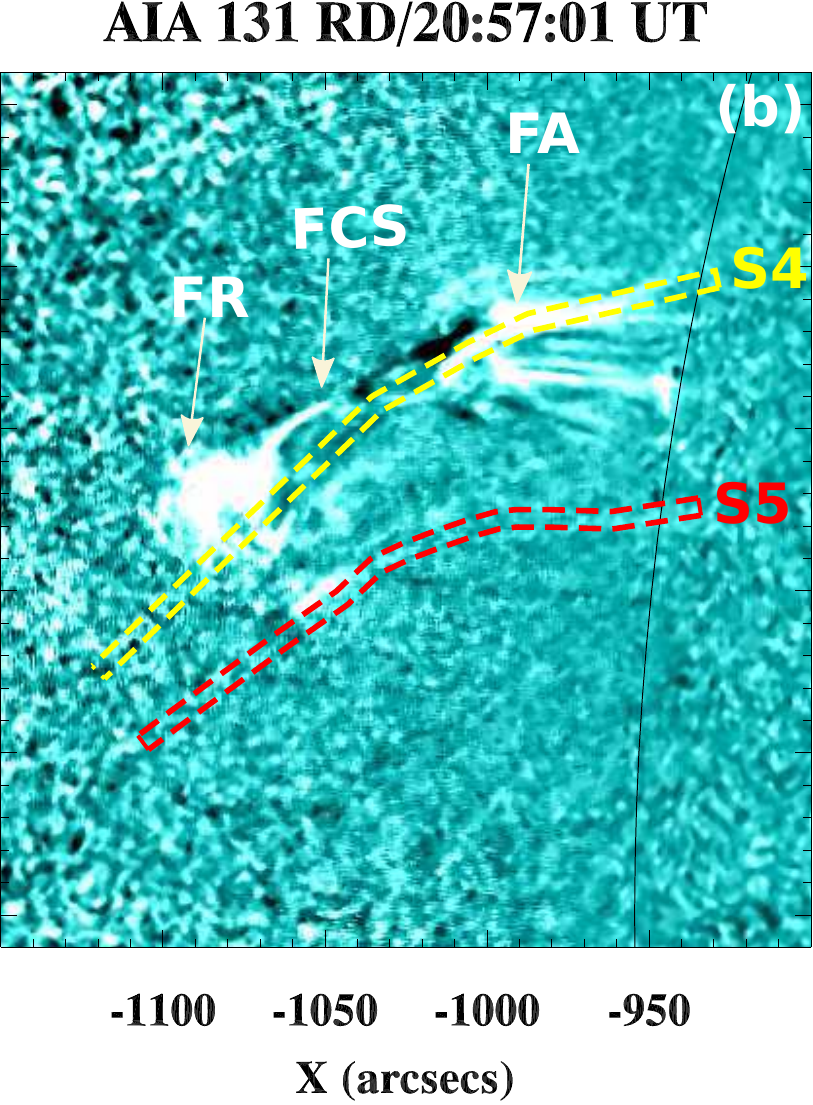}
\includegraphics[width=4.1cm]{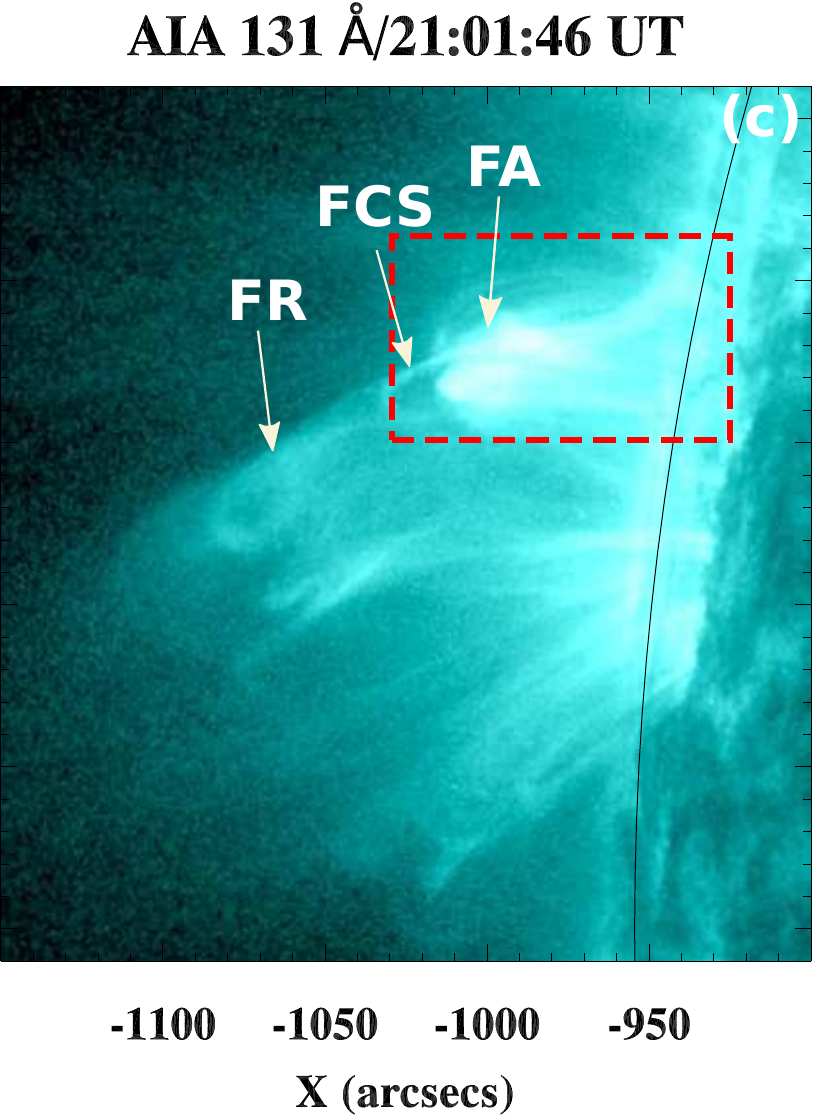}
\includegraphics[width=4.1cm]{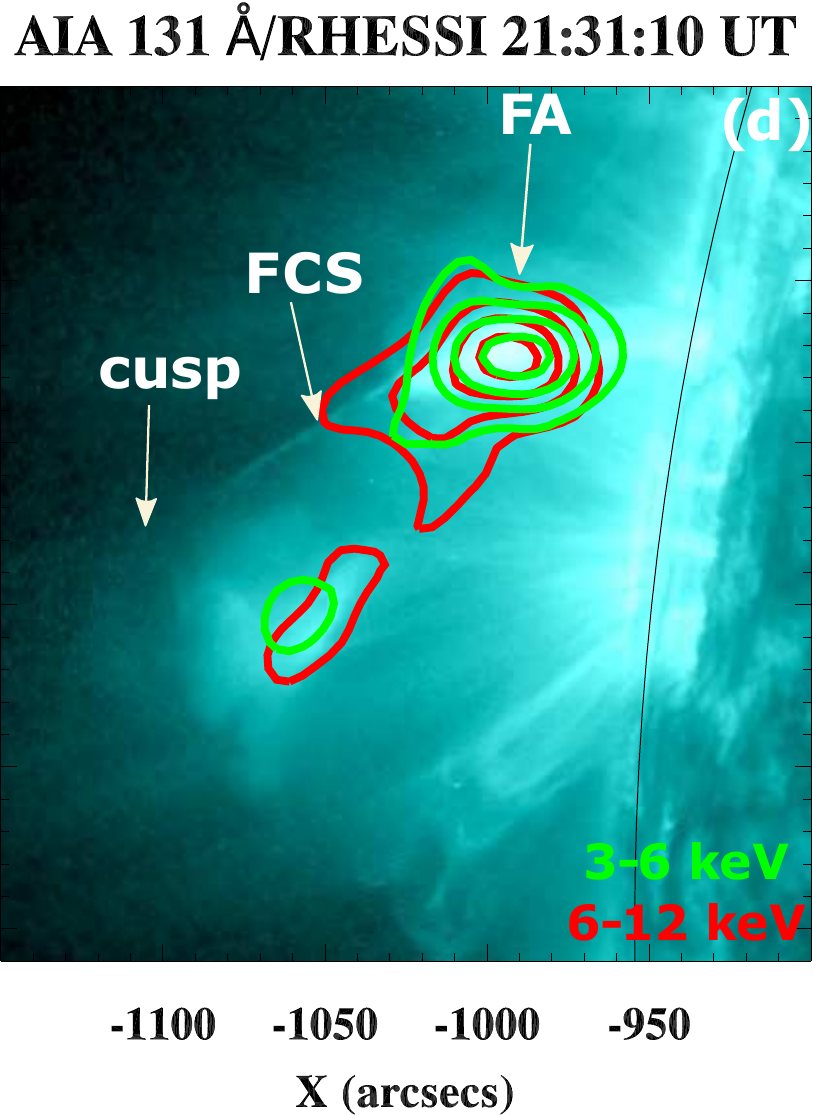}

\includegraphics[width=8.3cm]{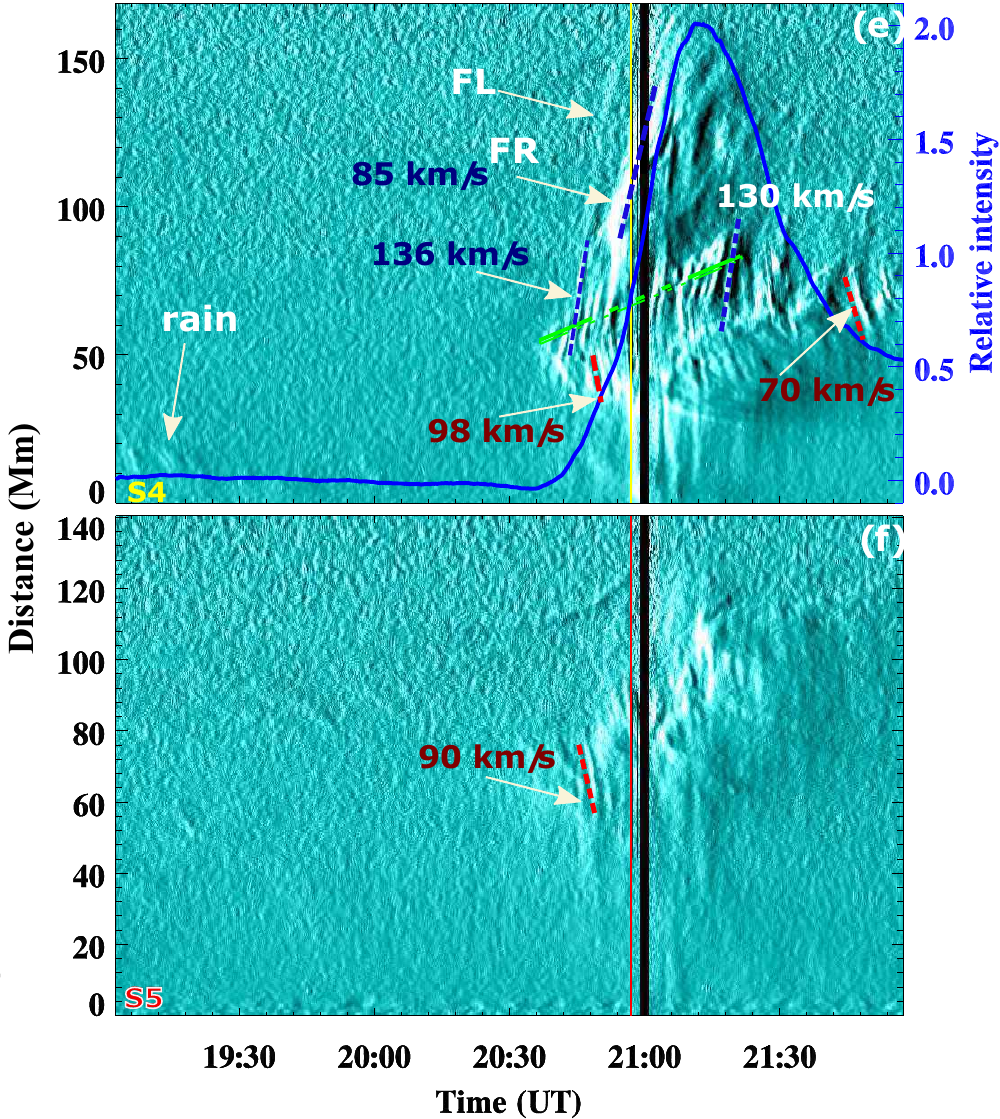}
\includegraphics[width=8.7cm]{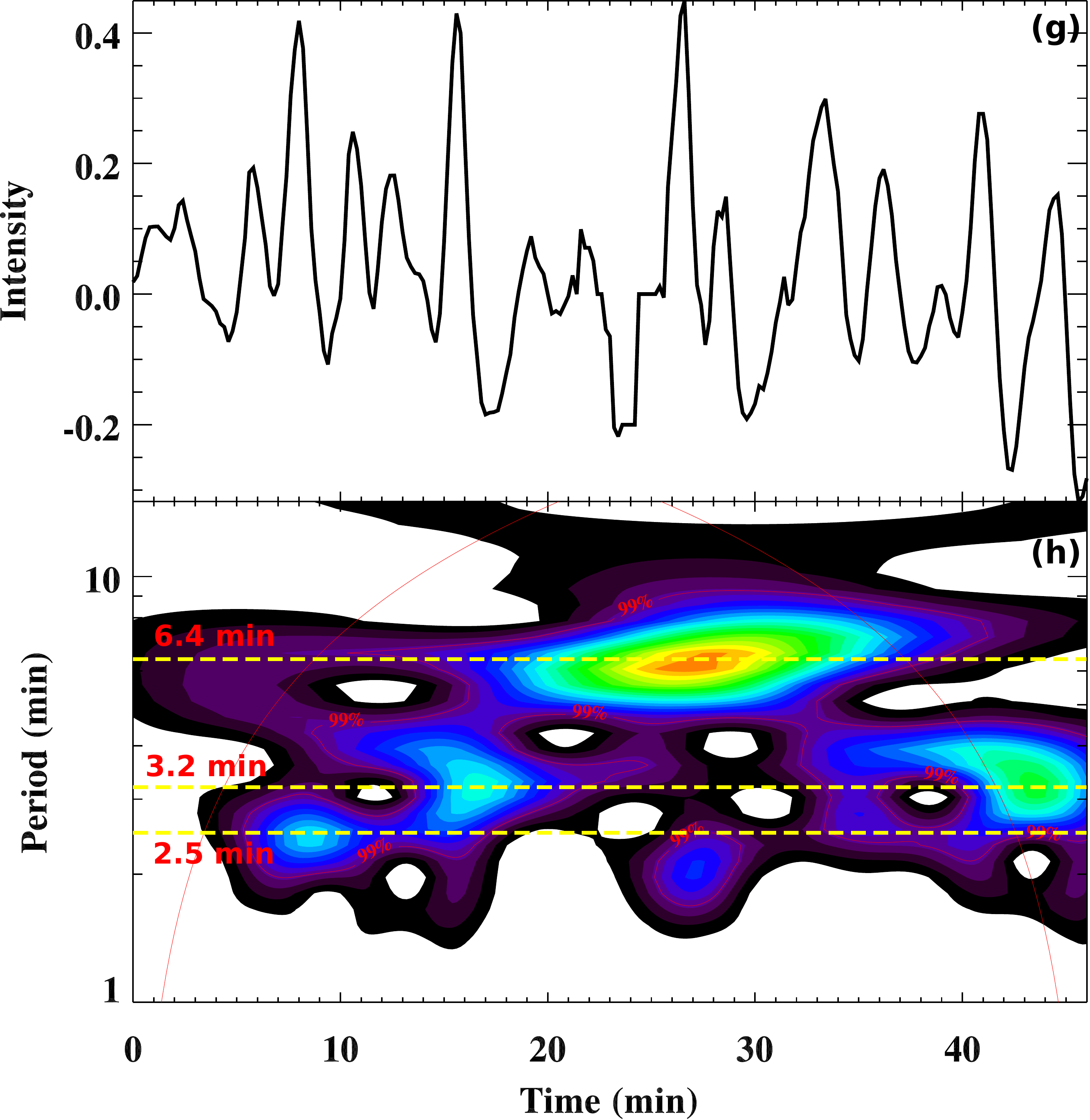}
}
\caption{(a,c,d) 131 \AA\ images before, during, and after the E1 eruption. Dimming region d2 contours ($-10\%$, $-20\%$ of peak intensity) observed in the 193 \AA\ channel are overplotted on the cotemporal 131 \AA\ image in panel a. Panel d is overlaid by the \emph{RHESSI} X-ray contours in 3-6 keV (green) and 6-12 keV (red) during the flare decay phase. The contour levels are $30\%$, $50\%$, $70\%$, and $90\%$ of peak intensity. H = plasma heating sites, FA = flare arcade, FCS = flare current sheet, FR = flux rope (b) 131 \AA\ running difference ($\Delta t = 1$ min) image. (e,f) Time-distance running-difference intensity plots along slices S4 and S5 in panel b using 131 \AA\ images. The vertical line indicates the timing of explosive breakout reconnection. The blue curve in panel e is the AIA 131 \AA\ relative intensity extracted from the dashed red box outlined in panel c. FL = frontal loop. (g) AIA 131 \AA\ running-difference intensity extracted from the green slit in panel e. (h) Wavelet power spectrum of the running-difference intensity. The start time is $\sim$20:36 UT. (An animation of this Figure is available online.)} 
\label{fig5}
\end{figure*}
%%%%%%%%%%%%%%%%%%%%%%%%%%%%%%%%%%%%%%%%%%%%%%%%%%%%%%%%%%%%%%%%%%%%%%
%%%%%%%%%%%%%%%%%%%%%%%%%%%%%%%%%%%%%%%%%%%%%%%%%%%%%%%%%%%%%%%%%%%%%%
\begin{figure*}
\centering{
\includegraphics[width=5.7cm]{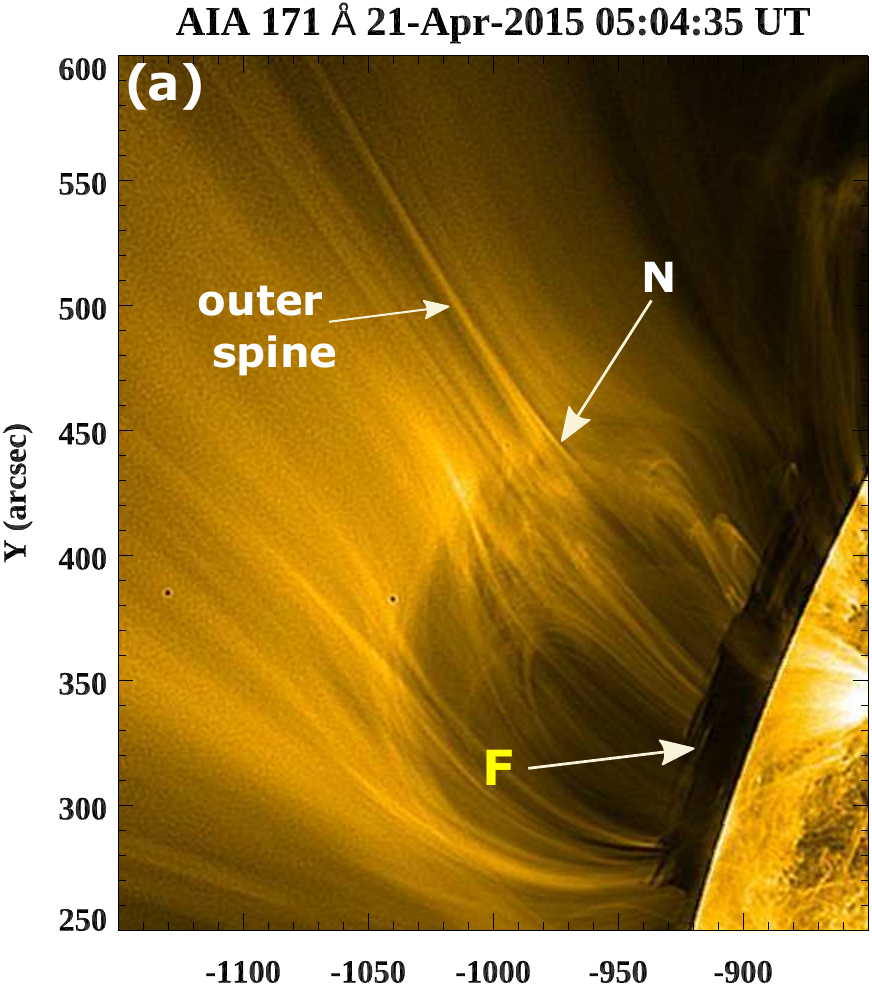}
\includegraphics[width=12cm]{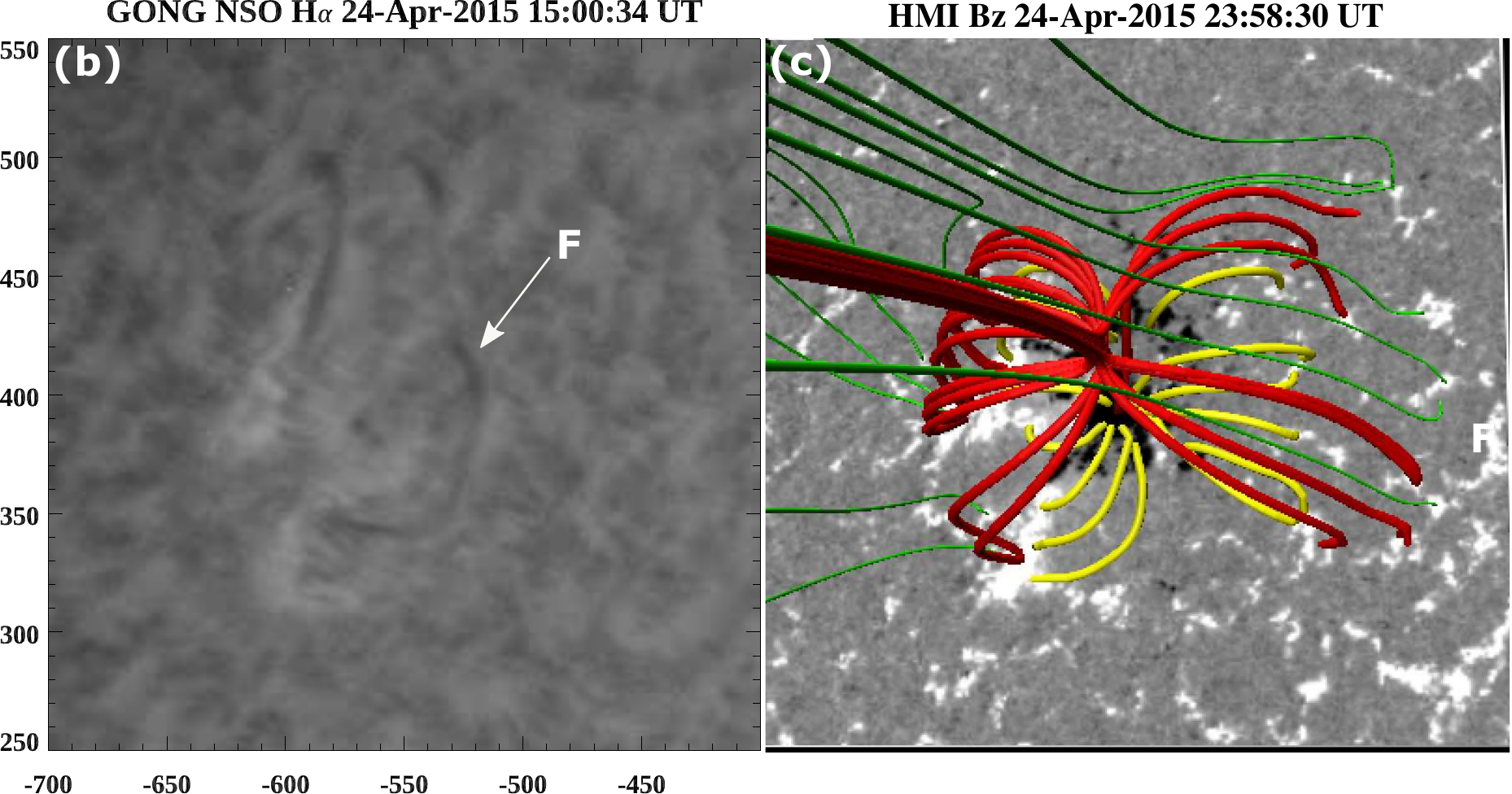}

\includegraphics[width=6.4cm]{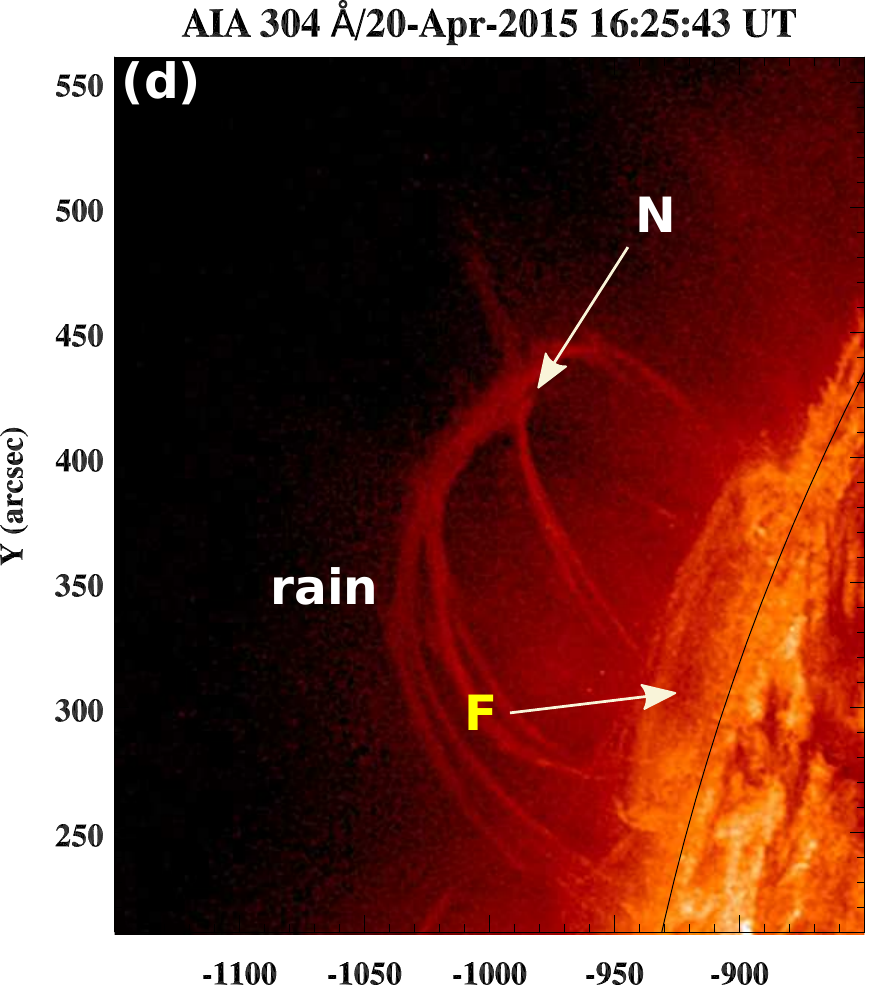}
\includegraphics[width=6.85cm]{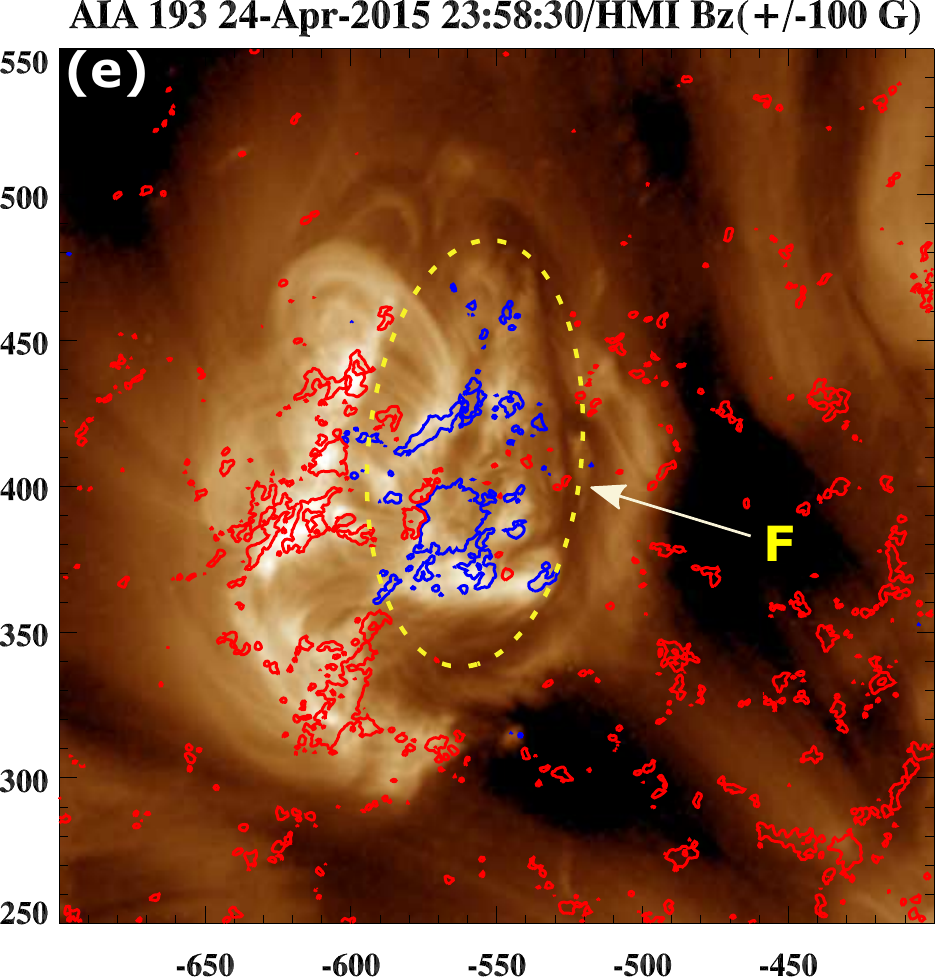}
}
\caption{(a) 171 \AA\ image of the E2 source region on 2015 April 21. N = null, F = filament. (b) H$\alpha$ image on 2015 April 24 showing filament F. (c) Potential-field extrapolation of the source region at 23:58:30 UT on 2015 April 24. (d) 304 \AA\ image showing coronal rain one day prior to the eruption. Plasma outflow is also seen along the outer spine. (e) 193 \AA\ image overlaid by HMI magnetogram contours; the filament is marked by yellow dashed line.} 

\label{fig6}
\end{figure*}

%%%%%%%%%%%%%%%%%%%%%%%%%%%%%%%%%%%%%%%%%%%%%%%%%%%%%%%%%%%%%%%%%%%%%%
\begin{figure*}
\centering{
\includegraphics[width=6.3cm]{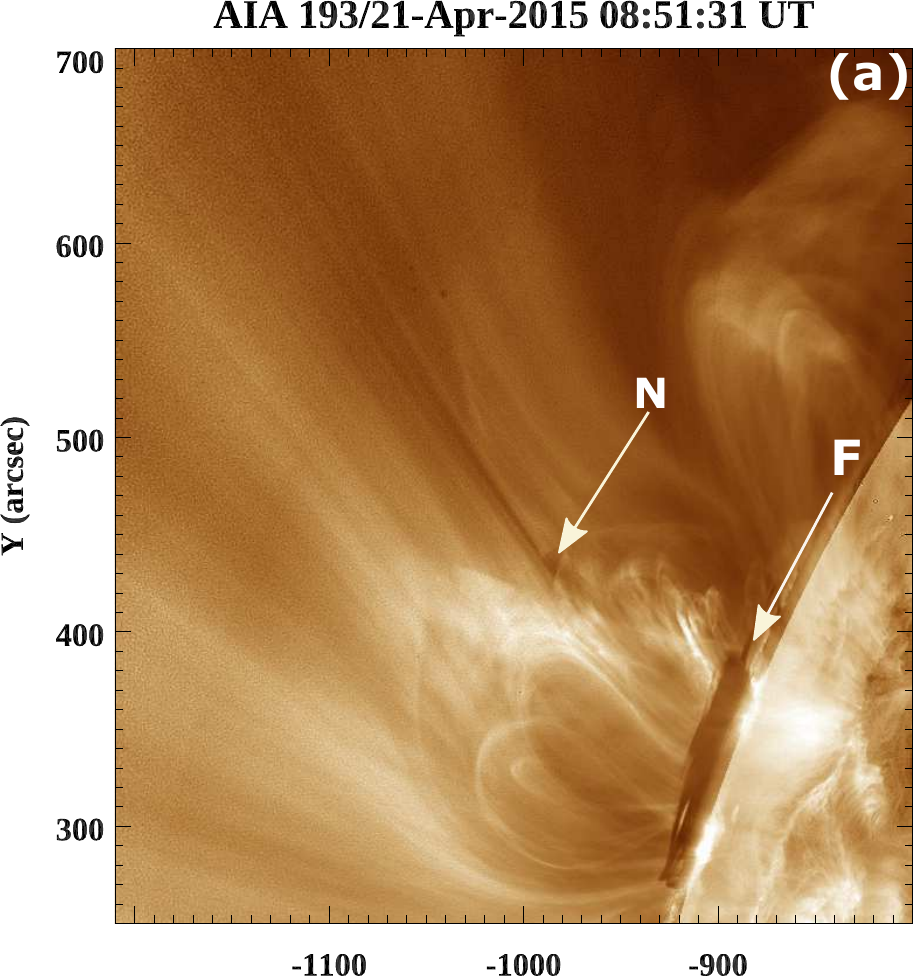}
\includegraphics[width=5.5cm]{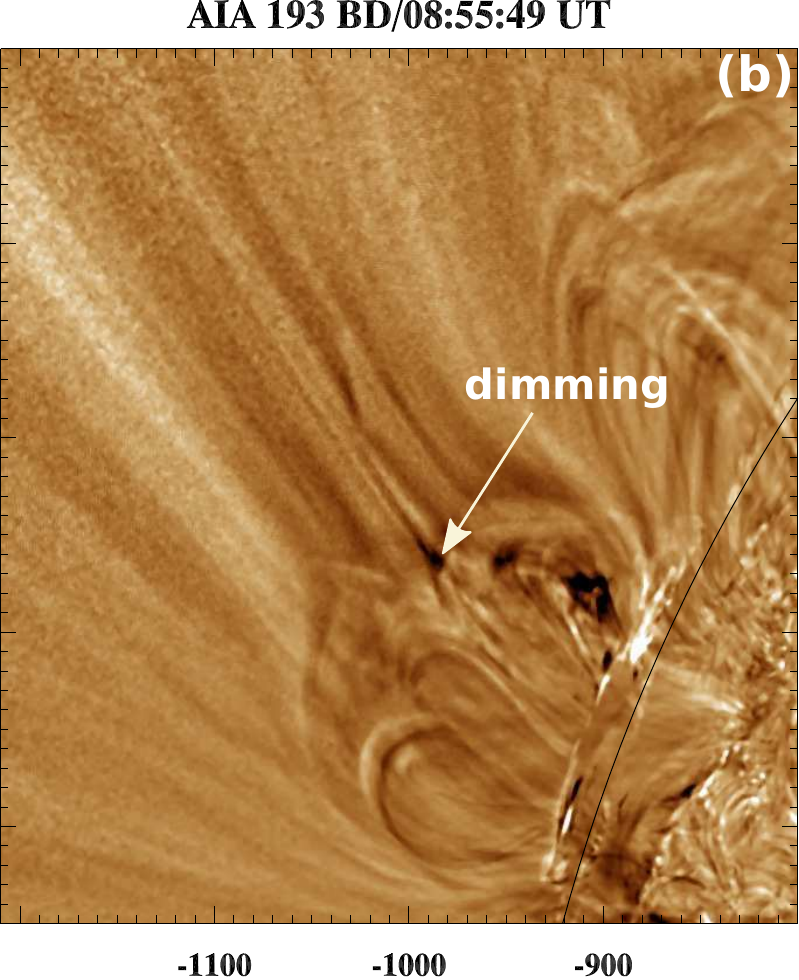}
\includegraphics[width=5.5cm]{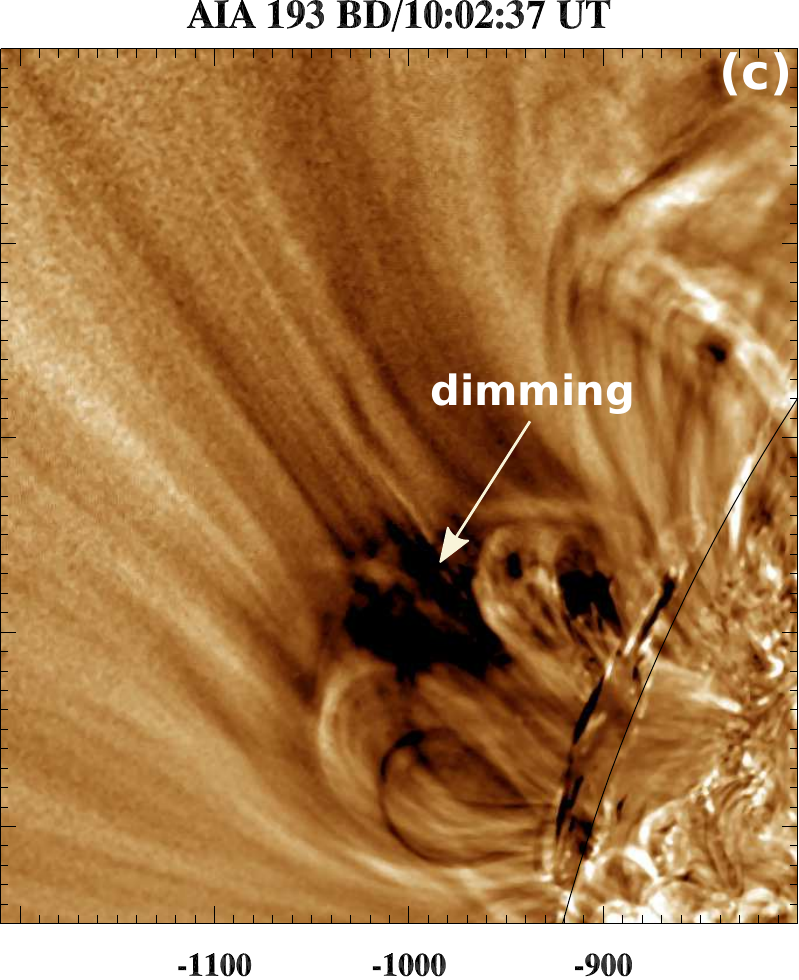}

\includegraphics[width=6.3cm]{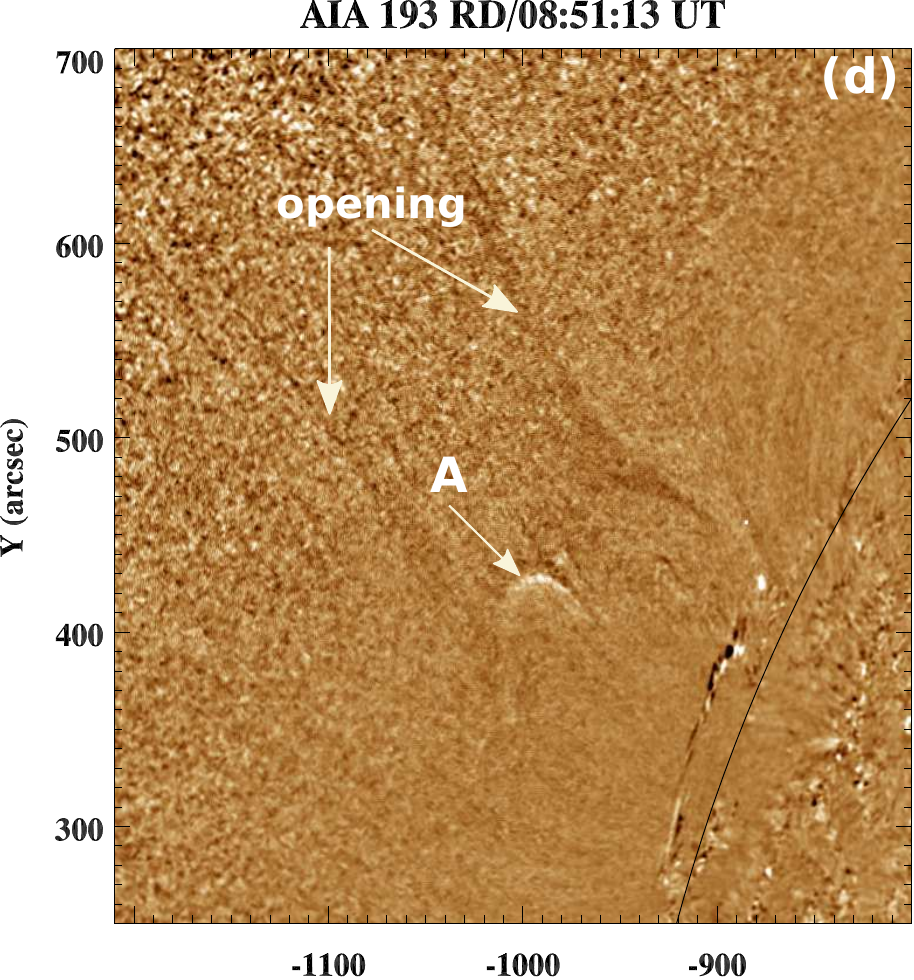}
\includegraphics[width=5.5cm]{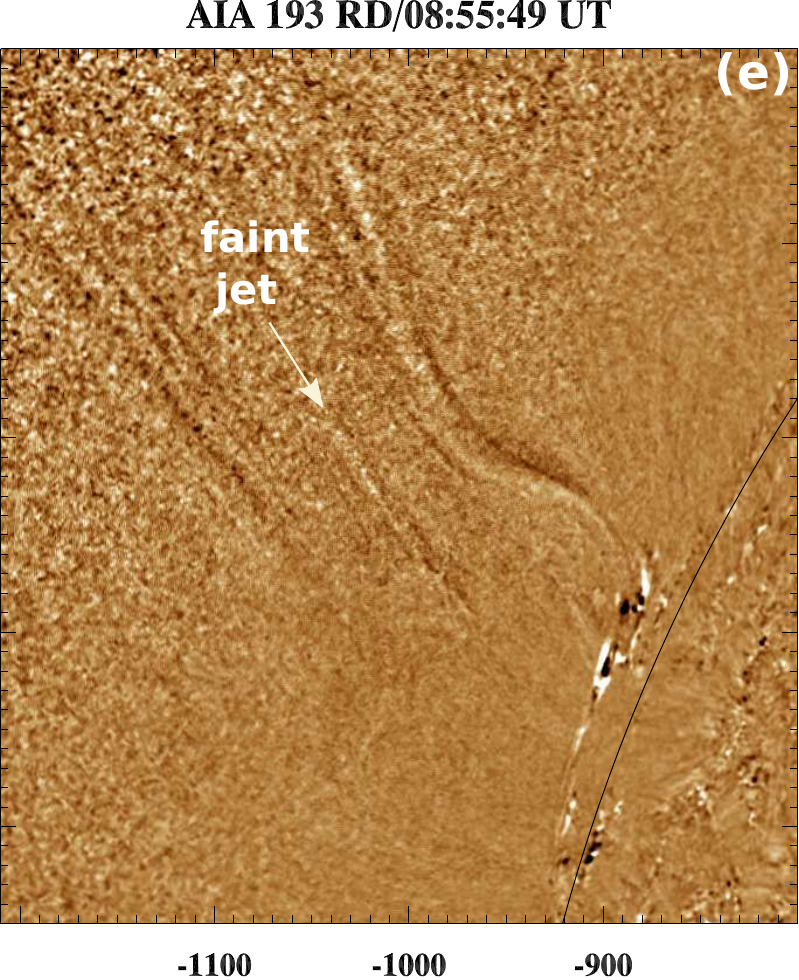}
\includegraphics[width=5.5cm]{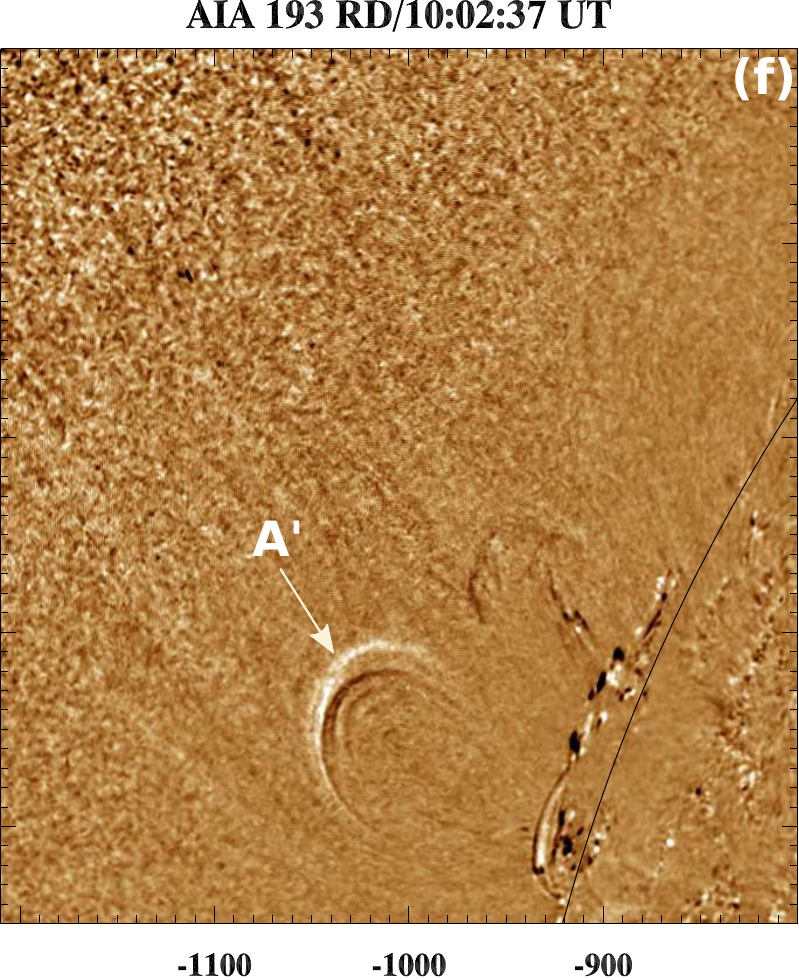}

\includegraphics[width=6.3cm]{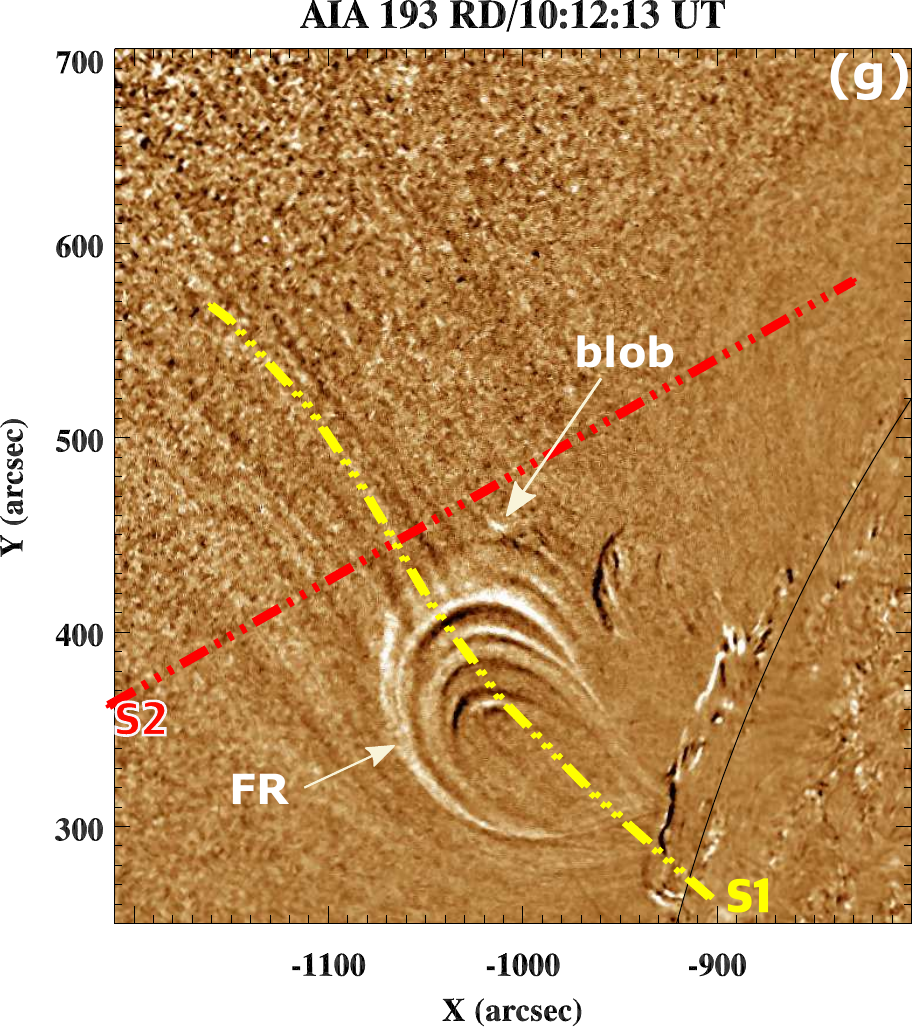}
\includegraphics[width=5.5cm]{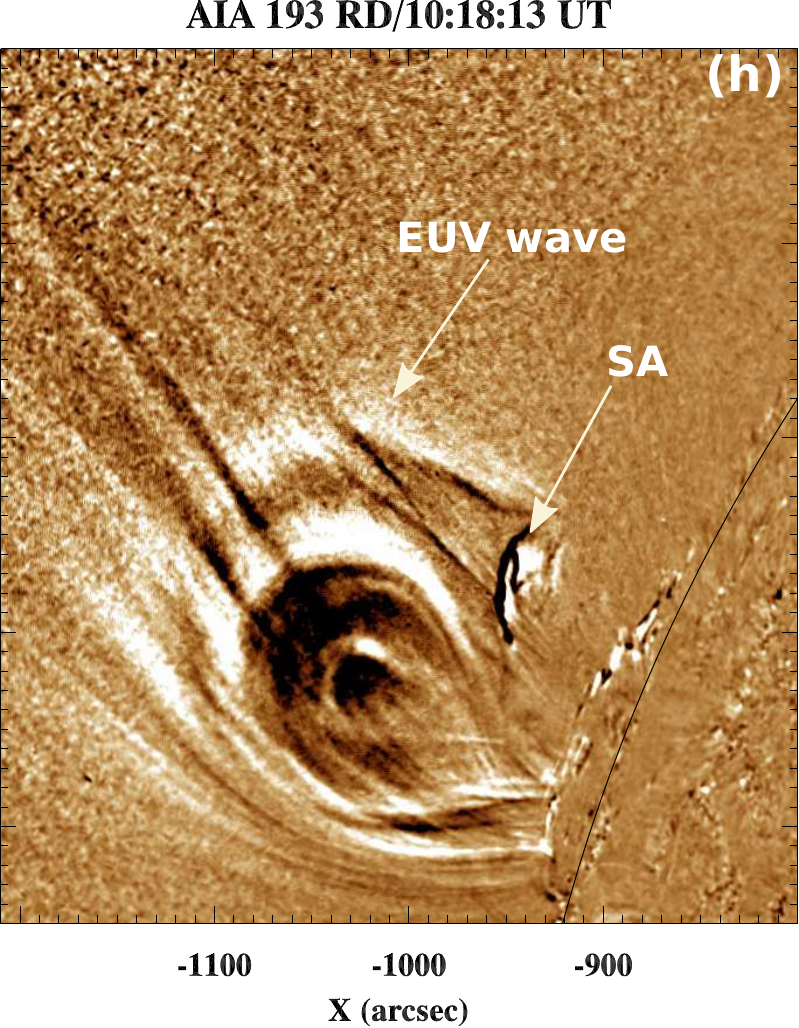}
\includegraphics[width=3.2cm]{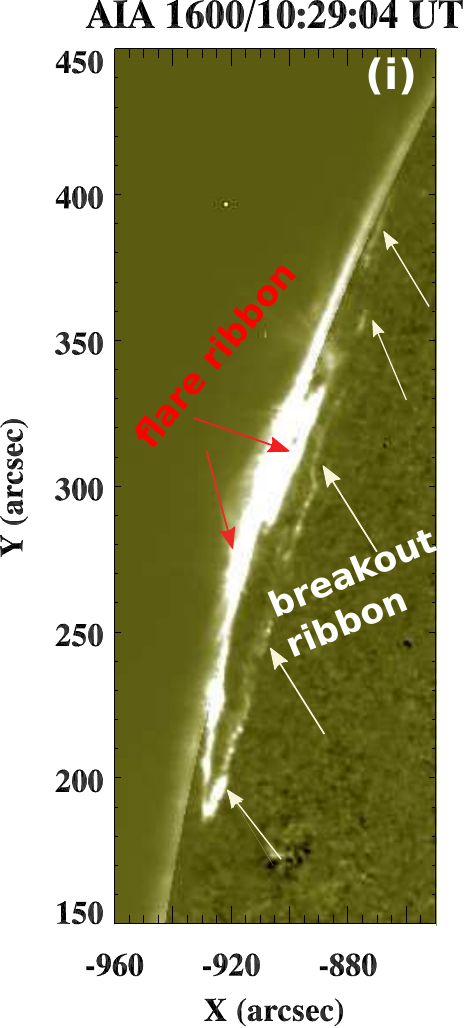}
}
\caption{(a-c) 193 \AA\ intensity and base-difference images during the pre-eruption phase of E2. N = null, F = filament. (d-f) 193 \AA\ running-difference images ($\Delta t = 1$ min) showing the pre-eruption opening. A and A$^\prime$ are rising arc-shaped structures. Note that panels in the top and middle rows are cotemporal. (g-h) The encounter between the outer loops of the rising flux rope and the breakout current sheet for the first eruption. S1 and S2 are slices used to create the time-distance plots in Figure \ref{fig8}, FR = flux rope, SA = side arcade. (i) 1600 \AA\ base-difference image during the first eruption showing flare and breakout ribbons. (An animation of this Figure is available online.)} 
\label{fig7}

\end{figure*}
%%%%%%%%%%%%%%%%%%%%%%%%%%%%%%%%%%%%%%%%%%%%%%%%%%%%%%%%%%%%%%%%%%%%%%
\begin{figure*}
\centering{
\includegraphics[width=5.85cm]{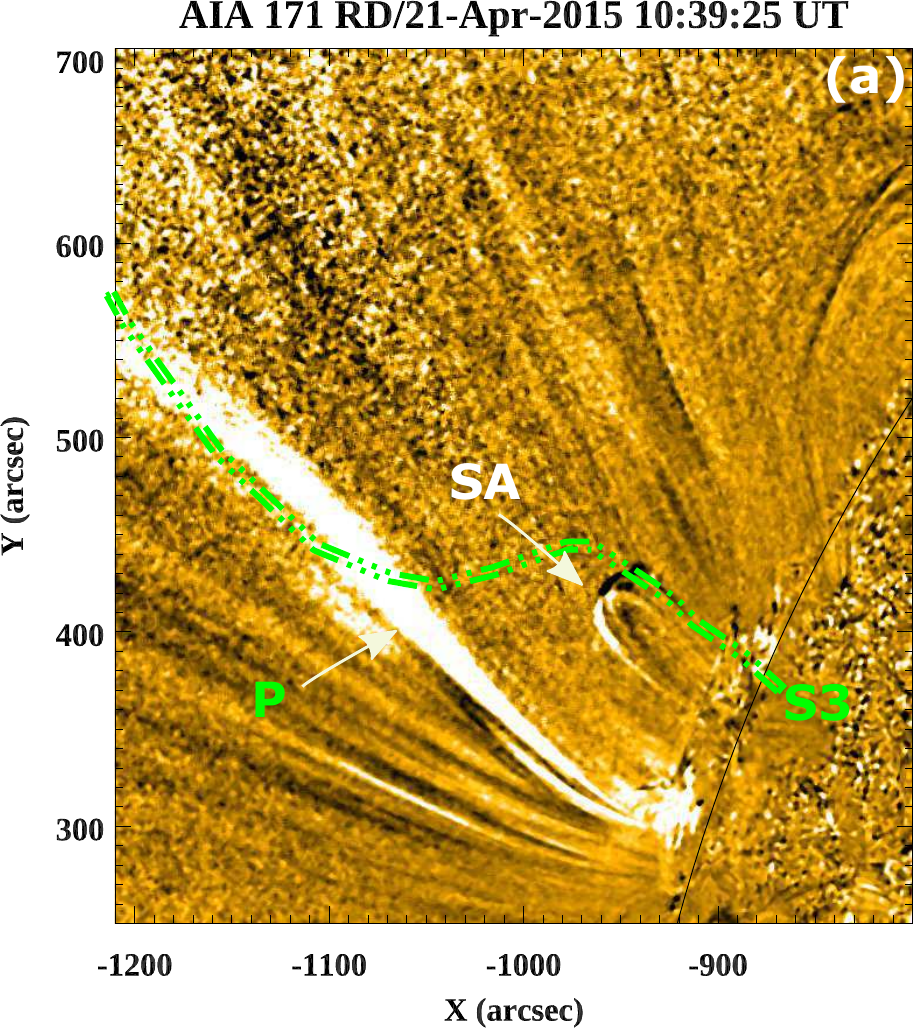}
\includegraphics[width=5.2cm]{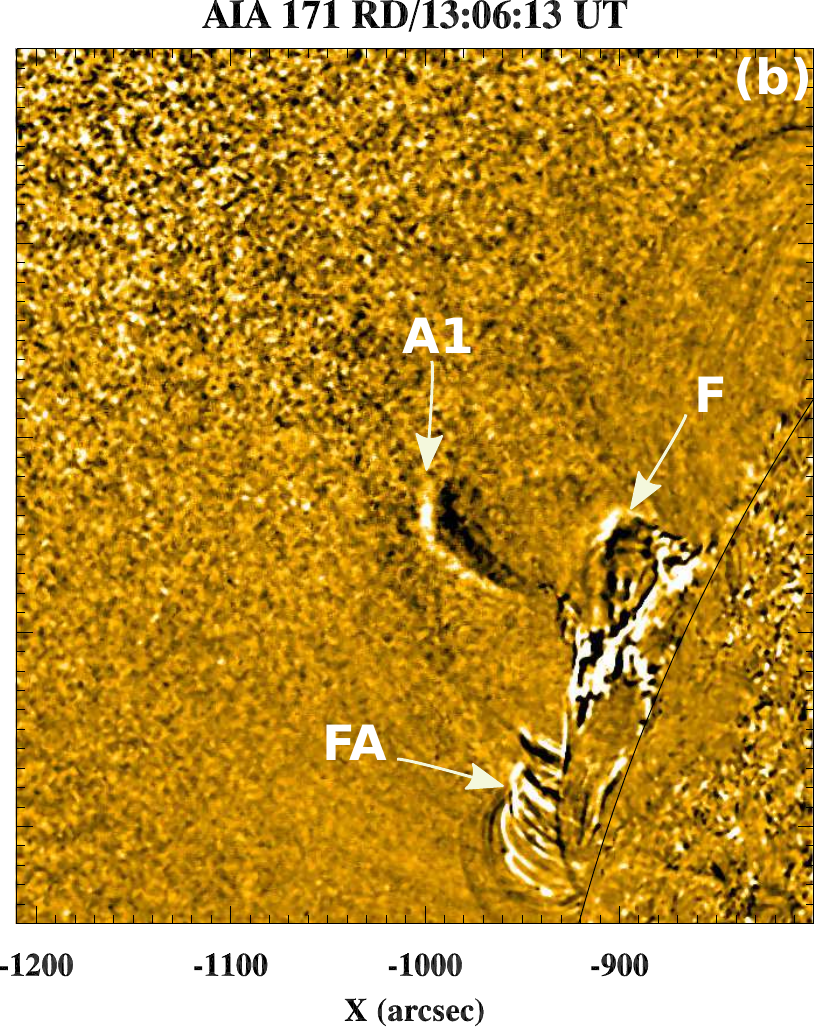}
\includegraphics[width=5.1cm]{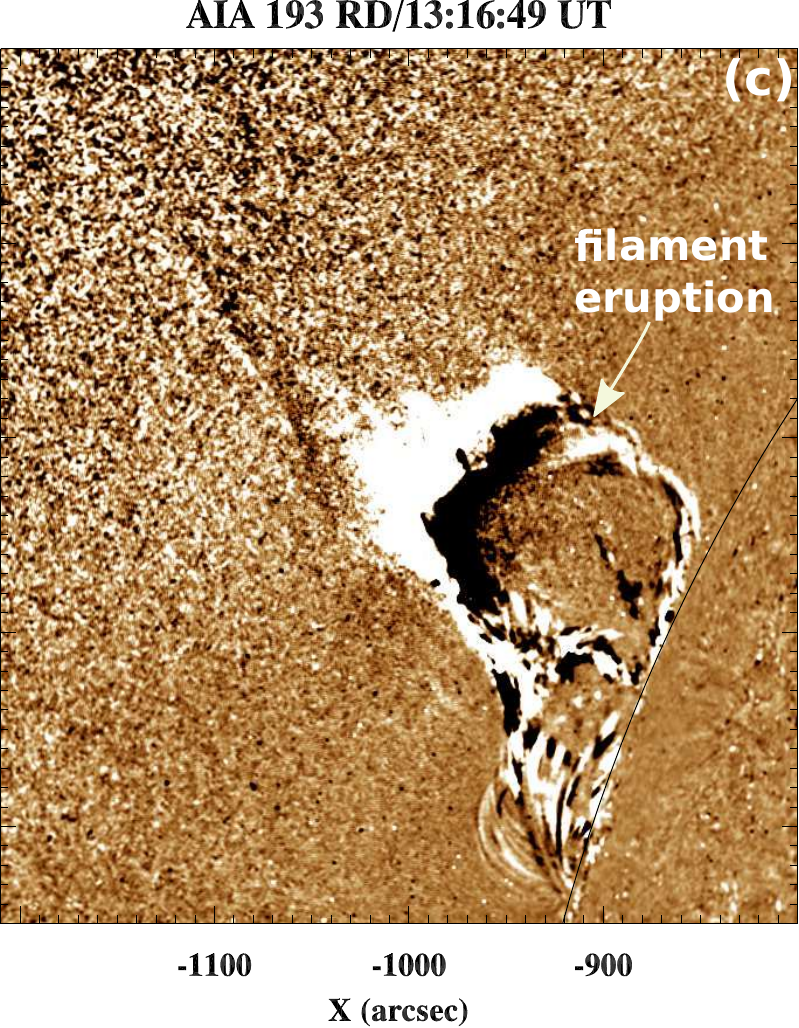}
\includegraphics[width=15cm]{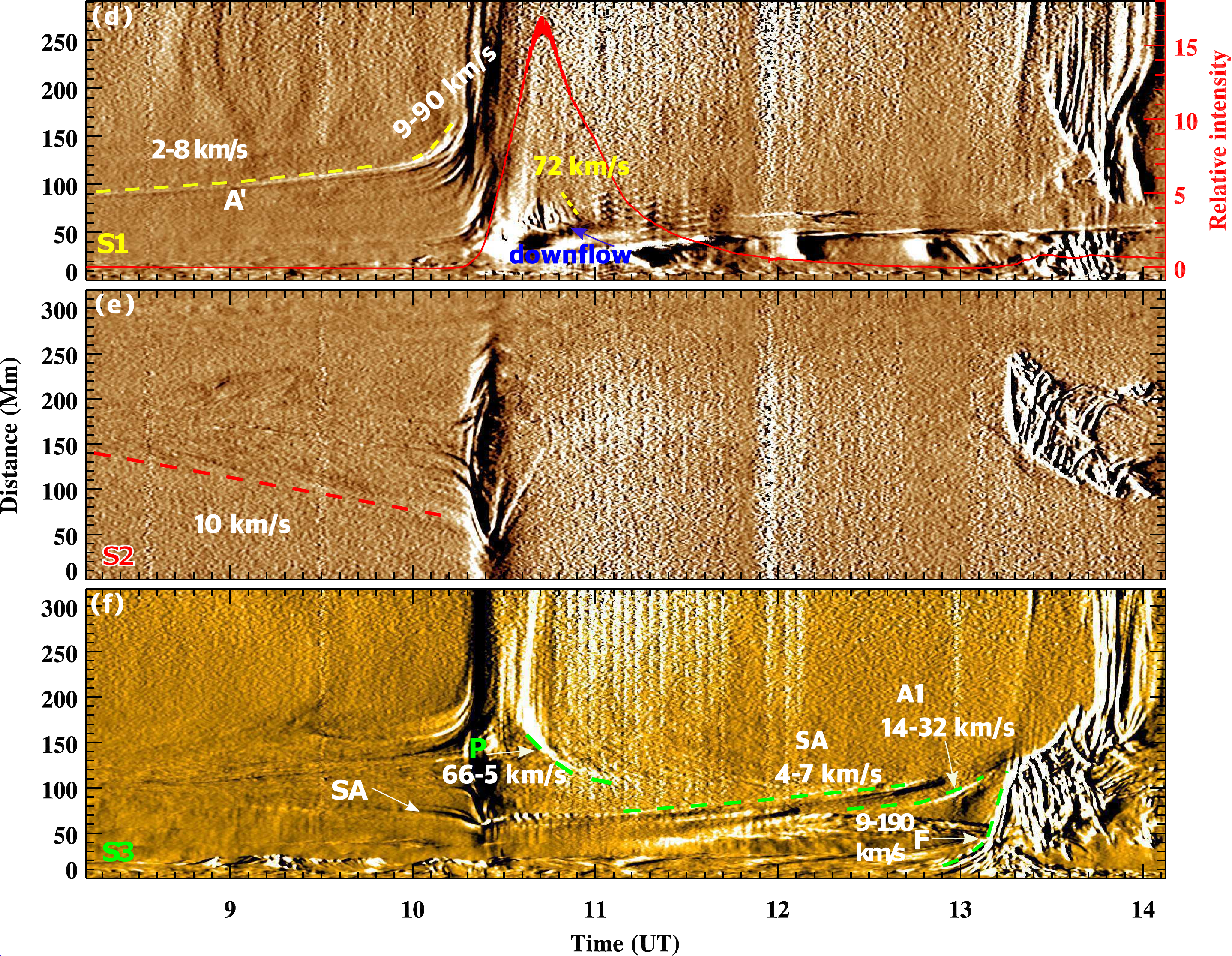}
}
\caption{(a,b) 171 \AA\ running-difference images ($\Delta t = 1$ min) after the first eruption and just before the onset of the second eruption in event E2. SA = side arcade, FA = flare arcade, F = filament, P is a bright open structure near the BCS, and A1 is a loop within the arcade above the filament. (c) Second eruption, of filament F. (d-f) Time-distance intensity plots along slices S1 and S2 (Fig.\ \ref{fig7}g) and S3 (panel a) using 193 \AA\ (panels d,e) and 171 \AA\ (panel f) running-difference images. (An animation of this Figure is available online.)} 
\label{fig8}
\end{figure*}

%%%%%%%%%%%%%%%%%%%%%%%%%%%%%%%%%
\begin{figure*}
\centering{
\includegraphics[width=6.4cm]{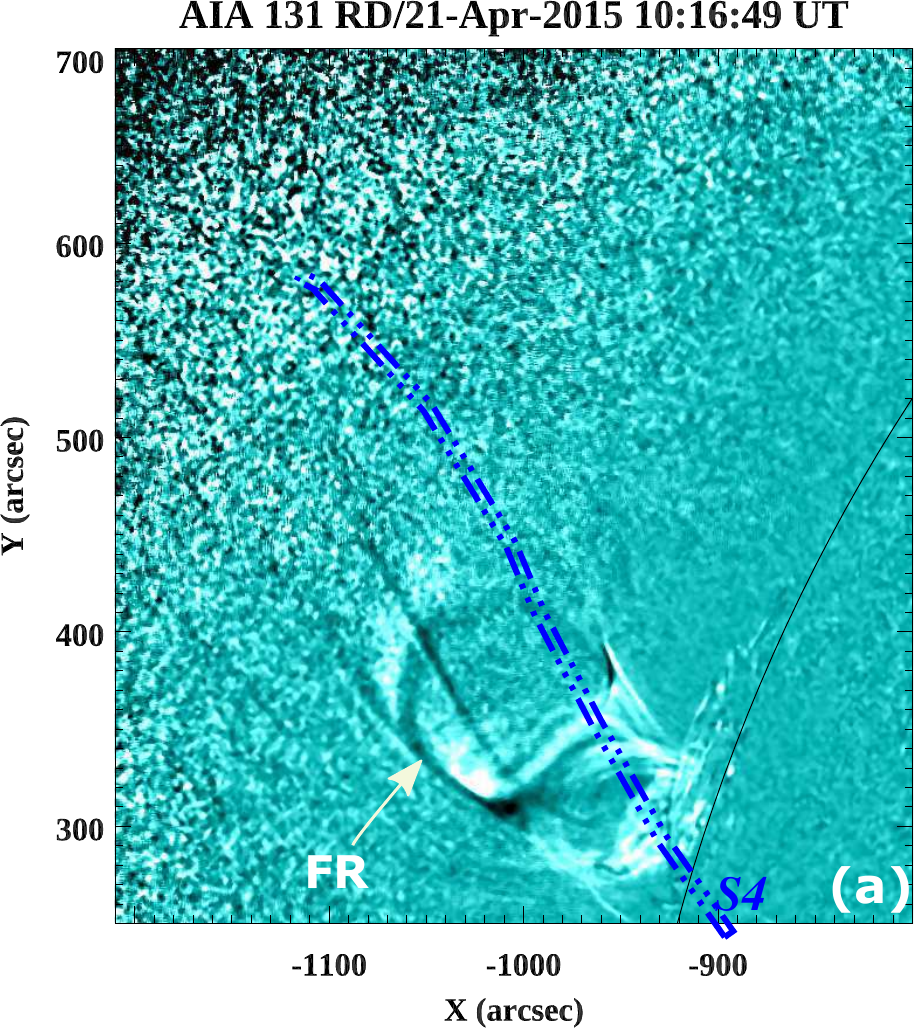}
\includegraphics[width=5.6cm]{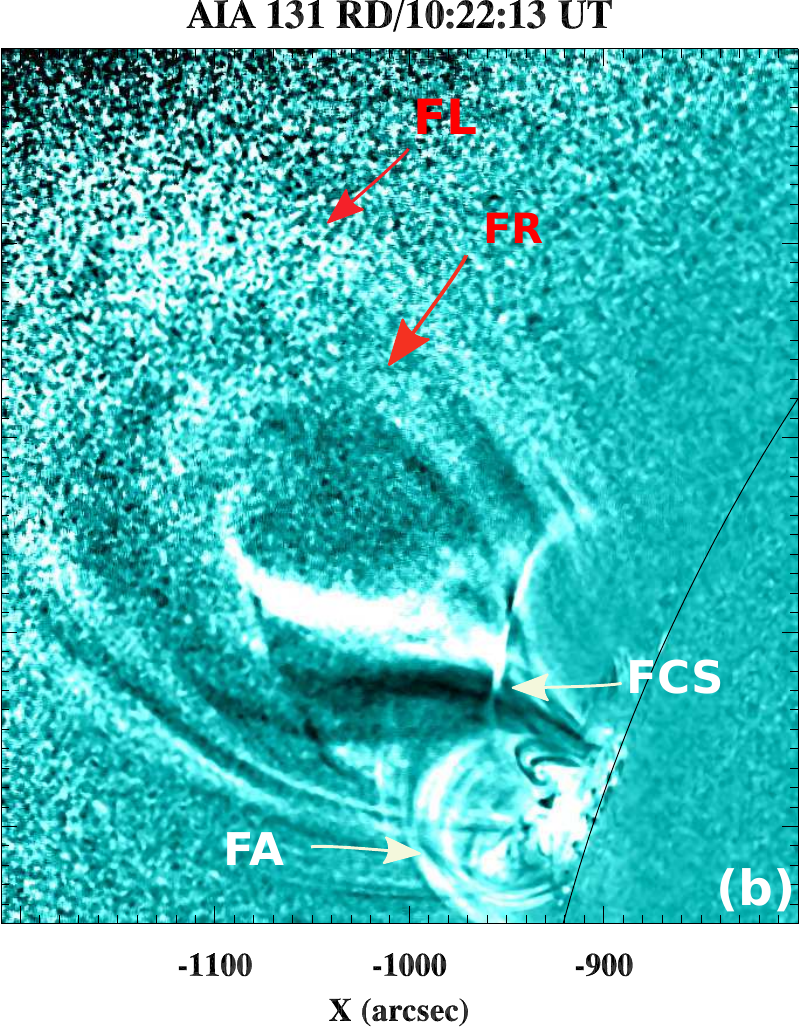}
\includegraphics[width=5.6cm]{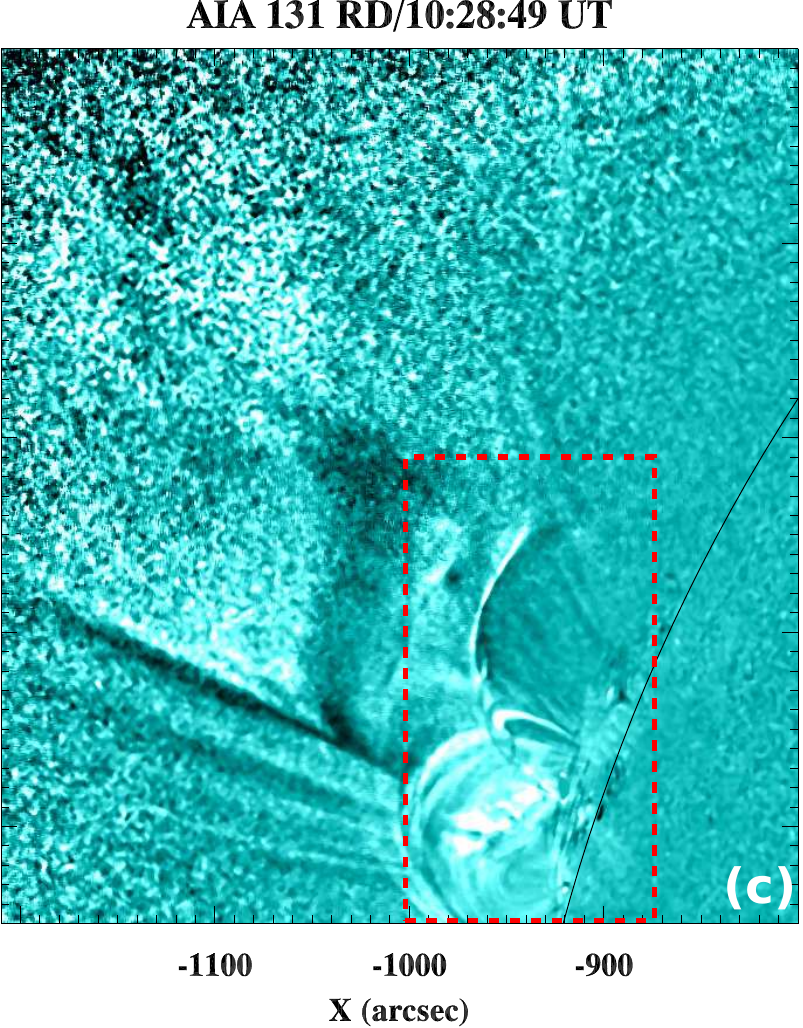}
\includegraphics[width=17cm]{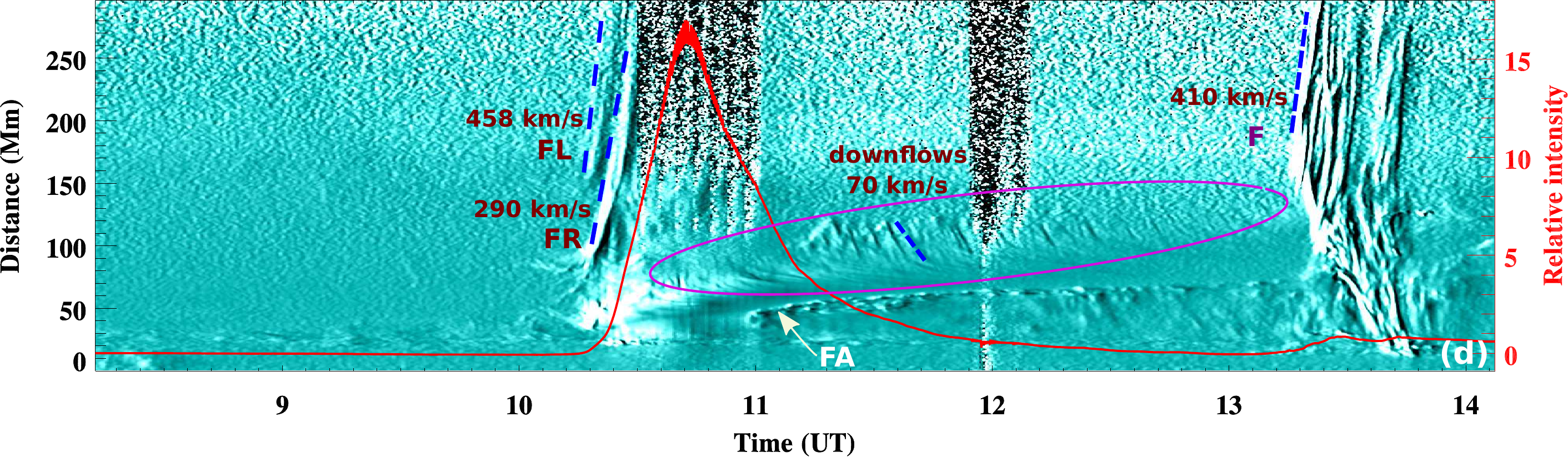}
}
\caption{(a-c) 131 \AA\ running-difference images ($\Delta t = 1$ min) at three times during the first E2 eruption. FR = flux rope, FA = flare arcade, FCS = flare current sheet, FL = frontal loop. (d) Time-distance intensity plot along slice S4 (shown in panel a) using 131 \AA\ running-difference images. The relative emission intensity (red solid line) was extracted from the dashed rectangular box in panel c. The magenta oval surrounds the quasiperiodic downflows below the first erupting flux rope. (An animation of this Figure is available online.)} 
\label{fig9}

\end{figure*}

%%%%%%%%%%%%%%%%%%%%%%%%%%%%%%%%%%%%%%%%%%%%%%%%%%%%%%%%%%%%%%%%%%%%%

\section{OBSERVATIONS}\label{obs}
We used \emph{Solar Dynamics Observatory} Atmospheric Imaging Assembly (\emph{SDO}/AIA; \citealt{lemen2012}) full-disk images of the Sun (field-of-view $\approx$ 1.3$R_\odot$) with a spatial resolution of 1.5$\arcsec$ (0.6$\arcsec$ pixel$^{-1}$) and a cadence of 12 s, in the following channels: 1600~\AA\ (\ion{C}{4}+continuum, $T \approx 5000$~K, 0.1 MK), 304~\AA\ (\ion{He}{2}, $T \approx 0.05$~MK), 171~\AA\ (\ion{Fe}{9}, $T \approx 0.7$~MK), 211~\AA\ (\ion{Fe}{14}, $T \approx 2$~MK), 193~\AA\ (\ion{Fe}{12}, \ion{Fe}{24}, $T \approx 1.2$~MK and $\approx 20$~MK), AIA 94~\AA\ (\ion{Fe}{10}, \ion{Fe}{18}, $T \approx 1$~MK, $T \approx 6.3$~MK), and 131~\AA\ (\ion{Fe}{8}, \ion{Fe}{21}, \ion{Fe}{23}, i.e., $T \approx 0.4$, $10$, and $16$~MK) images.  We cleaned and substantially sharpened all of the AIA images with an innovative noise-gating technique \citep{deforest2017}. Line-of-sight magnetic-field measurements obtained with \emph{SDO}'s Helioseismic and Magnetic Imager (\emph{SDO}/HMI; \citealt{scherrer2012}) were employed to establish the magnetic structure and environment of the source regions. We utilized a potential-field extrapolation code \citep{nakagawa1972} that is publicly available in the GX simulator package of SSWIDL \citep{nita2015}. Selected portions of the HMI magnetograms that cover the relevant source regions were used for the extrapolation, and the VAPOR visualization package was used to plot field lines.
We used data from the \emph{Reuven Ramaty High Energy Solar Spectroscopic Imager} (\emph{RHESSI}; \citealt{lin2002}) to ascertain the evolution of both soft and hard X-ray sources during the eruptive flares. The CLEAN algorithm \citep{hurford2002} reconstructed the \emph{RHESSI} images with an integration time of 12 s. Finally, we examined \emph{Solar and Heliospheric Observatory} Large-Angle and Spectrometric Coronagraph (\emph{SOHO}/LASCO; \citealt{brueckner1995,yashiro2004}) C2 white-light images (field-of-view $\approx$ 2--6$R_\odot$) to characterize the early jets and the subsequent CMEs in the low corona.

The source regions for these events are shown in \emph{SDO}/AIA 211 \AA\ images (Fig.\ \ref{fig2}), which have the best contrast between CH and AR. Source regions for limb events E1 and E2 are shown on 2015 April 24  (i.e., 4 days after the eruption), while the source of the disk event E3 is  shown shortly before eruption. 

\section{Results}\label{results}
\subsection{Event \#1 (E1)}

The E1 source region was located within a coronal hole at the east limb on 2015 April 20 (Fig.\ \ref{fig2}(a)). An AIA 171 \AA\ image shows the emission pattern of a typical pseudostreamer \citep[e.g.,][]{masson2014,karna2019} before eruption (Fig.\ \ref{fig3}(a)). To determine the magnetic topology, we used an HMI magnetogram on 2015 April 23, when the region had rotated onto the disk. The source was not classified as an active region by NOAA because it consisted of plage without a sunspot. The source region contains a filament (marked F in Fig.\ \ref{fig3}(a,b)) along the elliptical PIL, which remained stationary during the flare/eruption. The minority polarity (negative/blue) was surrounded asymmetrically by the majority polarity (positive/red) of the CH. The potential-field extrapolation reveals the fan/spine topology along with a 3D null (Fig.\ \ref{fig3}(c)). Three days after the eruption, when the source region was well on the disk, the peak negative (positive) line-of-sight magnetic-field strength was $-780$ G ($+974$ G).

\subsubsection{Pre-eruption activity}

During the pre-eruption stage, we detected persistent coronal rain in the AIA 304 \AA\ channel (Fig.\ \ref{fig3}(g)), starting with weak rain one day before the eruption that became stronger around 10:00 UT on April 20.  The rain was much more intense along the northern fan surface than along the southern side.  The accompanying 304 \AA\ movie reveals continual draining of cool plasma toward the northern footpoint until around 18:00 UT. Coronal-rain--associated downflows also were detected in the AIA 131 \AA\ bandpass during the early evolution and disappeared later, around 20:00 UT (Fig.\ \ref{fig5}(c)). We point out that this channel is sensitive to the cool Fe VIII line ($T \approx 0.4$ MK), as well as to hotter coronal emissions; the rain is most likely emitting in the Fe VIII line, coming from the transition region between the condensations and the adjacent coronal plasma. At warm coronal temperatures (AIA 193 \AA), linear structures outside the fan near the original null moved northward (Fig.\ \ref{fig3}(d)). As explained in \S\ref{discussion}, we refer to this transverse motion away from the initial null location as \emph{pre-eruption opening}. The accompanying Figure \ref{fig3} animation in the cool 304 \AA\ channel reveals an increase in rain emanating from the base of the pre-eruption opening and draining back to the solar surface along that open structure. Simultaneously, AIA 193 running-difference images show a bright loop underneath the dome (Fig.\ \ref{fig3}(e). These opening and closing structures are likely a consequence of interchange reconnection. Similar evolution (opening/closing) is also seen in MHD simulations of interchange reconnection at the null in pseudostreamers \citep{masson2014}. Cotemporal AIA 193 \AA\ images reveal the onset of coronal dimming (feature d1, Fig.\ \ref{fig3}(h)) coincident with the northward motions outside the fan (arrows, Fig.\ \ref{fig3}(e)). A more compact dimming region, d2, appeared near the null and became progressively darker and larger throughout the pre-eruption phase (Fig.\ \ref{fig3}(i)).  During this evolution, loop-like structures appeared beneath the initial null during 19:30-19:46 UT (Fig.\ \ref{fig3}(d,e) and accompanying movie).  A rising arc-shaped structure (A) was observed below the null point later, at 20:38:25 UT (Fig.\ \ref{fig3}(f)). There was no evidence of flare reconnection during this phase. 

To understand the dynamics and sequence of events before and during the eruption, we created time-distance intensity plots along slices S1 (yellow), S2 (red), and S3 (blue) in the AIA 193 \AA\ running-difference images (Fig.\ \ref{fig4}(a)). The S1 plot reveals the onset of slow ($v \approx 4$ \kms) pre-eruption opening around 19:30 UT and associated quasiperiodic downflows along the southern fan surface near the null (Fig.\ \ref{fig4}(b) and accompanying movie). Thereafter, the northern arcade A rose slowly and nonradially, accelerating from $v \approx 1$ to $v \approx 30$ \kms during 19:50-20:54 UT (Fig.\ \ref{fig4}(c)).   A bright, inclined, thin feature, which we interpret to be the breakout current sheet, became visible as A approached the initial null site.  A small blob appeared southwest of the BCS (Fig.\ \ref{fig4}(a)). The AIA 131 \AA\ intensity and running-difference images also reveal compression and/or heating near the BCS prior to the impulsive flare, as discussed below. LASCO C2 coronagraph images at 20:00 and 20:36 UT (Fig.\ \ref{fig4}(e,f)) show a jet associated with the pre-eruption opening. The identification of the jet is based on tracking it from low to high corona, i.e., from AIA 193 \AA\ to the LASCO C2 field of view (see AIA 193 \AA\ and LASCO C2 animations for the temporal evolution.)

\subsubsection{Eruption}

According to the GOES soft X-ray flux (1-8 \AA) and the AIA 131 \AA\ relative intensity (cyan curve in Fig.\ \ref{fig4}(b)) profiles, the eruption was associated with a C2.4-class flare that started at 20:40 UT, peaked at 21:16 UT, and ended at 22:15 UT.  A narrow, slow mass ejection reached $v \approx 360$ \kms as measured along S2 in the AIA 193 \AA\ channel at 21:05-21:08 UT, simultaneous with the disappearance of the bright feature marked BCS (Fig.\ \ref{fig4}(b)); shortly thereafter a slow CME ($v \approx 415$ \kms) was detected in the LASCO field of view (FOV) (Fig.\ \ref{fig4}(g)). Strong downflows ($v \approx 250$ \kms) streamed along the loops beneath the breakout current sheet around 21:03-21:09 UT (S3, Fig.\ \ref{fig4}(d); see accompanying animation). 

About 8 min prior to the start of the GOES flare, heating of plasma near the BCS was observed in AIA 131 \AA\, followed by the first sign of flare plasma heating in the lower corona at 20:37:22 UT (Fig.\ \ref{fig5}(a-c) and accompanying movie). The two heating sites are marked by H in Figure \ref{fig5}(a). They shifted outward slightly from their initial positions during the hour that elapsed between panels a and c, as is typical during eruptive flares. The dimming region d2 was located slightly above the upper hot spot H, showing that d2 was hotter and/or less dense than its surroundings (due to plasma evacuation), especially in its lower portions. Repetitive upflows and downflows, along with tiny blobs, propagated along a plasma sheet that formed above the bright loops of the flare arcade (FA, Fig.\ \ref{fig5}(c)) from about 20:40 UT onwards. To determine whether the blobs/flows moving upward along the FCS were quasiperiodic, we performed a Morelet wavelet analysis  \citep{torrence1998} of the 131 \AA\ running difference signal (Fig.\ \ref{fig5}(g)) extracted from the green slit in Figure \ref{fig5}(e). The wavelet power spectrum reveals the presence of statistically significant periods, with periods of $\approx$2.5, 3.2, and 6.4 min above the 99$\%$ significance level (Fig.\ \ref{fig5}(h)).  

A circular feature above the plasma sheet first appeared at 20:56 UT and expanded as it rose toward the BCS. We interpret the sheet and the circular feature as the flare current sheet (FCS) and flux rope (FR), respectively (Fig.\ \ref{fig5}(b,c)). The features FA, FCS, and FR closely resemble the corresponding structures that we reported in our previous analyses of coronal jets occurring in smaller fan/spine topologies \citep{kumar2018,kumar2019b}. The clearly seen FR ``bubble'' and the frontal loops (FL) ahead of it encountered the BCS, enhanced the heating there and the downflows beneath it, and then disappeared from the 131 \AA\ channel around 21:18 UT, leaving behind a stationary hot structure below the BCS (see Fig.\ \ref{fig5} animation). The strong downflows, the disappearance of the arc A and the BCS, and the increase in the flare intensity are all signatures associated with the onset of explosive flare and breakout reconnection. Consequently, we interpret the similar features in the studied pseudostreamer events according to the breakout scenario outlined in \S\ref{intro}.

\emph{RHESSI} observed only the decay phase of the flare, after 21:30 UT, but it detected two coronal thermal X-ray sources that coincided closely with the hottest features observed in the AIA 131 and 94 \AA\ channels. The \emph{RHESSI} intensity contours in the 3-6 keV (green) and 6-12 keV (blue) bands, overlaid on the 12-s cadence AIA 131 \AA\ image at 21:31 UT, reveal a strong lower coronal source (X=-1000$\arcsec$,Y=180$\arcsec$) at the top of the flare arcade and a weak upper coronal source (X=-1050$\arcsec$,Y=110$\arcsec$) at the bright cusp near the BCS (Fig.\ \ref{fig5}(d)). 

Time-distance intensity plots of slices from AIA 131 \AA\ running-difference images during 19:00-22:00 UT measure the dynamic evolution of the erupting FR, FCS, and FA (slice S4, Fig.\ \ref{fig5}(d)) and the downflows generated prior to and during the event (slice S5).  Intense, quasiperiodic, bidirectional plasma flows with intermittent substructure traveled along the BCS during 20:38 UT-22:00 UT (Fig.\ \ref{fig5}(e)). A strong downflow along the flare arcade was triggered when the flux rope interacted with the BCS (21:00-21:10 UT) (Fig.\ \ref{fig5}(e,f)). The typical speeds of some of the tracked upflows in S4 were $v \approx 130$-136 \kms (blue dotted line), whereas downflow speeds were $v \approx 70$-98 \kms (blue dotted line). The flux rope rose at $v \approx 85$ \kms prior to eruption, after the onset of flare reconnection. A similar evolution was seen in the AIA 94 \AA\ channel, which suggests the existence of hotter plasma ($T \sim $6-10 MK) there, in agreement with the \emph{RHESSI} observations. The multiple plasma blobs, quasiperiodic flows along the FCS, and rising FR were detected only in the AIA 131 and 94 \AA\ hot channels, while the rising arc structure A was observed only in the warm AIA 193 and 211 \AA\ channels. 

\subsection{Event \#2 (E2)}

The source region (NOAA AR \#12333) of E2 was located north of event E1 within the same coronal hole. The 171 \AA\ image several hours before the eruption displays a pseudostreamer above a dark filament (Fig.\ \ref{fig6}(a)).   
The H$\alpha$ image reveals the filament F lying along the elliptical PIL about 3 days after the eruption (Figure \ref{fig6}(b)).
As expected, the potential-field extrapolation shows a fan/spine topology with a 3D null (Fig.\ \ref{fig6}(c)). The peak negative (positive) line-of-sight magnetic-field strength was $-$1110 G ($+$850 G). Persistent coronal rain fell from the null (N) along the fan loops one day before the eruption (Fig.\ \ref{fig6}(d)). However, 3-4 hours before the first eruption, the rain was mostly concentrated toward the southern side of the dome where the first eruption occurred. The 193 \AA\ image three days after the eruption shows a reformed elliptical filament along the PIL (yellow dashed lines; Fig.\ \ref{fig6}(e,f)). This active region produced a pair of sympathetic eruptive flares. 

\subsubsection{Pre-eruption activity}

At 08:51:31 UT, a fan/spine configuration containing a dark filament is evident in the 193 \AA\ observations (Fig.\ \ref{fig7}(a)). The 193 \AA\ base- and running-difference images reveal a region of coronal dimming near the initial null from 08:30 onward, which grew as flare onset approached (Fig.\ \ref{fig7}(b,c) and accompanying animation).  As for E1, pre-eruption openings were visible on both sides of the distorted null/BCS (Fig.\ \ref{fig7}(d)), reaching transverse speeds $v \approx 10$ \kms (Fig.\ \ref{fig8}(e)) along the slice S2 shown in Figure \ref{fig7}(g). A faint jet appeared between the opening sides around 08:51 UT (Fig.\ \ref{fig7}(e)); subsequently two narrow jets were detected in the LASCO C2 coronagraph running-difference images after 09:36 (Fig.\ \ref{fig10}(a)). 

From 08:15 to 09:50 UT, a bright arc A rose in the southern side of the dome, followed by a series of slowly rising loops that reached the BCS and disappeared (Fig.\ \ref{fig7}(d) and accompanying animation). One of these bright loops, marked A$^\prime$ (Fig.\ \ref{fig7}(f)), rose at $v \approx 2$-8 \kms (Fig.\ \ref{fig8}(d)) but did not disappear. Instead, A$^\prime$ transformed into a quasi-circular feature during 10:02-10:12 UT (Fig.\ \ref{fig7}(f,g)), coincident with the onset of brightenings beneath it. We interpret this feature as a flux rope (FR) that was formed initially by slow flare reconnection. The rising FR accelerated to $v \approx 90$ \kms during 10:04-10:14 UT (Fig.\ \ref{fig8}(d), track A$^\prime$), signifying the transition from slow to fast flare reconnection. At the same time, a compressed arc that we identify as a side arcade (SA) became visible in the northern part of the dome (Figs.\ \ref{fig7}(h) and \ref{fig8}(a,f)).

\subsubsection{First Eruption}

According to the GOES soft X-ray light curves, the first E2 eruption was associated with an M2.2 class flare that started at 10:17 UT, peaked at 10:40 UT, and ended at 10:59 UT. We use the mean relative intensity profile extracted from a 131 \AA\ image (red box in Fig.\ \ref{fig9}(c)) as a proxy for the flare (red curves in Figs.\ \ref{fig8}(d) and \ref{fig9}(d)). The EUV intensity increased from 10:16 to 10:40 UT, starting a few minutes after the front of the FR encountered the BCS; at the same time, a tiny blob was ejected from the middle of the BCS (Fig.\ \ref{fig7}(g)).  We attribute this activity to the onset of explosive flare and breakout reconnection. A fast EUV wave appeared ahead of the FR (Fig.\ \ref{fig7}(h)) and propagated toward higher latitudes. The 1600 \AA\ base-difference image at 10:29:04 UT (Fig.\ \ref{fig7}(i)) reveals a very bright flare ribbon at the east limb and a larger, dimmer, extended elliptical ribbon (width $w \approx 200 \arcsec$) produced by the breakout reconnection.  Although rising arcade loops and the FR were visible in the 193 \AA\ channel, the slow flare reconnection under the evolving FR and associated downflows were best observed in the hotter channels (131 and 94 \AA). The first sequence of the animation accompanying Figure \ref{fig9} shows the onset of fast flare reconnection underneath the rising arcades at about 10:04 UT, leading to the formation and fast rise of the FR (Fig.\ \ref{fig9}(a,b)). The frontal loop (FL) ahead of the FR rose at $v \approx 458$ \kms in the AIA FOV, although the FR itself only reached $v \approx 290$ \kms (Fig.\ \ref{fig9}(d)).  The inclined linear feature between the FR and the flare arcade (FA) probably is the FCS (Fig.\ \ref{fig9}(b)), which moved southward during the eruption and became faint at around 11:15 UT (see animation accompanying Fig.\ \ref{fig9}). This eruption involved the southern portion of the filament channel, but some initial activation of the northern part is visible from about 10:00 UT on, e.g., the feature labeled SA (Fig.\ \ref{fig7}(f-h)). 

\begin{figure*}
\centering{
\includegraphics[width=15cm]{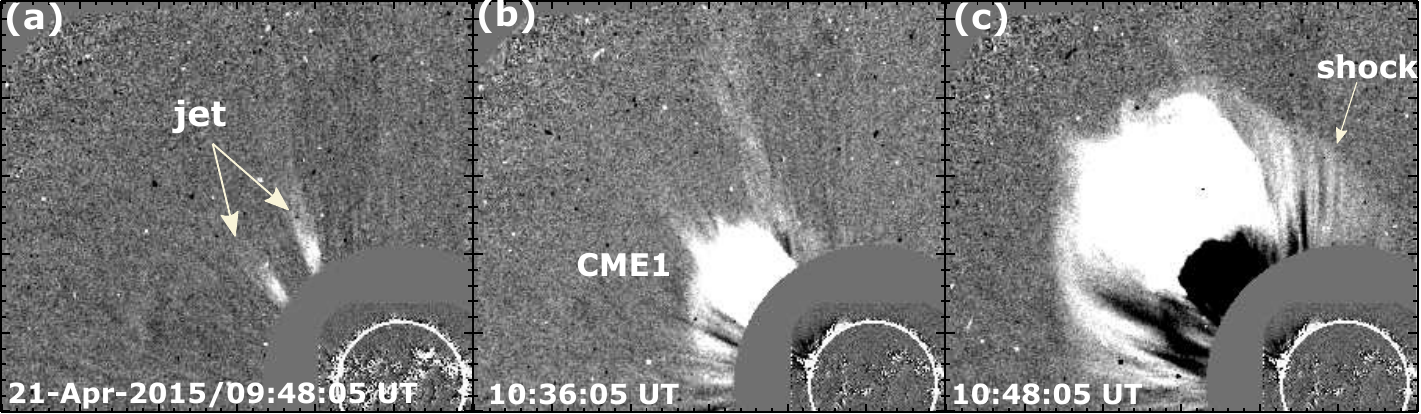}

\hspace{0.02cm}
\includegraphics[width=15cm]{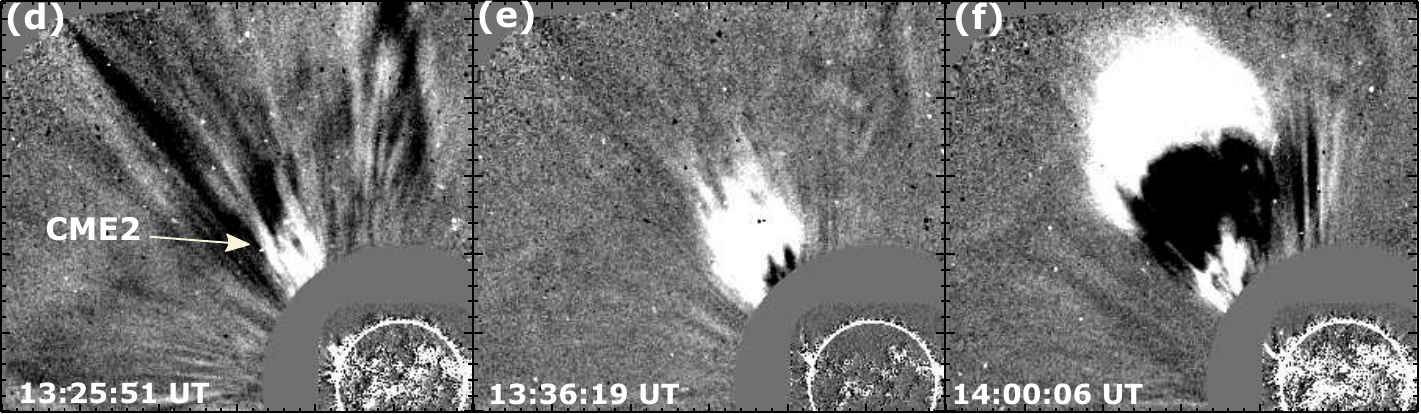}
}
\caption{\emph{SOHO}/LASCO C2 coronagraph running-difference images at 6 selected times during E2, showing (a) pre-eruption jets, (b,c) CME1 ($v = 2039$ \kms), (d-f) CME2 ($v = 1079$ \kms). (An animation of this Figure is available online.)} 
\label{fig10}

\end{figure*}

Quasiperiodic warm downflows ($v \approx 72$ \kms) were observed for several minutes during the peak and decay phase of the first eruptive flare (Fig.\ \ref{fig8}(d)). This eruption produced a fast partial-halo CME with $v \approx 2039$ \kms at 10:36:05 UT in the LASCO C2 field of view (Fig.\ \ref{fig10}(b,c)). A white-light front propagating laterally in the northern direction was correlated with the fast EUV wave. Although this eruption did not include a filament, the existence of a flare arcade indicates that a filament channel was present \citep[see, e.g.,][]{kumar2018}. 
%%%%%%%%%%%%%%%%%%%%%%%%%%%%%%%%%

%%%%%%%%%%%%%%%%%%%%%%%%%%%%%%%%%%%%%%%%%%%%%%%%%%%%%%%%%%%%%%%%%%%%%
\subsubsection{Second Eruption}
       
Persistent downflows of hot plasma with the same speed ($v \approx 70$ \kms) were observed for the next 2.5 hours between the two E2 eruptions (Fig.\ \ref{fig9}(d)). The second eruption started some three hours after the first, at about 13:15 UT, and was associated with a weak C-class flare and a filament eruption. The filament F rose slowly ($v \geq 9$ \kms) from about 13:12 UT onward (Fig.\ \ref{fig8}(f)). The time-distance plot along slice S3 (Fig.\ \ref{fig8}(a,f)) shows the slow rise of the arcades (SA, A1) above the filament and the appearance of adjacent bright open features (marked P). Shortly after the first eruption, P moved northward ($v \approx 65$ \kms during 10:30-11:10 UT), then arcade SA expanded slowly ($v \approx 4$-7 \kms) during 11:09-12:40 UT (see 171 \AA\ animation accompanying Fig.\ \ref{fig8}). A series of loops systematically approached the open structures (P and others) near the BCS and disappeared. SA opened up and disappeared around 12:50 UT. Below SA, another rising set of loops (A1) transitioned from $v \approx 14$-32 \kms during 12:45-13:08 UT, then accelerated starting at 12:55 UT. The filament F rose and accelerated from 9 \kms to 190 \kms during 12:54-13:15 UT. After A1 disappeared, the filament erupted with a clockwise untwisting motion and a projected radial velocity $v \approx 410$ \kms (Fig.\ \ref{fig9}(d)). This eruption also produced a fast CME ($v \approx 1079$ \kms), first seen in LASCO C2 at 13:25:51 UT (Fig.\ \ref{fig10}(d-f)). 

%%%%%%%%%%%%%%%%%%%%%%%%table%%%%%%%%%%%%%%%%%%%%%%%%%%%%%%
\begin{longrotatetable}
\begin{longtable}{c c c c c c c c c c c}

\caption{Characteristics of the selected events} \\
\hline \\

\label{tab1}
Event \#   &Active region    &Initial Dome  &Initial Height &Flare & Filament/ &Flux rope&Width & Pre-eruption &Pre-eruption  &CME \\
Eruption      &     &width & of the null$^c$/BCS &class  &Eruption & formation&of the  & opening   &coronal &properties:\\
date      & Photospheric    &(arcsec)        & (arcsec) &      &           &         &hot flux  &with faint &dimming & Linear speed\\
           &field strength     &       &           &     &           &      &rope & jets      & near the& (\kms)\\
           & (min/max values         &       &            &     &           &          &(arcsec) &  & null& Angular\\
           &in G)                   &       &            &      &           &         & &           &         &width\\
\hline
   
1.       &\# No sunspot    & 210       &155        &C2.4 & Yes/No  & Circular feature  &52-64$^\star$ & 1 hr before   & Yes&415, 29$^\circ$ \\    
2015    & (plage)           &         &           &     &        & observed                & &the flare   &&  \\
April 20       &          &           &           &      &       &  (94/131 \AA)         & &         &&  \\
           &($-$780 G, $+$974 G)$^b$                  &           &           &       &      &           & &&   &  \\
           &         &           &           &       &      &         & &   & & \\
           &       &           &           &       &      &         & &   & & \\
   
2.        &\# 12333 & 196       &112        &M2.2 & Yes/No (I) & Circular feature  &120-142&1.2 hr before &Yes&2039, 191$^\circ$(PH$^a$) \\    
2015   &       &           &           &C1.8     &Yes/Yes (II) & observed    &$\dagger$      &the flare   &&1079, 83$^\circ$\\
 April 21       & ($-$1110 G, $+$850 G)$^b$                     &         &           &      &       & (94/131 \AA)         & &         &&\\
 
Sympathetic &                      &           &           &       &      &             & &&   &  \\
eruptions    &            &           &           &       &      &         & &   & & \\
&       &           &           &       &      &         & &   & & \\

3.        &\# 12261 &200        &148        &C3.7 & Yes/Yes  & Inferred from &$\dagger$&2.5 hr before &Yes&1078, 210$^\circ$(PH) \\    
2015       &       &           &           &     &           & brightenings   &      &the flare   &&\\
 January 12   &($-$1025 G, $+$1460 G)                      &           &           &      &       & under filament          & &         && \\
 
    &                      &           &           &       &      &             & &&   & \\
 
\hline

\end{longtable}

\small
\noindent
${}^\star$ Width is estimated during formation and eruption of the circular feature using AIA 131 \AA\ images. Width range is due to expansion of the flux rope during eruption.\\
${}^\dagger$ Unable to measure the flux-rope width due to projection or nondetection of the circular feature.\\ 
${}^a$ PH=Partial halo CME.\\
${}^b$ For these limb events, the HMI photospheric magnetic field was measured three days after the eruption.\\
$^c$ The approximate initial height of the null was measured in AIA 193/171 \AA\ limb images for all three events. Null-point height (above the limb) ranges over $\approx$ 112-155 arcsecs $\approx$ 81-112 Mm $\approx$ 0.11-0.16 Rs.\\
Note: All events were associated with interplanetary type III radio bursts observed by \emph{Wind}/WAVES (1-14 MHz). Burst spectra can be found \href{https://www.dropbox.com/s/cax3rkluuipawsh/ip-typeiii-wind-waves.pdf?dl=0}{here}. 

\end{longrotatetable}

%%%%%%%%%%%%%%%%%%%%%%%%%%%%%%%%%
\begin{figure*}
\centering{
\includegraphics[width=6.2cm]{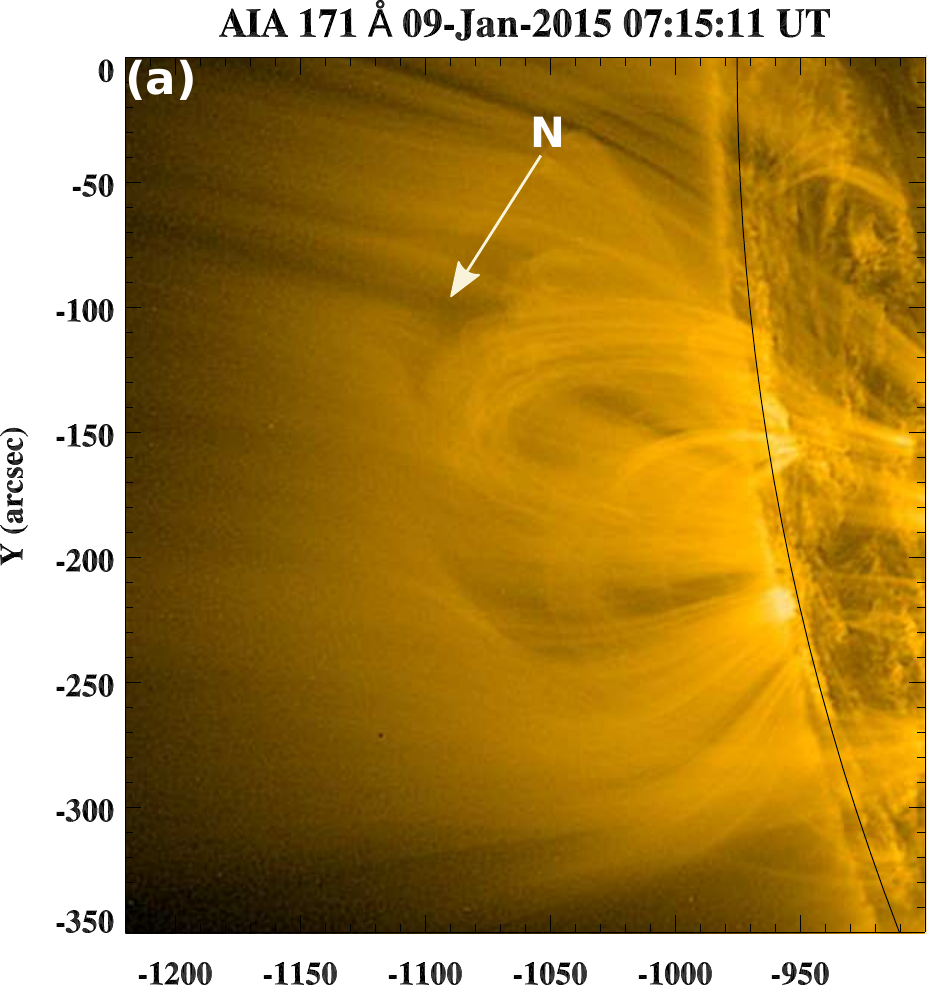}
\includegraphics[width=11.3cm]{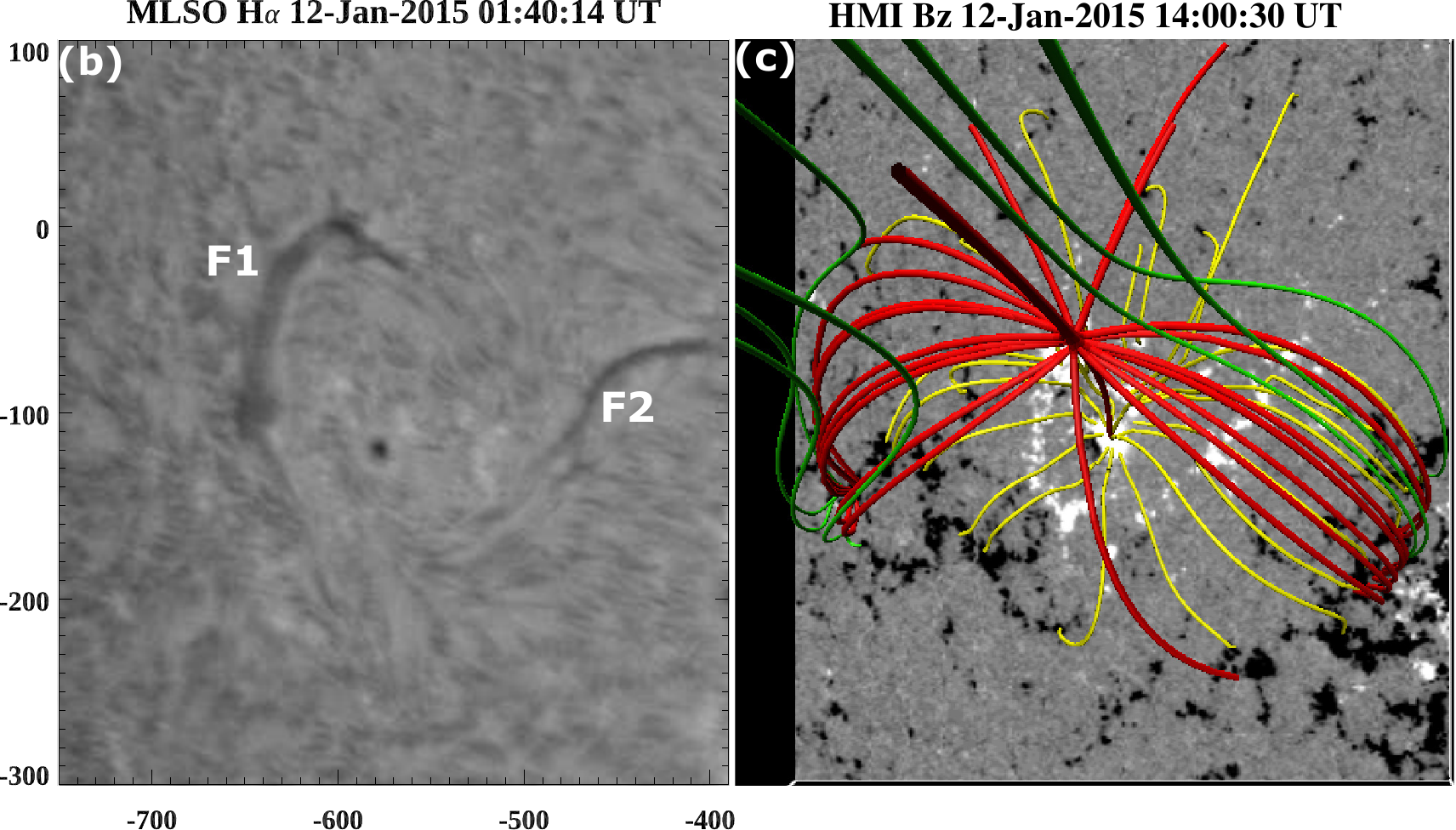}

\includegraphics[width=6.8cm]{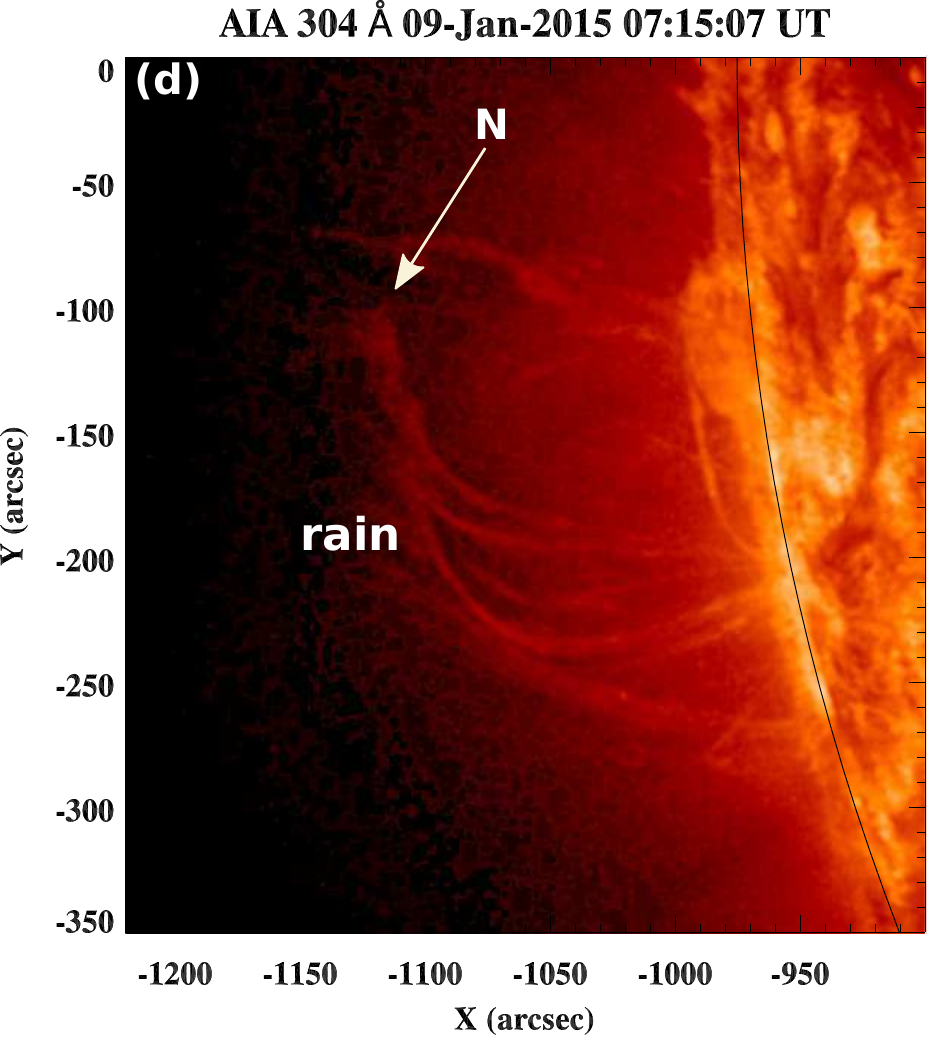}
\includegraphics[width=6.4cm]{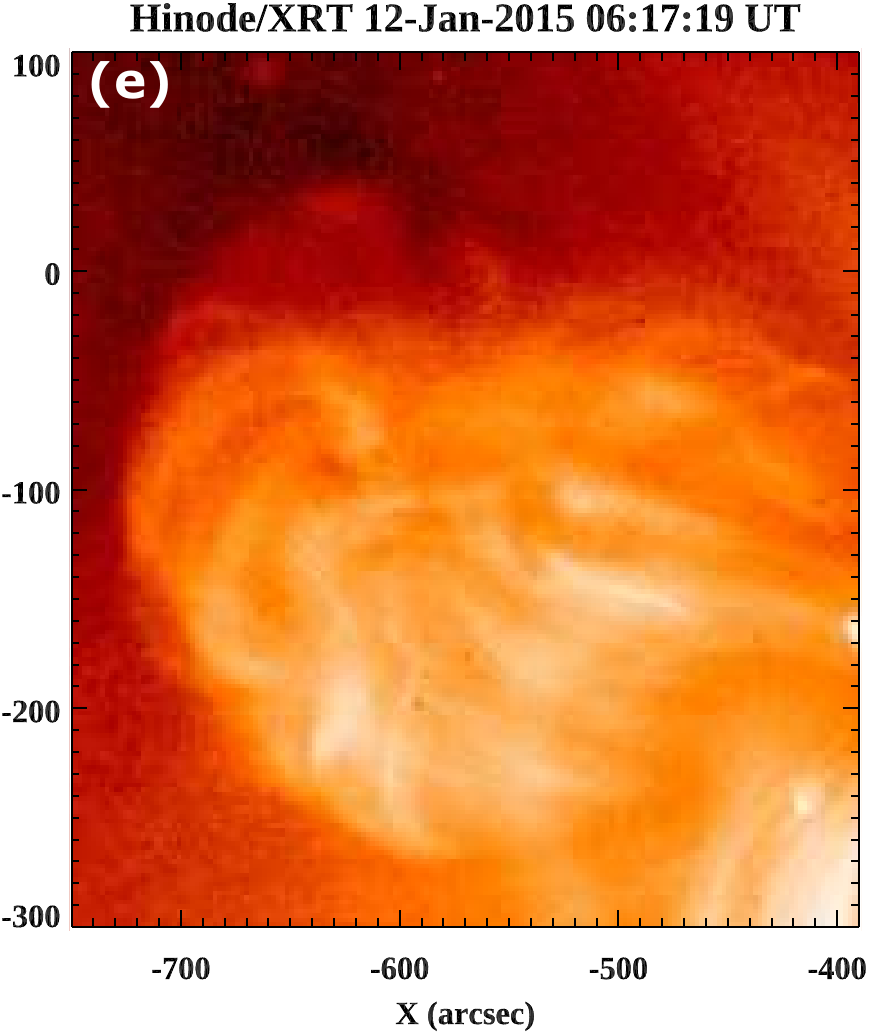}
}
\caption{(a,d) 171 and 304 \AA\ images of the source region for event E3 on 2015 January 09, three days prior to eruption. N = null. (b) H$\alpha$ image from Mauna Loa Solar Observatory at 01:40 UT on 2015 January 12 with filaments F1 and F2 marked. (c) Potential-field extrapolation of the source region at 14:00:30 UT on 2015 January 12. (d) 304 \AA\ image showing coronal rain. (e) Hinode/XRT (Be thin/open) image showing fan loops (T$\approx$2 MK) in the source region at 06:17:19 UT on 2015 January 12.}
\label{fig11}

\end{figure*}

%%%%%%%%%%%%%%%%%%%%%%%%%%%%%%%%%%%%%%%%%%%%%%%%%%%%%%%%%%%%%%%%%%%%%
%%%%%%%%%%%%%%%%%%%%%%%%%%%%%%%%%%%%%%%%%%%%%%%%%%%%%%%%%%%%%%%%%%%%%
%%%%%%%%%%%%%%%%%%%%%%%%%%%%%%%%%
\begin{figure*}
\centering{
\includegraphics[width=17cm]{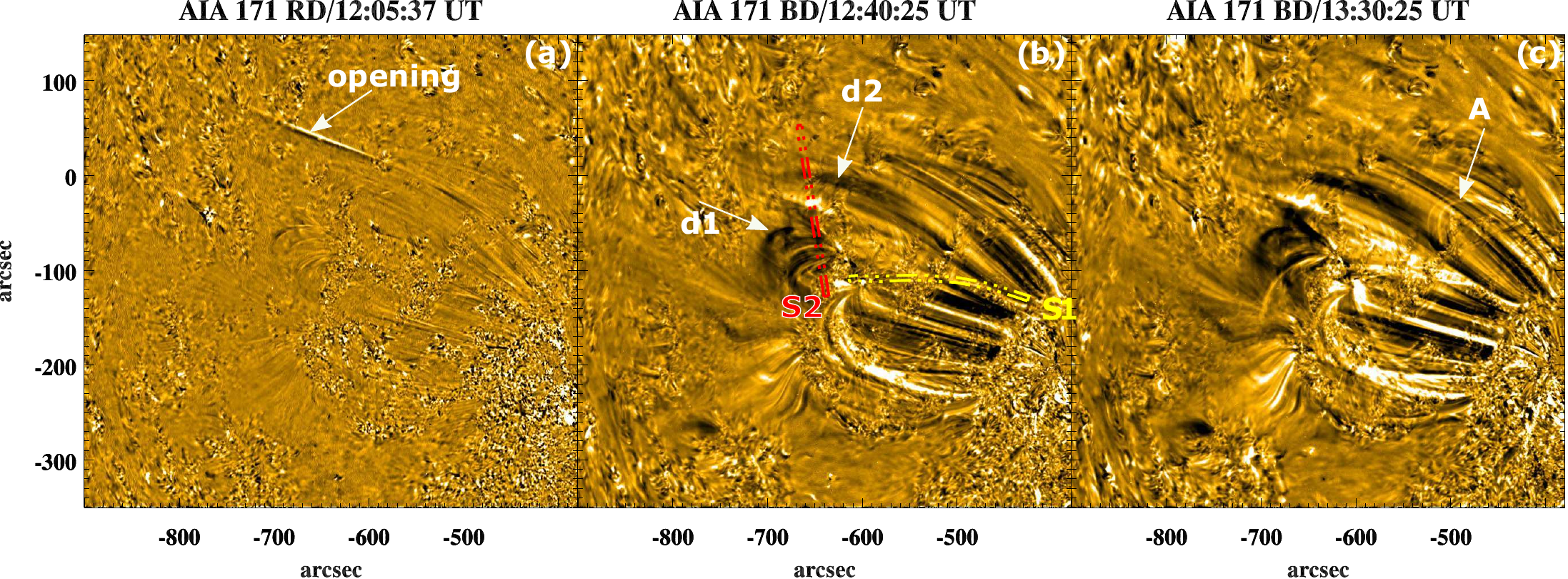}
\includegraphics[width=17cm]{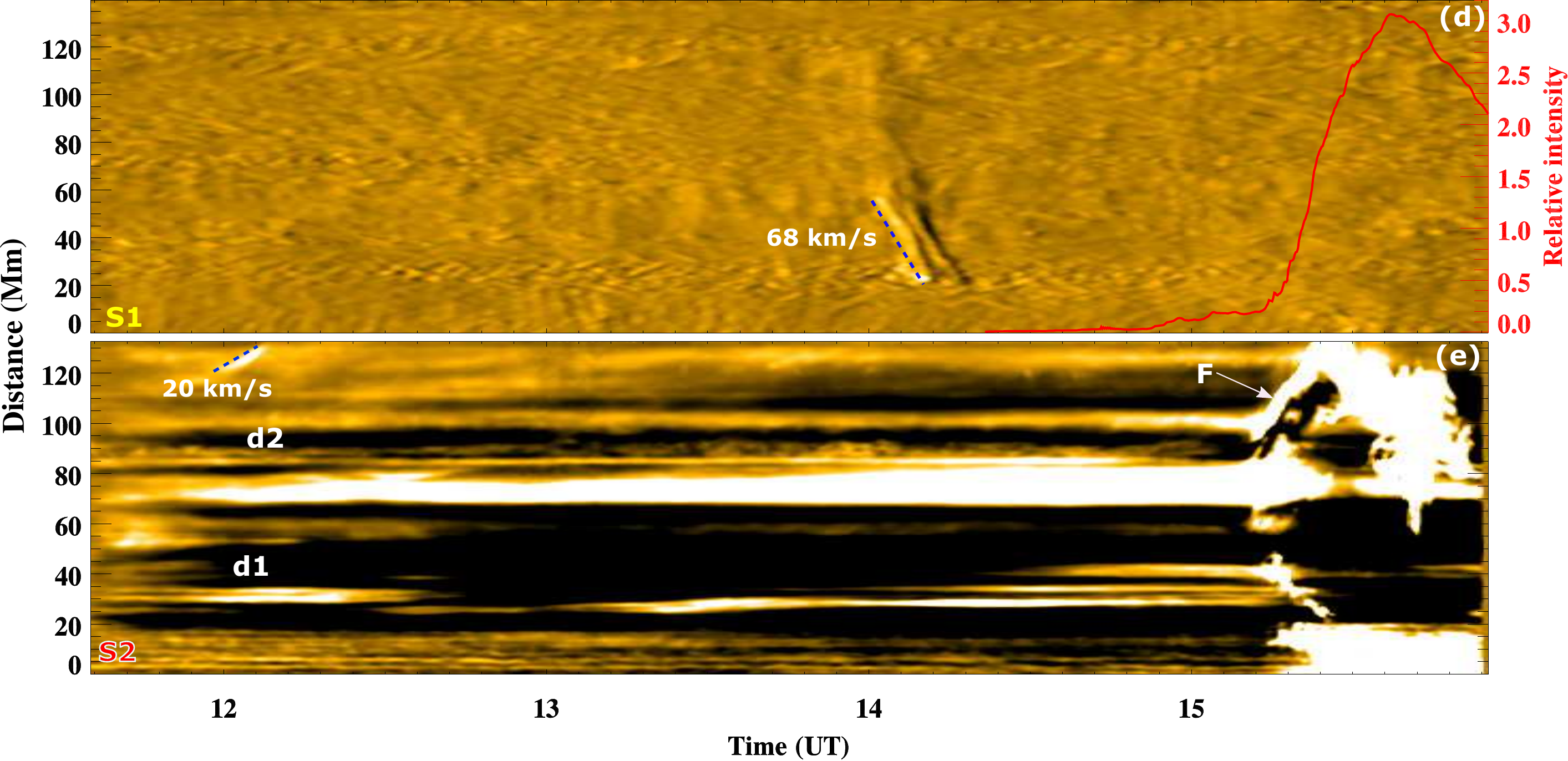}
}
\caption{(a-c) 171 \AA\ running- and base-difference images of the E3 source region, showing pre-eruption opening and dimming signatures d1, d2. A = bright arcade loops. (d,e) Time-distance intensity (running and base-difference) plots along slices S1 and S2 shown in panel b. Red curve in panel d is the 131 \AA\ relative intensity as a proxy of the flare emission. F1 = erupting filament. (An animation of this Figure is available online.)} 
\label{fig12}

\end{figure*}

%%%%%%%%%%%%%%%%%%%%%%%%%%%%%%%%%%%%%%%%%%%%%%%%%%%%%%%%%%%%%%%%%%%%%
%%%%%%%%%%%%%%%%%%%%%%%%%%%%%%%%%
\begin{figure*}
\centering{
\includegraphics[width=5.5cm]{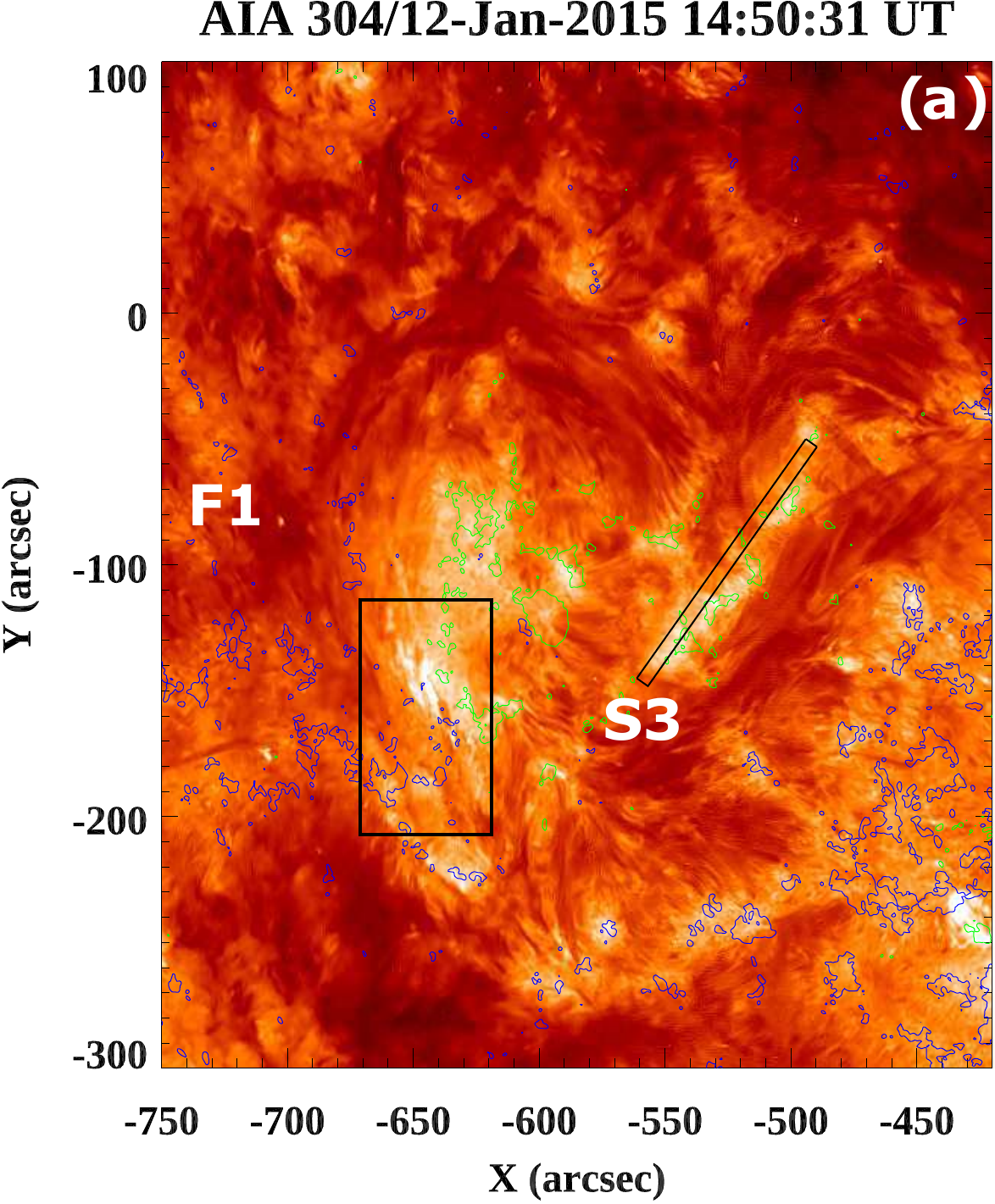}
\includegraphics[width=5.9cm]{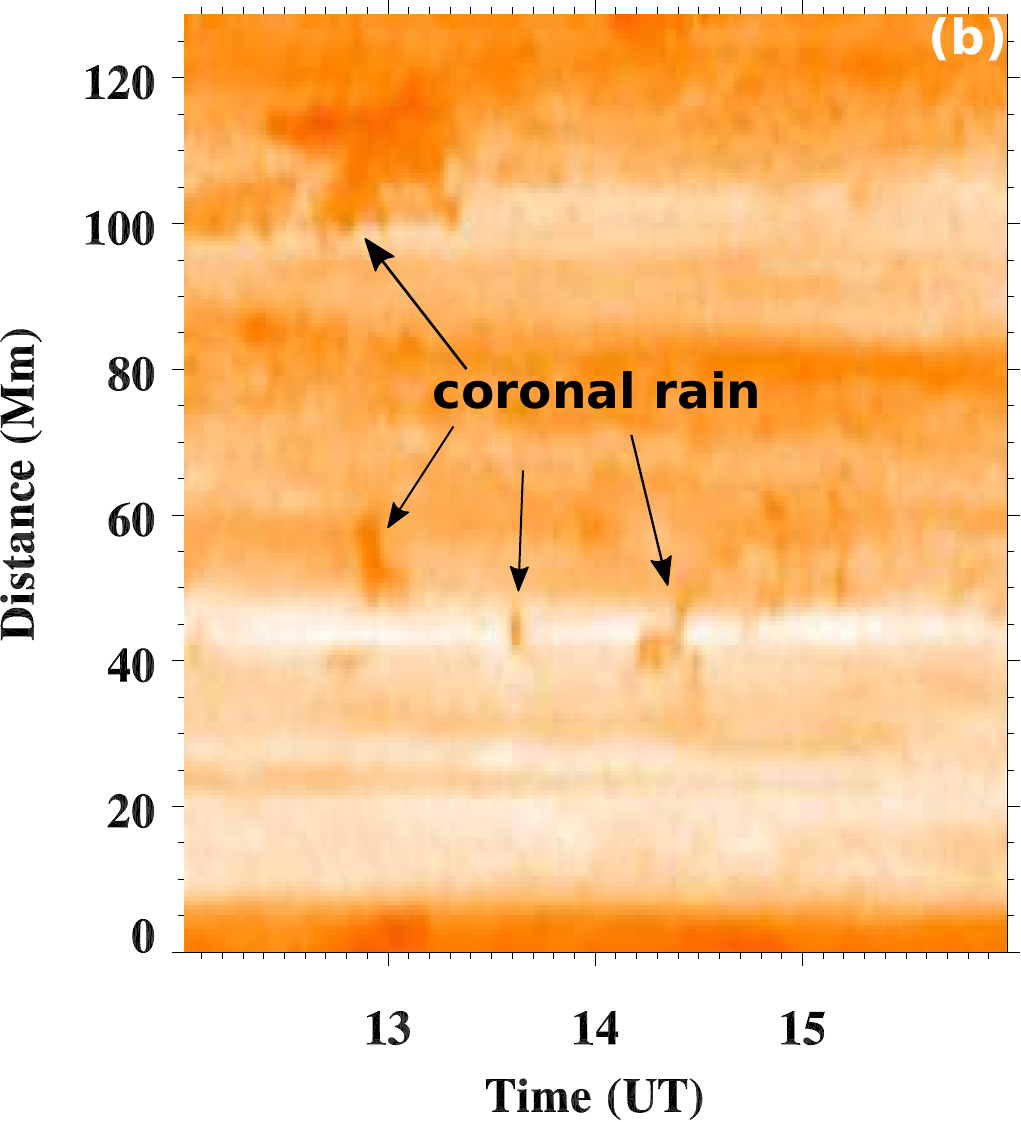}
\includegraphics[width=6.2cm]{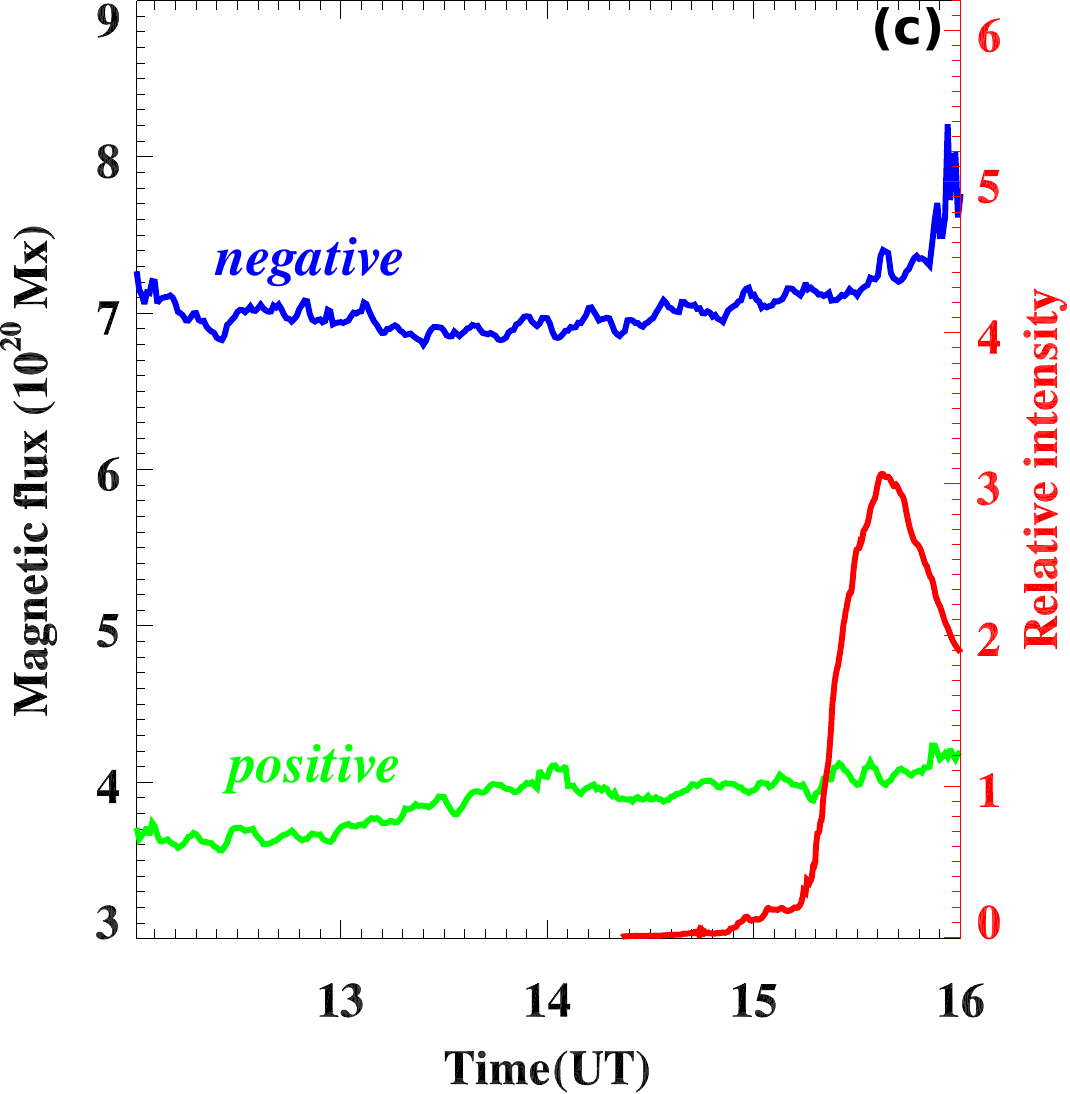}
}
\caption{(a) 304 \AA\ image during initiation of flare reconnection beneath filament F1 in event E3. (b) Time-distance intensity plot along slice S3 showing coronal rain prior to the eruption. (c) Temporal evolution of positive and absolute negative fluxes (within $\pm$50 G contours) from HMI magnetograms, extracted from the black box outlined in panel a. The red curve is the AIA 131 \AA\ relative intensity. (An animation of this Figure is available online.)} 
\label{fig13}

\end{figure*}
%%%%%%%%%%%%%%%%%%

\subsection{Event \#3 (E3)}
Our third event occurred on 2015 January 12 in NOAA AR \#12261 (location S11E32) inside a coronal hole. The 171 \AA\ image on January 9, three days before the eruption, reveals the typical warm emission structure of a pseudostreamer (Fig.\ \ref{fig11}(a)). 
Two inverse-J-shaped filaments (F1 and F2) were situated along the same elliptical PIL (Fig.\ \ref{fig11}(b)), but only F1 erupted in event E3.
 A potential-field extrapolation of the source region displays a clear null-point topology (Fig.\ \ref{fig11}(c)), with a dome width $w \approx 200 \arcsec$ and null height $h \approx 148 \arcsec$. The 304 \AA\ image on 2015 January 9 (3 days before the eruption) shows coronal rain apparently radiating from the null and falling down the fan surface (Fig.\ \ref{fig11}(d)).
The Hinode/XRT image (Fig.\ \ref{fig11}(e)) reveals hot ($\approx$2 MK), bright fan loops radiating from the center of the source region at least 8 hours before the eruption. Note that the loops on the east side above F1, i.e., at the subsequent eruption site, appear more highly sheared than those on the west side.     
  
\subsubsection{Pre-eruption activity}

Early signs of activation were observed over three hours before eruption onset. The 171 \AA\ running-difference images and a time-distance plot created from slice S1 (yellow outline in Fig.\ \ref{fig12}(b)) reveal pre-eruption opening of the field above F1 during 11:53-12:10 UT (Fig.\ \ref{fig12}(a,d) and accompanying animation). The newly opened structure moved northward at $v \approx 20$ \kms, accompanied by the appearance of strong coronal dimmings d1 and d2 above filament F1 (Fig.\ \ref{fig12}(b,e)).  Arcade loops inside the northern part of the dome brightened around 12:52 UT (feature A in Fig.\ \ref{fig12}(c)).  A bright linear feature between d1 and d2 appeared around 11:50 UT and gradually became wider prior to eruption, according to the time-distance plot created from slice S2 of base-differenced images (Fig.\ \ref{fig12}(b,e)). 

Strong downflows ($v \approx 68$ \kms) were observed in slice S1 from 14:00-14:18 UT (Fig.\ \ref{fig12}(d)). Brightenings indicative of slow flare reconnection under F1 started at about 14:33 UT (Fig.\ \ref{fig13}(a) and accompanying 304 \AA\ animation), 15 minutes after the downflows ended. Although coronal rain is difficult to detect on the disk, the Figure \ref{fig13} animation and a time-distance plot (Fig.\ \ref{fig13}(b)) created from slice S3 (Fig.\ \ref{fig13}(a)) clearly show coronal rain falling along the fan loops from 12:30 UT onward, coincident with the strong coronal dimming. Rain also was seen just before the onset of slow flare reconnection below F1 during 14:15-14:32 UT (Fig.\ \ref{fig13}(b)). 

To quantify the evolution of the photospheric magnetic field underneath F1, we extracted positive (green) and negative (blue) fluxes from HMI line-of-sight magnetograms during 12:00-16:00 UT using $\pm 50$ G contours within the black rectangular box in Figure \ref{fig13}(a). The magnetograms and derived flux profiles (Fig.\ \ref{fig13}(c) and animation) do not show any evidence of significant flux emergence or cancellation during a 3 hr interval before the flare. The magnetic-field evolution of our other events could not be measured because they occurred at the limb. 

\subsubsection{Eruption}

\begin{figure*}
\centering{
\includegraphics[width=6.4cm]{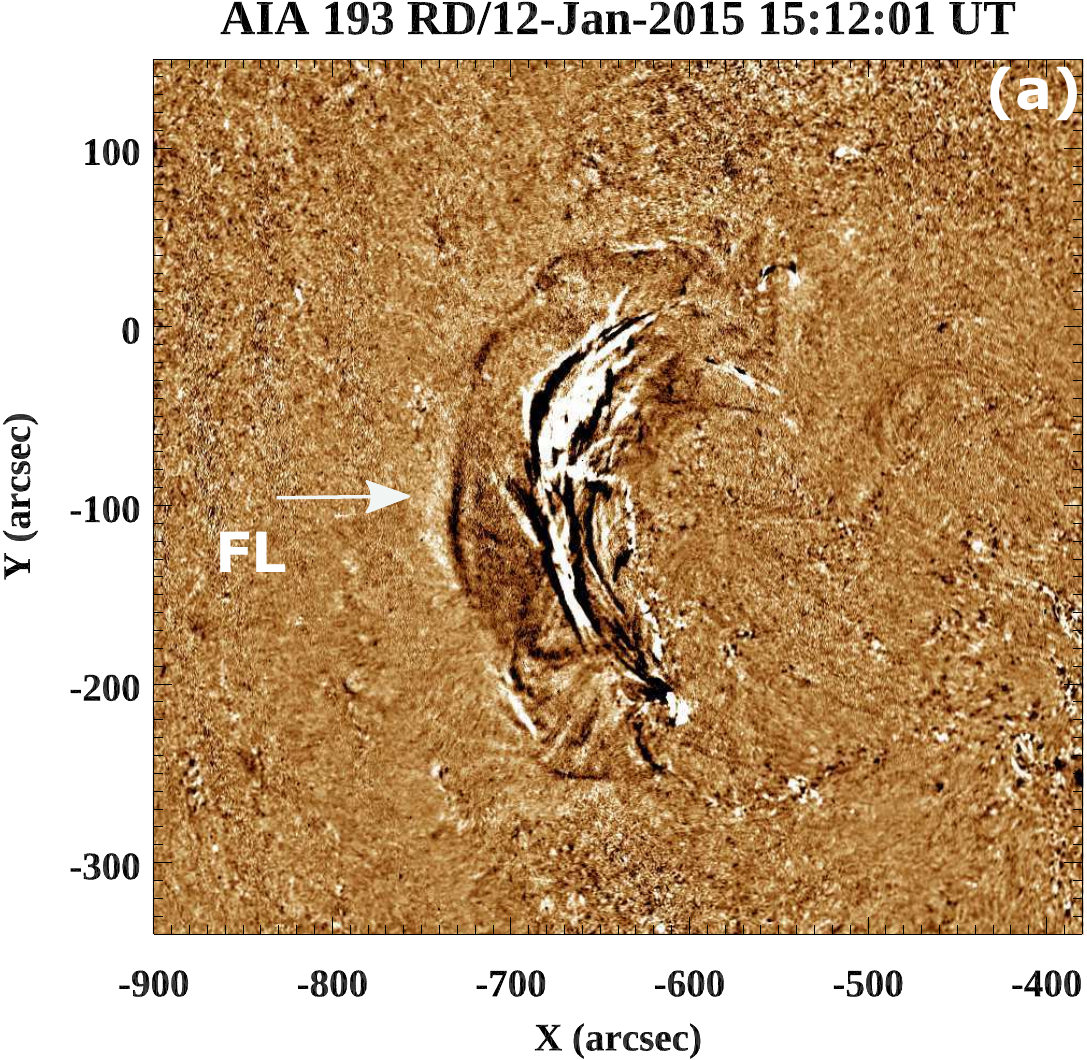}
\includegraphics[width=5.5cm]{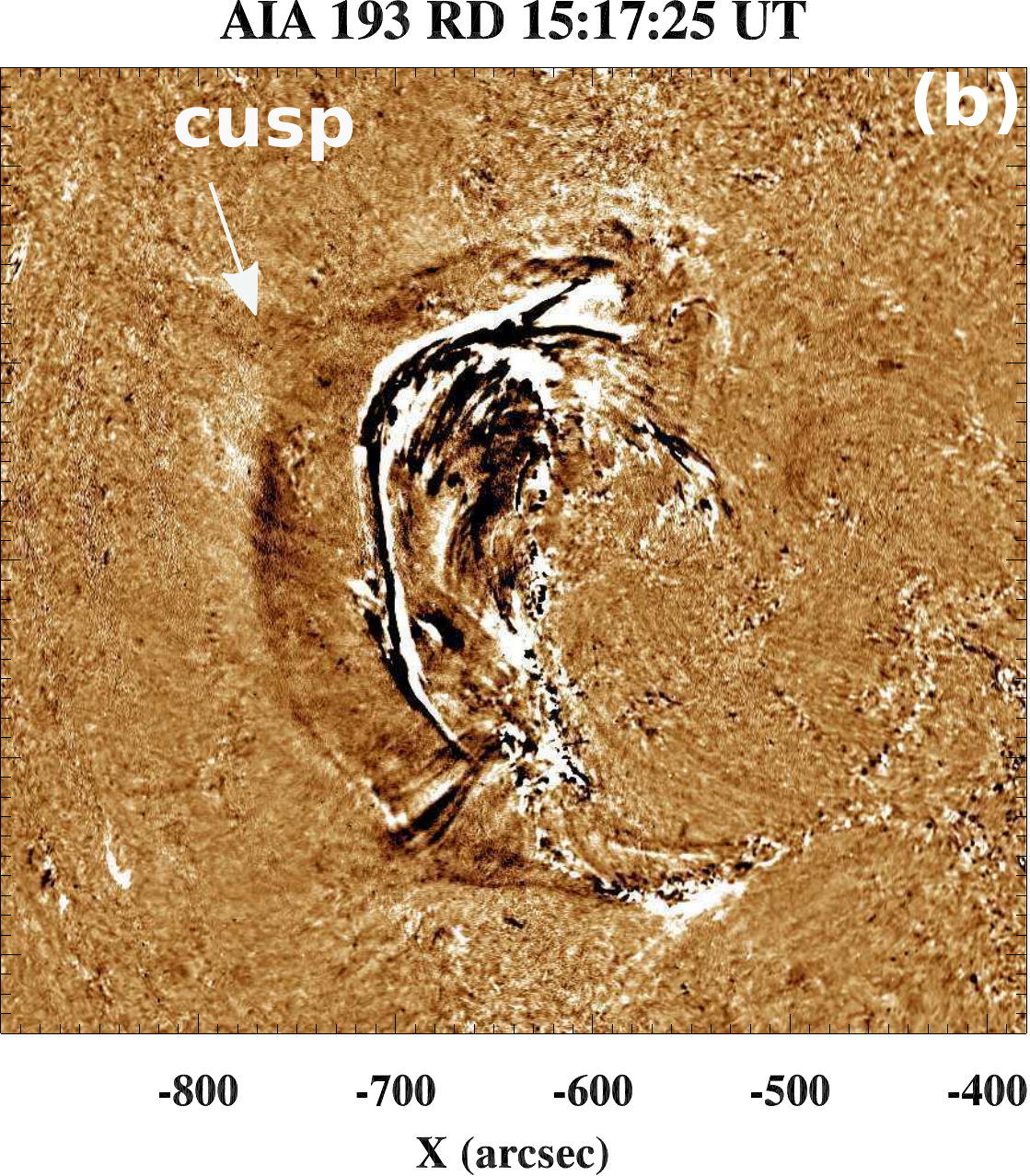}
\includegraphics[width=5.5cm]{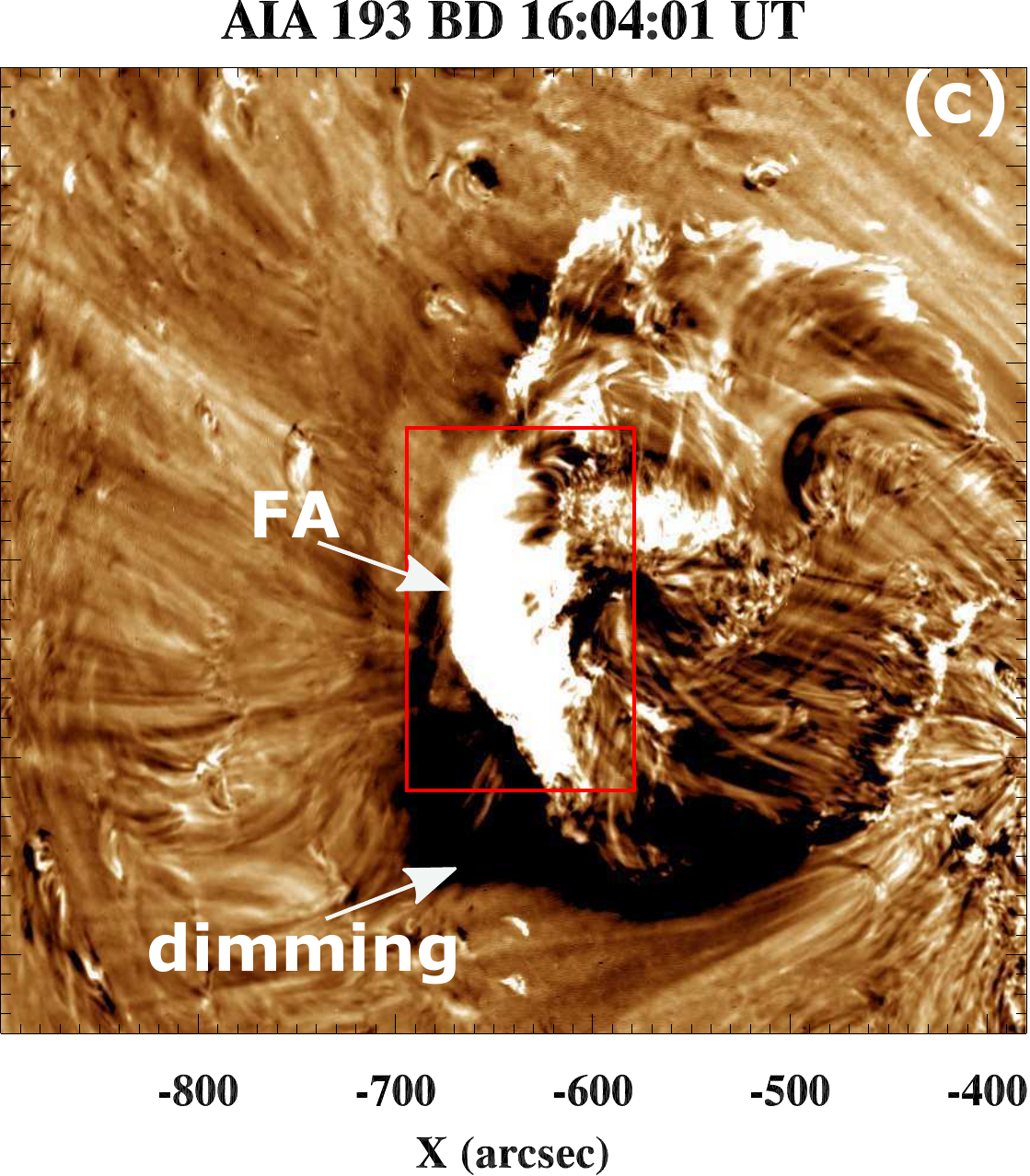}

\includegraphics[width=16cm]{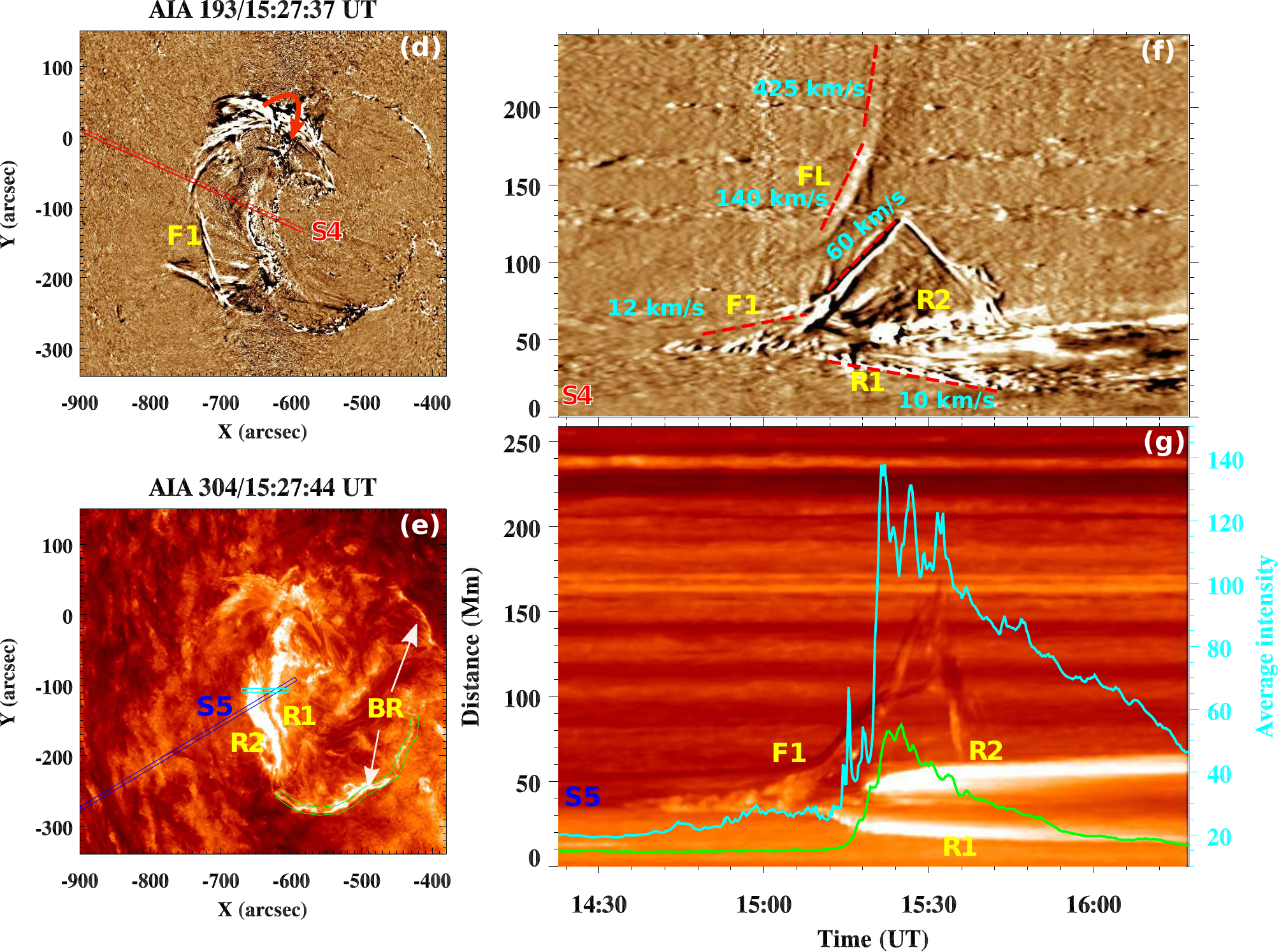}
}
\caption{(a-d) 193 \AA\ running- and base-difference images during and after the E3 eruption. FL = frontal loop, FA = flare arcade, F1 = filament. (e) 304 \AA\ image. The cyan (green) line surrounds the part of the flare ribbons (breakout ribbon) used to measure the average intensity vs time. (f) Time-distance plot created from slice S4, shown in panel d, in the 193 \AA\ running-difference images. (g) 304 \AA\ time-distance plot created from slice S5 shown in panel e. The cyan (green) curve is the average 304 \AA\ intensity in selected portions of the flare ribbons R1 and R2 (breakout ribbon BR) outlined in panel f. (An animation of this Figure is available online.)} 
\label{fig14}

\end{figure*}

From 14:42 to 15:06 UT, the filament F1 rose slowly with $v \approx 12$ \kms, then accelerated rapidly to $v \approx 60$ \kms during 15:07-15:25 UT (Fig.\ \ref{fig14}(d,e)). A frontal loop (FL) appeared ahead of the filament during the later interval, initially moving outward at $v \approx 140$ \kms then reaching $v \approx 425$ \kms before disappearing from slice S4 (red outline in Fig.\ \ref{fig14}(d)) around 15:22 UT (Fig.\ \ref{fig14}(a,e)). A cusp-shaped structure appeared above FL at 15:17:25 UT (Fig.\ \ref{fig14}(b)). 

Two flare ribbons, R1 and R2, separated slowly ($v \approx 10$ \kms) as the filament rose swiftly, until 15:55 UT (Fig.\ \ref{fig14}(f)). Both the flare ribbons and a breakout ribbon (BR) are displayed clearly in the 304 \AA\ images and a time-distance intensity plot along slice S5 from about 15:15 UT on (Fig.\ \ref{fig14}(f,g) and animation). The circular breakout ribbon was not visible during the pre-eruptive phase, but it brightened at the same time as the flare ribbons, indicating that fast flare and breakout reconnection were triggered roughly simultaneously. Figure \ref{fig14}(g) also reveals brightenings beneath the rising filament in slice S5, starting around 14:35 UT and continuing until the rising filament twists and exhibits signs of internal heating. Interestingly, the flare ribbon intensity (cyan curve in Fig.\ \ref{fig14}(g)) contains a quasi-periodic pulsation with a period of about 5 min. Over time, the narrow BR extended in the counterclockwise direction from south to north, evidently mapping the extent of flux transfer due to breakout reconnection. 

\begin{figure*}
\centering{
\includegraphics[width=13cm]{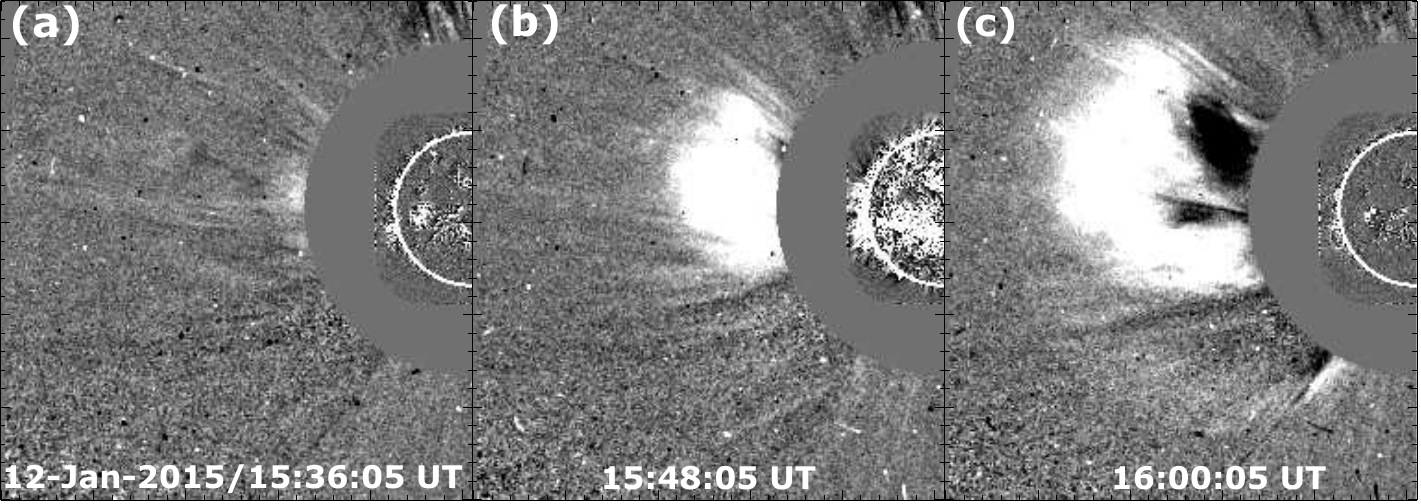}
}
\caption{LASCO C2 coronagraph images showing the CME ($v = 1078$ \kms) associated with E3.} 
%(An animation of this Figure is available online.)} 
\label{fig15}

\end{figure*}

This filament eruption was asymmetric: the northern leg of F1 (which appeared to rotate clockwise) mostly exhibited cool plasma draining toward the surface, while much of the plasma in the southern leg apparently was expelled outward. Strong coronal dimming was observed in the 193 \AA\ images southeast of the flare ribbons during the decay phase of the eruptive flare (Fig.\ \ref{fig14}(c)). The filament eruption accompanied a fast partial halo CME ($v = 1078$ \kms) observed by the LASCO C2 coronagraph (Fig.\ \ref{fig15}).  A faint shock appeared ahead of the CME front in the coronagraph images, which also show that the shock traveled much farther southward than the CME itself and deflected a streamer at high southern latitudes.
  %%%%%%%%%%%%%%%%%%%%%%%%%%%%%%%%%

%%%%%%%%%%%%%%%%%%%%%%%%%%%%%%%%%%%%%%%%%%%%%%%%%%%%%%%%%%%%%%%%%%%%%
%%%%%%%%%%%%%%%%%%%%%%%%%%%%%%%%%

%%%%%%%%%%%%%%%%%%%%%%%%%%%%%%%%%%%%%%%%%%%%%%%%%%%%%%%%%%%%%%%%%%%%%
    
\section{DISCUSSION}\label{discussion}
Event E1 exemplifies a hybrid jet-CME event: the jet initially resulted from slow breakout reconnection, which enabled the formation of a hot flux rope ($T \approx 10$ MK) via flare reconnection and the ultimate expulsion of a narrow, slow CME. The opening of field lines at the BCS started about one hour prior to the flare reconnection, accompanied by coronal dimmings above the BCS. At the same time, a localized region of hot plasma appeared near but not coincident with the dimming regions, indicating that the plasma in the dimming regions was depleted and not merely heated out of the warm bandpasses. Pre-eruption opening and pre-flare heating near the BCS are consistent with breakout reconnection playing a key role in triggering the eruption. The LASCO C2 images reveal a clear association between the jet and the pre-eruption openings on either side.  

During the impulsive phase, explosive breakout reconnection above the flux rope and flare reconnection below it occurred almost simultaneously, as predicted by the MHD simulations \citep{lynch2013}. Both current sheets (BCS and FCS) were detected in this event; multiple hot plasmoids propagated bidirectionally in the FCS, but only a single plasmoid was observed in the BCS. Quasiperiodic downflows below the BCS were observed prior to and during the encounter of the flux rope with the BCS, while quasiperiodic upflows and downflows were detected in the FCS (periods: 2.5, 3.2, and 6.4 min) for about 1 hour during the flare reconnection. Another distinct signature of bursty reconnection, small magnetic islands, appeared in the FCS \citep[e.g.,][]{guidoni2016}. This source region contained a stationary filament that did not erupt: the flux rope was seen only in the AIA hot channels (94/131 \AA), while the AIA 304/171 \AA\ images did not show any signs of filament eruption. Two thermal X-ray sources appeared in the \emph{RHESSI} images: the lower source was situated at the flare arcade loop-top, while the upper source most likely originated in heated plasma near the BCS. The eruption produced a narrow, slow CME that appeared to transition smoothly from the initial jet.  These characteristics indicate that the flux rope did not survive its escape intact, but rather lost a significant amount of flux during breakout reconnection.

Event E2 is a perfect example of sympathetic eruptions with interesting differences. The first eruption occurred on the eastern side of the elliptical PIL, while the second eruption disrupted the remainder of the filament channel. We detected the following pre-eruption activities before the first of the paired eruptions: (i) a series of rising loops (A) that reached the null and disappeared, simultaneous with the onset of faint jets seen in the 193 \AA\ and coronagraph images; (ii) pre-eruption opening about 1.2 hours prior to flare onset; (iii) coronal dimming above the BCS; (iv) slowly rising loops (A$^\prime$) below A that did not disappear, beginning about 2 hours prior to the flare; and (v) jets and tiny plasmoids along the outer spine. These phenomena are definitive signatures of slow breakout reconnection and its immediate consequences, before fast energy release sets in. 

Warm frontal loops started to rise as slow flare reconnection and associated heating occurred in the eastern part of the filament channel,  consistent with the creation and acceleration of the circular feature identified as a flux rope (FR). The series of frontal loops ahead of the FR did not heat up, indicating that they were not incorporated into the flux rope through flare reconnection. As explosive breakout reconnection above the flux rope produced a large elliptical ribbon (diameter $d \approx 200 \arcsec$ along the limb), explosive flare reconnection at the FCS produced strong, long-lasting (about 2.5 hours) downflows and a slowly growing arcade below the erupting flux rope.  The first E2 eruption produced a fast partial-halo CME (2039 \kms) without disturbing the underlying filament segment, so no filament material was ejected. During this interval, however, slowly rising arcades (SA) north of the first eruption reconnected with the open structures near the first BCS, removing flux above the other part of the filament and setting the stage for the sympathetic eruption. 

The second E2 eruption started when sufficient overlying flux was eaten away by breakout reconnection, about three hours after the first eruption. The filament became activated (flows along the threads) and started accelerating approximately one hour before flare onset. Interchange reconnection between the flux rope and the adjacent open flux enabled untwisting motions and eruption of the filament. Similar motions are frequently detected in helical jets during the explosive breakout reconnection phase, when the filament-supporting flux rope reconnects with the external field through the BCS \citep{wyper2017,wyper2018,kumar2018,kumar2019a,kumar2019b}. The second filament eruption produced another fast CME, traveling at approximately half the speed of the first.  E2 closely follows the sympathetic eruption scenario shown in Figure 1, without the idealized symmetry of the MHD simulation: here only part of the filament erupted during the second eruption, indicating that the flare reconnection site in the first eruption was probably located above the northern segment of the filament. This also emphasizes that filament channels are not necessarily continuous structures with a single axis, but rather consist of segments that can act independently, in agreement with prominence seismology results \citep{luna2014}.

Our E2 observations are inconsistent with the simulation by \citet{torok2011} of sympathetic pseudostreamer eruptions. That simulation began with highly twisted, pre-existing flux ropes underneath the pseudostreamer lobes, and the null-point (breakout) reconnection was triggered by a nearby CME. In the E2 events, as in the simulation by \citet{lynch2013}, the pre-eruption configuration did not exhibit or require high twist, the breakout reconnection was driven by expansion of the conjugate pseudostreamer lobes, and no nearby CME was observed or required. 

Event E3 first exhibited pre-eruption activities (e.g., opening) around 2.5 hours prior to flare onset. The opening was followed by strong dimmings near the BCS and brightening of arcade loops inside the dome. Interestingly, during the onset of coronal dimming, persistent coronal rain was observed along the fan loops. All of these observational signatures are associated with slow breakout (interchange) reconnection. In addition, the HMI magnetogram did not show any significant flux emergence or cancellation at the pre-flare brightening site during three hours before the eruption. Therefore, flux emergence or cancellation did not trigger the eruption, in agreement with our previous studies of coronal jets \citep[e.g.,][]{kumar2019b}. Transient brightenings below the filament indicate that reconnection built the twisted flux rope around it prior to eruption. The clockwise rotation of the north leg of the filament suggests that the flux rope was right handed, which is consistent with the chirality deduced from the filament barbs in the H$\alpha$ image (Fig.\ \ref{fig11}(b)). Fast reconnection underneath the flux rope produced two diverging ribbons connected by a flare arcade, while fast breakout reconnection above the flux rope produced a circular breakout ribbon at the footpoints of the separatrix. The flux rope surrounding the filament erupted as a fast CME, but only the southern part of the filament was ejected with it; the plasma in the northern leg and part of the southern leg drained back toward the surface.  
 
All three events have several characteristics in common. As shown in Table 2, their initial pseudostreamer configurations have similar null heights, dome widths/circular ribbon diameters, and radial photospheric field strengths. Persistent coronal rain fell over part or all of the fan for 1--3 hours before the eruptions, although the relationship between coronal rain and subsequent eruption remains unclear. During the main impulsive phase, flare arcades and ribbons appeared and metric and interplanetary type III radio bursts were detected. The sequence of pre-eruptive interchange (breakout) reconnection, elevation of core flux, flux rope formation through flare reconnection, accelerated flare and breakout reconnection when the flux rope encountered the BCS, and eruption of part or all of the flux rope is equally applicable to all, with the added interest of a sympathetic eruption in event E2.  Non-radial propagation and untwisting of the hot flux rope within the dome are consistent with our simulations of breakout jets/CMEs \citep{lynch2016,wyper2017}, and have been observed in large-scale pseudostreamer eruptions \citep{wang2018}. In summary, a similar sequence of activity is seen in small-scale jets and these large-scale jets and CMEs, strengthening our contention that breakout is a universal mechanism for eruptions on all scales \citep{wyper2017,wyper2018}.      

However, there appears to be a threshold for CME formation, in that all of the small embedded bipoles with line-of-sight photospheric field strengths below 100 G studied in our earlier investigations formed jets rather than structured bubble-shaped CMEs. Coronal-hole jets have been found to extend into the LASCO C2 coronagraph field of view, i.e., white-light jets with angular width 3--7$^\circ$ \citep{wang1998,nistico2010,moore2015,kumar2018}.
For coronal-hole jet source regions, the line-of-sight photospheric magnetic-field strengths are a factor of 20 or more weaker than for typical ARs, and the initial null heights are about 5 times lower than the CME-producing regions discussed here. On the other hand, many active-region periphery jets are also associated with CMEs; these source regions have much stronger magnetic fields than the CH jets. In one example \citep{kumar2016}, an eruption originating in a large fan/spine configuration began with a broad jet, followed by a classic 3-part CME containing an erupting filament.  Our prior analyses concentrated on either jets or CMEs, however, and did not include transitional events such as E1. Therefore, the present study is focused on narrowing down the options dictating which source regions could produce jets or CMEs, or both. 

For all three events, the coronal rain started many hours before, increased in intensity, and disappeared shortly before the eruption. In the limb events (E1 and E2), intense rain was mainly detected along the separatrix and fan loops located on the eruption side (no rain in the other half side of the dome) during the early pre-eruption phase. The rain continuously decreased during the slow rise of the flux rope, and completely disappeared shortly before the explosive breakout/flare reconnection. In E3, coronal rain disappeared during the slow rise of the filament, which is presumably supported by the rising flux rope. In all events, moreover, the pre-eruption dimming and intense coronal rain occur simultaneously. These observations suggest that pre-eruption opening via interchange reconnection causes a rapid increase in the coronal rain. However, a dedicated statistical study of a large sample of events is required to firmly establish the physical connection between coronal rain and the onset of eruption.
         
\section{CONCLUSIONS}\label{conclusions}
For the first time, we report clear observational evidence of pre-eruption opening and coronal dimming (plasma depletion) as signatures of slow breakout reconnection prior to the onset of eruptive flares. A rapid increase in coronal rain at the eruption side of the dome also was detected during the pre-eruption phase, which supports the hypothesis that slow interchange reconnection at the breakout current sheet can create coronal rain while opening the closed pseudostreamer flux \citep{mason2019}.
 
Observations of distinct breakout signatures are extremely important for testing 3D models of jet and CME initiation in multipolar flux systems. \citet{wang2015} studied 10 pseudostreamer eruptions and concluded that all are essentially fan-like jets that were not produced by interchange reconnection. In our case studies, however, pseudostreamers (the simplest multipolar system) produced jets as well as slow vs fast, narrow vs wide CMEs via the breakout mechanism. In all three events, the eruption was triggered by slow breakout reconnection well before the onset of explosive flare reconnection. The new diagnostics of pre-eruptive breakout complement the previously identified set of characteristics that signal the explosive phase of breakout reconnection: circular/elliptical breakout ribbons \citep[e.g.,][]{lim2017,doyle2019,lee2020}, magnetically connected remote EUV brightenings \citep[e.g.,][]{sterling2001}, interplanetary type III radio bursts from energetic electrons escaping along open field lines \citep{masson2013,masson2019,kumar2017}, and meter-wave sources higher in the corona than the associated flare emission \citep[e.g.,][]{aurass2013}. The observational results also confirm the predictions of our high-resolution MHD simulations of breakout CMEs \citep{lynch2013,dahlin2019}. Similar to our previous jet studies \citep{kumar2018,kumar2019a,kumar2019b}, we found that flux emergence, submergence, and cancellation did not play key roles in triggering these observed eruptions. 

Finally, we return to our initial question regarding the factors determining whether eruptions will be jets or CMEs.  All three events started with faint jets during the pre-eruption phase; Event 1 produced a stronger jet/narrow CME during the explosive phase, while Events 2 and 3 continued as fast CMEs. Out of four CMEs, two were associated with filament eruptions and the eruption of a hot flux rope formed during the flare reconnection, while the remaining two lacked cool filament material but were otherwise similar. The eruptions that included filaments (second eruption in E2, E3) clearly show rotation and disconnection of one leg of the filament in the 304 \AA\ images. For the eruptions without filaments (E1, first eruption in E2), circular features (flux ropes) are visible in the hot channels (131/94 \AA), but leg disconnection is difficult to detect unambiguously because the flux rope is seen only in cross-section. 

By analyzing eruptions from pseudostreamers (large embedded bipoles) for which most key properties and surroundings were comparable, we eliminated several factors as discriminants between sources of jets and CMEs: source size, photospheric field strength $\left\vert B_{radial}\right\vert$, environment (e.g., CH location and average surrounding CH field strength), and pre-eruptive evolution. However, we are not implying that these factors play no role in determining whether an eruption will be a jet or a CME. For example, if an eruption originates from a tiny embedded bipole in a coronal hole, then of course it will manifest as a jet; conversely, if it originates from a giant active region with a filament channel that is many Mm long, then it will inevitably become a CME. The key to our study is that we deliberately chose intermediate-sized regions capable of producing either jets or CMEs in order to isolate the salient property determining the type of event. 

Because we did not observe any clear features distinguishing between the sources of our jets vs CMEs, we conclude that presently unobservable aspects of the competing magnetic forces dictate the ultimate fate of the erupting filament-channel flux. From both observations and theory, however, we deduce two critically important factors: the amount of flux in the erupting flux rope, and the speed of the eruption.  According to the present study and earlier simulations, both breakout jets and CMEs with filament channels form flux ropes. However, faint jets characterize the initial slow breakout phase, while stronger jets/narrow CMEs appear during the explosive phase when the flux ropes are mostly or completely destroyed \citep{lynch2013,wyper2017,kumar2018}. The important point is that, {\it prior to eruption, removal of restraining flux through breakout and flux-rope formation through flare reconnection basically compete for the same closed flux}. If the overlying flux inside the separatrix is reconnected away through breakout, a small flux rope consisting primarily of filament-channel flux is formed. Such a small flux rope is likely to be destroyed by partial or complete interchange reconnection, yielding a jet. If, on the other hand, a large amount of flux is left in the rope after the outer layers undergo breakout reconnection, then the rope can retain its integrity and erupt as a CME. 

Even if a large flux rope does make it through the BCS, it must survive subsequent propagation in order to produce an observable CME. Because pseudostreamers are situated in unipolar open field, the background flux will always be oppositely directed to the flux rope on one side, and thus susceptible to reconnection there \citep[e.g.,][]{masson2013,masson2019}.  Such reconnection will be of the interchange type, so if it occurs beyond the Alfv\'en point (10-20 R$_\odot$), it may disrupt the integrity of the flux rope but not the amount of closed flux escaping into the heliosphere. If the eruption is slow, of order the solar-wind speed, then interchange reconnection is likely to eat away much of the flux rope before it can reach the Alfv\'en point. For a flux rope to survive, then, its speed must equal or exceed the local Alfv\'en speed, as was the case for the three fast CMEs in our events ($ v >1000$ km s$^{-1}$) but not for the jet/narrow CME in event E1. 

The arguments above imply that, in order to observe a typical ``bubble-like'' CME, the ejected flux rope must be both large and fast, requiring a substantial release of magnetic energy. Consequently, a critical quantity for determining the type of eruption (jet or CME) is the amount of free magnetic energy stored in the sheared filament channel. Also important, however, is the energy that must be overcome by this free energy: specifically, the energy in the overlying closed flux rooted outside the filament channel, and the open flux that bends over the separatrix dome. We conclude that the key criterion determining the type of eruption is the ratio of the energy in the sheared filament channel to the energy of the overlying ``strapping'' field. This conclusion emphasizes that knowledge of the detailed flux distributions are required to determine whether a CME or jet will result from intermediate-scale pseudostreamers. Knowing the width of the filament channel would provide a partial basis for determining the energy ratio, but this requires high-resolution chromospheric observations that are not routinely available for most of the Sun (including our three events), as well as measurements of the transverse component of the filament-channel field responsible for the free energy.  At present, the vector magnetic field can be measured only on-disk at the photosphere, so the field in the corona must be obtained by extrapolations, which are poorly constrained \citep[e.g.,][]{klimchuk1992,derosa2009}. We anticipate that DKIST observations will be able to test our hypothesis by measuring the coronal magnetic fields in and outside the flux rope for selected pseudostreamer eruptions. In parallel, the above conclusion can be tested by a series of numerical simulations in which the fluxes in the key zones are set, measured, and compared, and the results categorized according to the jet-to-CME nature of the eruption.  We are currently planning such an investigation. By narrowing down the factors that could lead to differences in eruptions from similar source regions, we have set the stage for further observational and computational studies that will definitively answer the question: why are some eruptions jets, while others are CMEs?

%%%%%%%%%%%%%%%%%%%%%%%%%%%%%%%%%%%%%%%%%%%%%%%%%%%%%%%%%%%%%%%%%%%%

\acknowledgments
We are grateful to the referee for insightful comments that have improved this paper, and we thank Joel T. Dahlin for stimulating discussions. \emph{SDO} is a mission for NASA's Living With a Star program. This research was supported by PK's appointment to the NASA Postdoctoral Program at the Goddard Space Flight Center, administered by the Universities Space Research Association through a contract with NASA, and by grants from NASA's Heliophysics Supporting Research (now part of an ISFM work package) and Guest Investigator (\#80NSSC20K0265) programs. Wavelet software was provided by C. Torrence and G. Compo, and is available at URL: http://atoc.colorado.edu/research/wavelets/. Magnetic-field extrapolations were calculated and visualized by VAPOR (www.vapor.ucar.edu), a product of the Computational Information Systems Laboratory at the National Center for Atmospheric Research.

%%%%%%%%%%%%%%%%%%%%%%%%%%%%%%%%%%%%%%%%%%%%%%%%%%%%%%%%%%%%%%%%%%%%
%%%%%%%%%%%%%%%%%%%%%%%%%%%%%%%%%%%%%%%%%%%%%%%%%%%%%%%%%%%%%%%%%%%%

\bibliographystyle{aasjournal}
\bibliography{reference.bib}

\end{document}